\documentclass[hidelinks,journal]{new-aiaa}

\usepackage[utf8]{inputenc}
\usepackage{gensymb}
\usepackage{mwe}
\usepackage{graphicx}
\usepackage{amsmath}
\usepackage[version=4]{mhchem}
\usepackage{siunitx}
\usepackage{verbatim}
\usepackage{longtable,tabularx}
\usepackage{xcolor}
\usepackage{booktabs}
\usepackage{caption}
\usepackage{subcaption}

\usepackage{placeins}

\usepackage{bm}
\usepackage{mathrsfs}
\newcommand{\grad}{\mbox{$\nabla$}}

\def\chapterautorefname~#1\null{Chapter~#1\null}
\def\sectionautorefname~#1\null{Section~#1\null}
\def\subsectionautorefname~#1\null{Section~#1\null}
\def\subsubsectionautorefname~#1\null{Section~#1\null}
\def\equationautorefname~#1\null{Eq.~#1\null}
\def\figureautorefname~#1\null{Fig.~#1\null}	
\def\tableautorefname~#1\null{Table~#1\null}

\newcommand{\Autoref}[1]{
	\begingroup \renewcommand\figureautorefname{Figure}
	\autoref{#1}
	\endgroup
}

\newcommand{\RN}[1]{%
\textup{\uppercase\expandafter{\romannumeral#1}}%
}

\definecolor{blue}{rgb}{0,0,0.7}
\newcommand{\revadd}[1]{#1} 

\title{Transitional Fluid-Ablation Interactions on a \\ Compression Ramp at Mach 8} 

\author{Ata Onur Başkaya$^{1}$\footnote{These authors are considered co-first authors contributing equally to this work.}\footnote{\raggedright Ph.D. Candidate, Aerospace Engineering, High Speed Aerodynamics Lab, A.O.Baskaya@tudelft.nl (Corresponding~Author); Student~Member~AIAA}}
\affil{$^1$Delft University of Technology, 2629HT Delft, The Netherlands}

\author{Sean Dungan$^{2}$$^{\text{*}}$\footnote{Ph.D. Candidate, Aerospace Engineering, 3179 Glenn L. Martin Hall Bldg; sdungan@umd.edu; Student Member AIAA}}
\affil{$^2$University of Maryland, College Park, MD, 20742, USA}

\author{Stefan Hickel$^{1}$\footnote{Full Professor, Aerospace Engineering, High Speed Aerodynamics Lab; S.Hickel@tudelft.nl; Member AIAA}}
\affil{$^1$Delft University of Technology, 2629HT Delft, The Netherlands}

\author{Christoph Brehm$^{2}$\footnote{Associate Professor, Aerospace Engineering, 3179 Glenn L. Martin Hall Bldg; cbrehm1@umd.edu; Member AIAA}}
\affil{$^2$University of Maryland, College Park, MD, 20742, USA}

\begin{document}

\maketitle

\begin{abstract}
Direct numerical simulations (DNS) are performed for a Mach 8 flow over a 15$\degree$ compression ramp undergoing ablation to analyze fluid-ablation interaction (FAI) effects on instability growth and transition of the high-speed boundary layer. Wind-tunnel experiments at these conditions show characteristic heating streaks which are first replicated for laminar flow over an inert ramp. Perturbations superimposed on the laminar baseflow are informed by stability calculations. The ramp is then replaced by a low-temperature ablator in our DNS and the interaction of the streaks with the ablating surface is examined. 
Ablation models implemented in two independently-developed fluid dynamics solvers are used to study this problem. Overall, both solvers have shown an excellent agreement regarding heat fluxes, mass blowing rates, and recession patterns. Numerical experiments are performed to assess the impact of accelerating recession and of higher amplitude perturbations. We demonstrate accurate predictions of ablative recession and present novel findings regarding its influence on the perturbation evolution for this configuration.
\end{abstract}

\section*{Nomenclature}

{\renewcommand\arraystretch{1.0}
\noindent\begin{longtable*}{@{}l @{\quad=\quad} l@{}}
$C_p$ & pressure coefficient \\
$c_p$ & specific heat at constant pressure, J/(kg$\,$K) \\
$\mathcal{D}$ & binary diffusion coefficient, m$^2$/s \\
$E$ & total specific energy, J/kg \\
$e$ & thermodynamic internal energy, J/kg \\
$\text{F}$ & ablation speed-up factor \\
$\mathbf{F}$ & total flux vector \\
$h$ & enthalpy, J/kg \\
$\mathbf{J}$ & diffusion flux vector, kg/(m$^2\,$s) \\
$k$ & surface thermal conductivity, W/(m$\,$K) \\
$M$ & Mach number \\
$M_c$ & molar mass of camphor, kg/mol \\
$M_w$ & mixture average molar mass at the wall, kg/mol \\
$\dot{m}$ & mass blowing rate, kg/(m$^2\,$s) \\
$N$ & number of gaseous species \\
$p$ & mixture pressure, Pa \\
$\mathbf{q}$ & total heat flux vector, J/(m$^2\,$s) \\
$\dot{q}$ & incident heat flux, W/m$^2$ \\
$R$ & specific gas constant of the wall mixture, J/(kg$\,$K) \\
$\mathcal{R}$ & universal gas constant, J/(mol$\,$K) \\
$Re$ & Reynolds number \\
$\mathbf{S}$ & source term vector \\
$St$ & Stanton number \\
$T$ & temperature, K \\
$t$ & time, s \\
$\mathbf{U}$ & conserved fluid variables vector \\
$u, v, w$ & velocity components in Cartesian frame, m/s \\
$x, y, z$ & spatial coordinates in Cartesian frame, m \\
$y_i$ & species mass fraction \\

\multicolumn{2}{@{}l}{{Greek Letters}} \\
$\alpha$ & vaporization coefficient \\
$\Delta h_\text{reac}$ & heat of reaction, J/kg \\
$\Delta h_\text{solid}$ & enthalpy required to heat the solid from initial to wall temperature, J/kg \\
$\Delta h_\text{subl}$ & enthalpy of sublimation, J/kg \\
$\delta_{99}$ & boundary layer height, m \\
$\varepsilon$ & surface emissivity \\
$\eta$ & wall-normal coordinate, m \\
$\lambda_z$ & spanwise wavelength, m \\
$\mu$ & viscosity, kg/(m$\,$s) \\
$\rho$ & mixture density, kg/m$^3$ \\
$\sigma$ & Stefan-Boltzmann constant, W/(m$^2\,$K$^4$) \\
$\sigma_y$ & Gaussian width, m \\
$\bm{\tau}$ & viscous stress tensor, Pa \\
$\dot{\omega}$ & species surface production rate, kg/(m$^2\,$s) \\

\multicolumn{2}{@{}l}{{Subscripts}} \\
$0$ & initial value \\
$c$ & camphor \\
$\text{cond}$ & conduction \\
$\text{diff}$ & diffusion \\
$i$ & species index \\
$\text{inv}$ & inviscid \\
$s$ & solid \\
$v$ & vapor \\
$\text{vis}$ & viscous \\
$w$ & wall \\
$\infty$ & free-stream conditions \\
\end{longtable*}}
\addtocounter{table}{-1} 

\section{Introduction} \label{introduction}

Vehicles traveling at hypersonic speeds must operate in extreme flight environments. Severe mechanical and thermal loads are the primary concern for aerothermodynamicists, who seek to ensure the survival of the onboard payload. In addition to strategic shaping of the outer mold line, ablative heat-shield materials have been and continue to be deployed as an effective means for redirecting heat away from the payload. When exposed to the high heating and temperatures of the surrounding flow field, the ablative surface material will typically first undergo endothermic chemical reactions with the atmosphere. Subsequently, the ablative material will then be blown away from the surface and into the boundary layer formed around the body. These two mechanisms of surface reactions and blowing rarely occur in a spatially uniform manner and even less so when the surrounding flow is transitioning from a laminar to a turbulent state.
Turbulent flow enhances heat transfer and consequently the ablation rate. Larger heat and mass transfer results in rapid evolution of the shape, constructing a feedback cycle, which persists throughout the high-temperature flight duration.
Such phenomena are critical since delaying or promoting earlier transition holds important implications for the thermal and mechanical loads imposed on the vehicle in this flow regime.

Accurate simulations of atmospheric entry flows require research in many diverse fields of study. 
Parts of this highly multi-disciplinary problem have been tackled from different fronts by many research groups over the last six decades.
Review articles on this past research have been compiled for studying laminar to turbulent transition~\cite{schneider2007erratum,zhong2012direct,reed2015verification,yang2022research} with roughness~\cite{schneider2008summary} as well as with ablation~\cite{canning1967review,schneider2010hypersonic}.
As for more recent studies from the past few years, notable contributions have been made in 
modeling gas-surface interactions (GSI)~\cite{bellas2017development,capriati2021development, needels2022sensitivity},
performing laminar hypersonic simulations with ablation coupling~\cite{chen2018multidimensional,schroeder2021coupled, rotondi2022numerical},
investigating stability and transition to turbulence in hypersonic flows~\cite{wheaton2021bolt,araya2022assessment,lugrin2022transitional,he2021hypersonic,melander2022numerical,gokcen2022characterization} with roughness~\cite{zhong2019dns,saikia2022effect} or with blowing~\cite{egorov2022numerical,kerth2022gas}
and further with ablation~\cite{zhong_2016,miro2021decoupling,karimi2022direct,rogers2022analysis,baskaya2025baroclinic},
and in performing turbulent hypersonic simulations~\cite{passiatore2021finite,urzay2021engineering}.
Moreover, there has also been a recently renewed interest for studying ablative shape change with low-temperature ablators~\cite{trevino2021,rotondi2022numerical,patel2023,baskaya2025baroclinic} as they exhibit characteristics similar to that of real flight ablators at low-enthalpy conditions. 
There still exist large uncertainties in the predictions and interactions of various effects, and some phenomena such as transition control with roughness~\cite{zhong2019dns} is yet to be understood properly.
Among all of these studies, use of direct numerical simulations (DNS) has been greatly influential in many of the aforementioned works due to the lack of reliable data in this field as ground testing cannot simultaneously replicate all aspects of atmospheric entry environments, and flight tests are rare.

Fluid-ablation interactions (FAI) are technically a coupled problem between fluid and solid dynamics, though more work is needed to investigate how important the ablation coupling is between the transitional boundary layer and material response (MR). Four recent works have made strides in this direction. Trevino~\cite{trevino2021} implemented a camphor subliming boundary condition with implicit time integration into US3D~\cite{candler2015} to perform DNS of fluid-ablation interactions. Their solver was able to generate the progression of smooth surfaces to narrow grooves at moderate Reynolds numbers. An acceleration factor of $10^4$ was applied to the recession rates that does not impact the surface balances, but was used to observe significant topological changes within the simulated fluid time. Patel~\cite{patel2023} presented a similar investigation of camphor ablation in the presence of a transitional flow over a backward-facing step. The separation of time scales was handled by artificially speeding up ablation rates by considering the Damköhler number defined as the ratio of post-shock velocity to wall velocity. Moving boundary ablation simulations showed considerable generation of vorticity over the camphor surface, however, ablation affecting specific boundary layer instability modes was not an objective of that research and, therefore, was not investigated. Dungan et al.~\cite{dungan2023} investigated camphor FAI over a straight cone at Mach 5.3. Using a loose coupling approach with curvilinear meshes and one-dimensional ray material response solvers, the authors showed only minor destabilization of first mode disturbances due to ablation. However, it was shown that significant fluid-solid coupling between the heat transfer and recession distributions is possible.
Başkaya and Hickel~\cite{baskaya2025baroclinic} studied effects of baroclinic vorticity production in Mach 8 boundary layer over a camphor ramp and identified intriguing observations attributed to diabatic cold walls and strong perturbations.
In light of all this, much work remains to exact the FAI effect on high-speed transitional boundary layers, as well as the most effective means to numerically simulate the problem.

This work seeks to make progress in this direction by studying FAI for Mach 8 transitional flow over a 15$\degree$ compression ramp. Experiments performed using identical geometry and flow conditions \revadd{over an inert ramp} exhibited significant heat transfer streaks at reattachment~\cite{chuvakhov2017}. Numerical work by Dwivedi et al.~\cite{Dwivedi2019} showed the separated flow selectively amplifies steady G\"ortler-like vortices that possess a spanwise wavelength similar to those observed in the experiments. Thus, this case should serve as a good proxy for studying FAI in the presence of realistic surface roughness and blowing. 
After numerically replicating the experimental conditions, the rigid ramp will be replaced by solid camphor to study the ablation of this low-temperature ablator.
Two separate numerical frameworks (INCA and CHAMPS) will be used as a means of validation and exploration of different techniques for solving the same problem.
As such, this work was initiated as a collaborative effort between two nations in the NATO STO Research Task Group AVT-346.

This manuscript is organized as follows: \autoref{governing_equations} enumerates the governing equations for the problem at hand and presents the physical modeling approaches, \autoref{sec:2D_ramp} presents the 2-D base flow solutions to compare agreement in fundamental flow quantities, then \autoref{sec:3D_ramp} considers the 3-D compression ramp under perturbed flow conditions to produce steady streaks, which interact with the ramp undergoing ablative recession. Conclusive remarks are made in~\autoref{sec:conclusion}.

\clearpage

\section{Governing Equations and Modeling} \label{governing_equations}

The compressible Navier--Stokes equations for a multispecies chemically-inert fluid can be written as
\begin{equation} \label{eq:navier_collapsed}
\frac{\partial \mathbf{U}}{\partial t} + \grad \cdot \mathbf{F}(\mathbf{U}) = \mathbf{S} \: ,
\end{equation}
where $ \mathbf{U} = [\rho_i, \rho u, \rho v, \rho w, \rho E]^\text{T} $ is the vector of conserved variables, and $ \mathbf{F} = \mathbf{F}_\text{inv} + \mathbf{F}_\text{vis} $ is the sum of inviscid and viscous fluxes. Along the $ \text{x} $ axis, they take the following forms:
\begin{equation}
\mathbf{F}_\text{inv} =
\begin{bmatrix}
\rho_i u 
\\ \rho u^2 + p \\ \rho uv \\ \rho u w \\ u(\rho E+p)
\end{bmatrix} , \quad
\mathbf{F}_\text{vis} =
\begin{bmatrix}
J_{\text{x},i} 
\\ -\tau_{\text{xx}} \\ -\tau_{\text{xy}} \\ -\tau_{\text{xz}} \\ -(\tau_{\text{xx}} u + \tau_{\text{xy}} v + \tau_{\text{xz}} w) + q_\text{x}
\end{bmatrix} \: ,
\end{equation}
where $ \rho_i $ is the species partial density for the $ i^\text{th} $ species, $ \mathbf{u} $ is the mixture average velocity, $ \rho $ is the mixture density, $ p $ is the mixture pressure, and $ E = e + 0.5(u^2+v^2+w^2) $ is the specific total energy, which is the sum of the thermodynamic internal energy $ e $ and the kinetic energy.
Note that the species source term within $\mathbf{S}$ is taken as zero as no homogeneous gas-phase reactions are occurring at the conditions studied in this work. Heterogeneous reactions due to ablation govern the species concentrations and approaches taken to model them are discussed in~\autoref{sec:fai_approach}.
Modeling approaches for the species diffusion flux $ \mathbf{J}_i $, the viscous stress tensor $ \bm{\tau} $, and the total heat flux $ \mathbf{q} $ are mentioned in~\autoref{sec:numerics}.

Specific heats and enthalpies for air and camphor are derived from NASA polynomials~\cite{mcbride2002nasa}. This is particularly important for camphor since unlike air, its properties vary considerably across the studied conditions. The viscosity and thermal conductivity of air are obtained from Blottner's curve fits~\cite{blottner1970chemically}, whereas those for camphor follow the models of Haskins presented by Zibitsker et al.~\cite{zibitsker2024_camphorIJHMT}. Diffusivity is similarly evaluated using the Haskins model, and Wilke's mixing rule is applied to compute the mixture properties. More details are provided in the~\autoref{sec:app_trans_coefs} for completeness.

\subsection{Numerical Methods}
\label{sec:numerics}

Two independently developed solvers are considered in this work. 
INCA is a high-fidelity finite-volume solver for direct numerical simulations (DNS) and large eddy simulations (LES) of the compressible chemically reacting Navier--Stokes equations and provides a large number of different discretization schemes on three-dimensional block-Cartesian grids with adaptive mesh refinement (AMR)~\cite{hickel2014subgrid,muller2016large}.
For the purposes of this study, a fifth-order weighted essentially non-oscillatory (WENO) scheme~\cite{jiang1996efficient} with HLLC flux function~\cite{toro2009riemann} is selected to discretize the inviscid terms.
Second-order centered differences are used for the viscous terms and the explicit third-order Runge-Kutta scheme of Gottlieb and Shu~\cite{gottlieb1998total} is selected for time integration.
In INCA, the ablating wall is represented by a cut-element IB method~\cite{orley2015cut}, which is a consistent and conservative extension of the finite-volume flux balance. This approach accurately accounts for cells being split by fluid/solid interfaces by considering the cut-elements that form from the intersection of the Cartesian fluid grid and the triangulation of the interface geometry. Details of this method and its extension to catalytic and ablative immersed boundaries have been presented by Başkaya et al.~\cite{baskayaETMM,baskaya2023assessment}.

The second solver used in this work is the CHAMPS framework being developed at the University of Maryland. Similar to INCA, the computational domain is divided up into a Cartesian, block-structured mesh, which supports AMR. Regular fluid cells (ones for which their differentiation stencil does not touch a solid boundary) compute inviscid fluxes using a Rusanov flux function and were reconstructed using a fifth-order WENO scheme~\cite{Brehm2015CAF}. The viscous fluxes were computed using a second-order accurate centered scheme. More special care is required when fluid cells have stencils that include cells inside the geometry (irregular cells). This is handled with a locally-stabilized finite difference stencil detailed by~\citet{BREHM2015}. Simulations without ablation were obtained with backward differentiation formula (BDF1) implicit time integration, but once ablation was included a strong-stability preserving 4-stage, 3rd-order explicit Runge-Kutta time integration scheme was used.

\subsection{Fluid-Ablation Interaction}
\label{sec:fai_approach}

Both solvers were developed independently, employing distinct methodologies to incorporate fluid-ablation interactions.
INCA considers a gas-surface interaction (GSI) boundary condition, while CHAMPS is coupled with a material response (MR) solver. Both the GSI~\cite{baskayaETMM,baskaya2023assessment} and MR~\cite{zibitsker2024_camphorIJHMT} methods have been described and successfully demonstrated in several reference works.
However, for this project we have developed and implemented a common interface to have exactly the same FAI models in both flow solvers.

Our FAI model follows the steady-state surface ablation approach of Bianchi et al.~\cite{bianchi2021numerical} and Zibitsker et al.~\cite{zibitsker2024_camphorIJHMT}.
We iteratively solve three coupled conservation equations to obtain the wall state: species mass fractions $y_{i,w}$, blowing flux $\dot{m}_w$, and temperature $T_w$.

The species mass balance for each gaseous species $i$ at the wall reads
\begin{equation}\label{eq:mass_balance}
\underbrace{\rho \mathcal{D} \frac{\partial y_i}{\partial \eta}}_{\text{diffusion}}
+ \underbrace{\dot{\omega}_i \vphantom{\frac{\partial y_i}{\partial \eta}}}_{\text{production}}
= \underbrace{\dot{m}_w \, y_{i,w} \vphantom{\frac{\partial y_i}{\partial \eta}}}_{\text{blowing}} \:,
\qquad i = 1,\ldots,N \:,
\end{equation}
where $\mathcal{D}$ the binary diffusion coefficient coming from the model of Haskins~\cite{zibitsker2024_camphorIJHMT} (see~\autoref{sec:app_trans_coefs}), $y_i$ the species mass fraction, $\eta$ the wall-normal coordinate, $\dot{\omega}_i$ the surface production rate of species $i$, and $N$ the number of gaseous species.
\revadd{For this study, we have considered a binary mixture of air and camphor.}
Summing~\autoref{eq:mass_balance} over all species gives the net mass blowing flux
\begin{equation}\label{eq:mdot}
    \dot{m}_w = \sum_{i=1}^{N} \dot{\omega}_{i,w} = \rho_w \, u_w = \rho_s \, v_{\mathrm{wall}} \:,
\end{equation}
where $\rho_w$ is the mixture density at the wall, $u_w$ is the blowing speed, $\rho_s$ is the solid density, and $v_{\mathrm{wall}}$ the surface recession speed. \revadd{Note that $u_w$ is the speed at which ablation products blow out of the surface and $v_{\mathrm{wall}}$ is the speed at which the surface is moving due to recession.}

The momentum balance at the surface couples the blowing flux to the local thermodynamic state of the gas:
\begin{equation}\label{eq:momentum}
    p_{\eta} = p_f + \rho_f u_f^2 = p_w + \rho_w u_w^2 \:,
\end{equation}
where $u$ is the wall-normal velocity component, and subscripts $f$, $w$, and $\eta$ refer to the near-wall fluid cell, the wall, and the net (combined) condition, respectively.
Together with the ideal gas law $p_w = \rho_w R T_w$ and the blowing-flux definition $\dot{m}_w = \rho_w u_w$, \autoref{eq:momentum} yields a closed-form expression for $p_w$, $\rho_w$, and $u_w$ given $T_w$ and $\dot{m}_w$ as
\begin{align}
    p_w     &= \frac{p_{\eta} + \sqrt{p_{\eta}^2
               - 4 R T_w\,\dot{m}_w^2}}{2} \:, \label{eq:pw}\\[4pt]
    \rho_w  &= \frac{p_w}{R T_w} \:, \label{eq:rhow}\\[4pt]
    u_w     &= \frac{\dot{m}_w}{\rho_w}
             = \frac{2 R T_w\,\dot{m}_w}
                    {p_{\eta} + \sqrt{p_{\eta}^2
                    - 4 R T_w\,\dot{m}_w^2}} \:, \label{eq:uw}
\end{align}
where $R = \mathcal{R}/M_w$ is the specific gas constant of the wall mixture with $\mathcal{R}$ the universal gas constant and $M_w$ the mixture average molar mass at the wall, and $p_\eta = p_f + \rho_f u_f^2$ is the net momentum flux evaluated in the first fluid cell.

The full surface energy balance equates all incoming energy fluxes to those leaving the wall surface~\cite{bianchi2021numerical},
\begin{equation}\label{eq:energy_full}
\underbrace{k\frac{\partial T}{\partial \eta} \vphantom{\sum_{i}^{N} \frac{\partial y_i}{\partial \eta}}}_{\text{conv. heating}}
+
\underbrace{\sum_{i}^{N} h_{i,w} \rho \mathcal{D} \frac{\partial y_i}{\partial \eta}}_{\text{diffusion heating}}
+
\underbrace{\dot{m}_w h_{s,w} \vphantom{\sum_{i}^{N} \frac{\partial y_i}{\partial \eta}}}_{\text{material enthalpy}}
=
\underbrace{q_{\text{cond}} \vphantom{\sum_{i}^{N} \frac{\partial y_i}{\partial \eta}}}_{\text{solid cond.}}
+
\underbrace{\varepsilon\sigma \left( T_w^4 - T_\infty^4 \right) \vphantom{\sum_{i}^{N} \frac{\partial y_i}{\partial \eta}}}_{\text{re-radiation}}
+
\underbrace{\dot{m}_w h_w \vphantom{\sum_{i}^{N} \frac{\partial y_i}{\partial \eta}}}_{\text{blowing}} \:,
\end{equation}
where $k$ is the surface thermal conductivity, $h_{i,w}$ is the enthalpy of species $i$ at $T_w$, and $h_{s,w}$ and $h_w$ are the solid and gas-mixture enthalpies at the wall. 
Multiplying the species mass balance~\autoref{eq:mass_balance} by $h_{i,w}$ and summing over all species yields
\begin{equation}\label{eq:enthalpy_aux}
\underbrace{\sum_{i}^{N} h_{i,w} \rho \mathcal{D} \frac{\partial y_i}{\partial \eta}}_{\text{diffusion heating}}
+
\underbrace{\sum_{i}^{N} h_{i,w}\,\dot{\omega}_i \vphantom{\sum_{i}^{N} \frac{\partial y_i}{\partial \eta}}}_{\text{heat release}}
=
\underbrace{\dot{m}_w h_w \vphantom{\sum_{i}^{N} \frac{\partial y_i}{\partial \eta}}}_{\text{blowing enthalpy}} \:.
\end{equation}
Substituting~\autoref{eq:enthalpy_aux} into~\autoref{eq:energy_full}
cancels the diffusion heating and blowing enthalpy terms, giving
\begin{equation}\label{eq:energy_compact}
\underbrace{k\frac{\partial T}{\partial \eta}}_{\text{conv. heating}}
=
\underbrace{\dot{m}_w\,\Delta h_{\text{reac}} \vphantom{\frac{\partial T}{\partial \eta}}}_{\text{heat of reaction}}
+
\underbrace{\dot{m}_w\,(h_{s,w} - h_{s,0}) \vphantom{\frac{\partial T}{\partial \eta}}}_{\text{solid conduction}}
+
\underbrace{\varepsilon\sigma \left( T_w^4 - T_\infty^4 \right) \vphantom{\frac{\partial T}{\partial \eta}}}_{\text{re-radiation}} \:,
\end{equation}
where $\Delta h_{\text{reac}} = \sum_i h_{i,w}(\dot{\omega}_i/\dot{m}_w) - h_{s,w}$ is the heat of reaction, $h_{s,0}$ is the solid enthalpy at the initial temperature $T_0$, and the solid conduction term $q_{\text{cond}} = \dot{m}_w(h_{s,w}-h_{s,0})$ follows from the steady-state ablation assumption~\cite{bianchi2021numerical}. 
For a single sublimating species with no gas-phase reactions, $\sum_i h_{i,w}\dot{\omega}_i = \dot{m}_w h_{g,w}$, so
$\Delta h_{\text{reac}}$ reduces to the enthalpy of sublimation $\Delta h_{\text{subl}} = h_{g,w} - h_{s,w}$.
Equation~\eqref{eq:energy_compact} then takes its final form,
\begin{equation}\label{eq:energy_final}
\underbrace{k\frac{\partial T}{\partial \eta}}_{\text{conv. heating}}
=
\underbrace{\dot{m}_w\,\Delta h_{\text{subl}} \vphantom{\frac{\partial T}{\partial \eta}}}_{\text{sublimation}}
+
\underbrace{\dot{m}_w\,\Delta h_{\text{solid}} \vphantom{\frac{\partial T}{\partial \eta}}}_{\substack{\text{steady-state}\\\text{solid conduction}}}
+
\underbrace{\varepsilon\sigma \left( T_w^4 - T_\infty^4 \right) \vphantom{\frac{\partial T}{\partial \eta}}}_{\text{re-radiation}} \:,
\end{equation}
where $\Delta h_{\text{solid}} = h_{s,w} - h_{s,0}$ is the enthalpy required to heat the solid from $T_0$ to $T_w$, and $\Delta h_{\text{subl}} + \Delta h_{\text{solid}}$ is the total heat of gasification. The left-hand side represents the convective heat flux delivered by the flow, while the three terms on the right are the energy sinks due to phase change, in-depth heating of the solid, and surface re-radiation, respectively.

The sublimation source term $\dot{\omega}_c$ in \autoref{eq:mdot} is evaluated from the Knudsen--Langmuir equation
\begin{equation}\label{eq:KL}
    \dot{\omega}_c
    = \alpha\,(p_v - p_c)
      \sqrt{\frac{M_c}{2\pi \mathcal{R} T_w}} \:,
\end{equation}
where $\alpha = 0.18$ is the vaporization coefficient, $p_v(T_w)$ the equilibrium vapor pressure of camphor from an Antoine-type correlation~\cite{zibitsker2024_camphorIJHMT}, $p_c$ the local partial pressure of camphor in the gas mixture, and $M_c$ its molar mass.
As camphor accumulates in the boundary layer, the difference in partial pressures $p_v - p_c$ drives the sublimation process and regulates the blowing rate.

The species densities $\rho_{i,w}$, blowing velocity $u_w$, and wall temperature $T_w$ obtained from this iterative procedure are imposed as boundary conditions for the Navier--Stokes equations, while $v_{\mathrm{wall}}$ is used to evolve the surface geometry over time. Since the material time scales are much slower than the flow time scales, surface recession is commonly accelerated by a speed-up factor. We discuss the impact of this speed-up factor in~\autoref{sec:pre-speed-up} and in~\autoref{sec:speed-up}.

\section{Base Flow Analysis}
\label{sec:2D_ramp}

We first consider a 2-D compression ramp to establish the base flow for the 3-D case, that will be discussed in~\autoref{sec:3D_ramp}.
Results obtained with INCA and CHAMPS are compared to assess how well the predictions from these independently-developed solvers agree.
Following a comparison with an inert isothermal wall, an ablative boundary condition is imposed on the ramp to compare the predictions of mass blowing rates.

\subsection{Geometry and Flow Conditions} \label{geometry_flow}

\begin{figure}[!hbt]
	    \begin{center}
		\includegraphics[clip,trim = 0.5cm 12cm 2cm 11.5cm, width=0.7\textwidth]{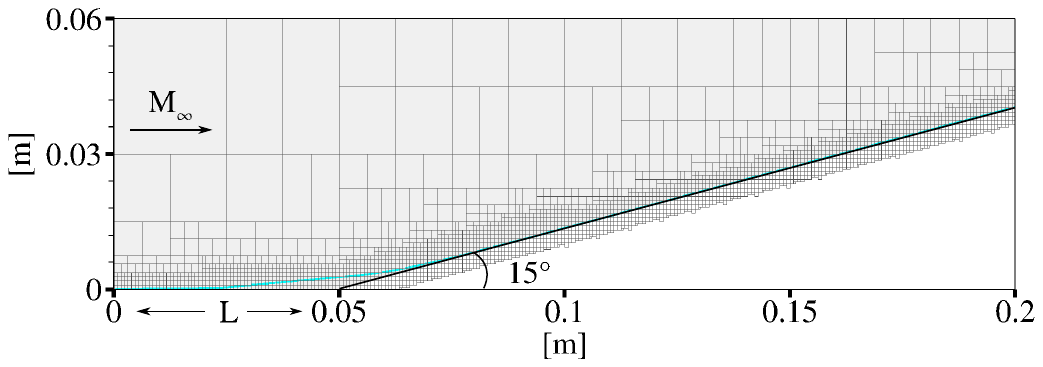}
		    \caption{\textbf{Sketch of the compression ramp geometry with the AMR block structure of the grid. The boundary layer edge is indicated by the colored line.
            }}
		    \label{fig:ramp_schematic}
	   \end{center}
\end{figure}

The geometry under consideration here is a 15$\degree$ compression ramp, which is shown schematically in \autoref{fig:ramp_schematic}. 
The computational setup numerically replicates the experiments of Chuvakhov et al.~\cite{chuvakhov2017}. 
A flat plate with a sharp leading edge of length $L=0.05$ m is followed by a ramp inclined by 15$\degree$ with respect to the horizontal (i.e., freestream) extending to $x=0.2$ m. 
The sketch also shows the separating and reattaching boundary layer edge with the colored line and the AMR block structure of the grid used by INCA.
The freestream conditions imposed at the leading edge of the flat plate are listed in~\autoref{table:freestream}. The reported Reynolds number based on length ($Re_L$) might be more readily contextualized by reporting it per unit length, or $Re_1 = 5.1 \times 10^6$/m. For comparison with reference results in~\autoref{sec:base_results}, single-species air is considered with inert isothermal wall boundary conditions ($T_w = 293 $\,K) over the flat plate and the ramp. An ablative boundary condition over the \revadd{deflected section of the} ramp is set in~\autoref{sec:base_abla}, where camphor is also added to the gas mixture. Remaining edges of the domain are set as non-reflecting zero-gradient outflow.
\revadd{Following results are obtained with a CFL number of 0.5.}
\begin{table}[h!]
\centering
\caption{\textbf{Freestream conditions for the considered compression ramp.}}
\begin{tabular}{c c c c c} 
 \hline
 $M_{\infty}$ & $P_{\infty}$ (Pa) & $T_{\infty}$ (K) & $T_w$ (K) & $Re_{L} (10^5)$\\
 \hline
 8 & 355 & 55 & 293 & 3.7\\
\hline
\end{tabular}
\label{table:freestream}
\end{table}

\subsection{Base Flow with Inert Wall} \label{sec:base_results}

\begin{figure} [tb]
\centering
\includegraphics[clip,trim = 1.5cm 6.2cm 1.5cm 10cm,clip,width=0.5\textwidth]{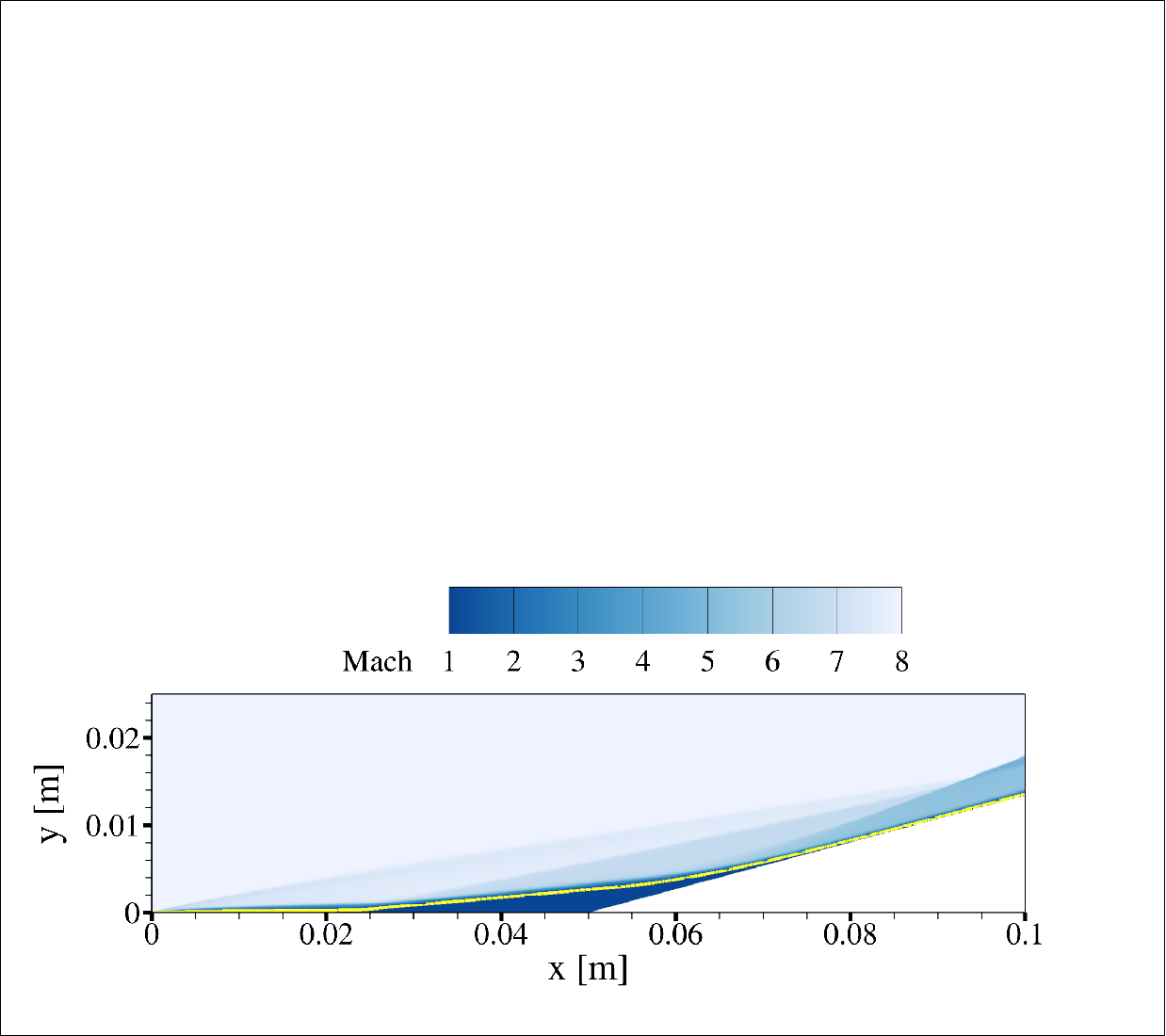} \\
	\begin{subfigure}[t]{.49\textwidth}\centering
		\includegraphics[clip,trim = 0.1cm 0.1cm 2cm 12.2cm,clip,width=0.999\textwidth]{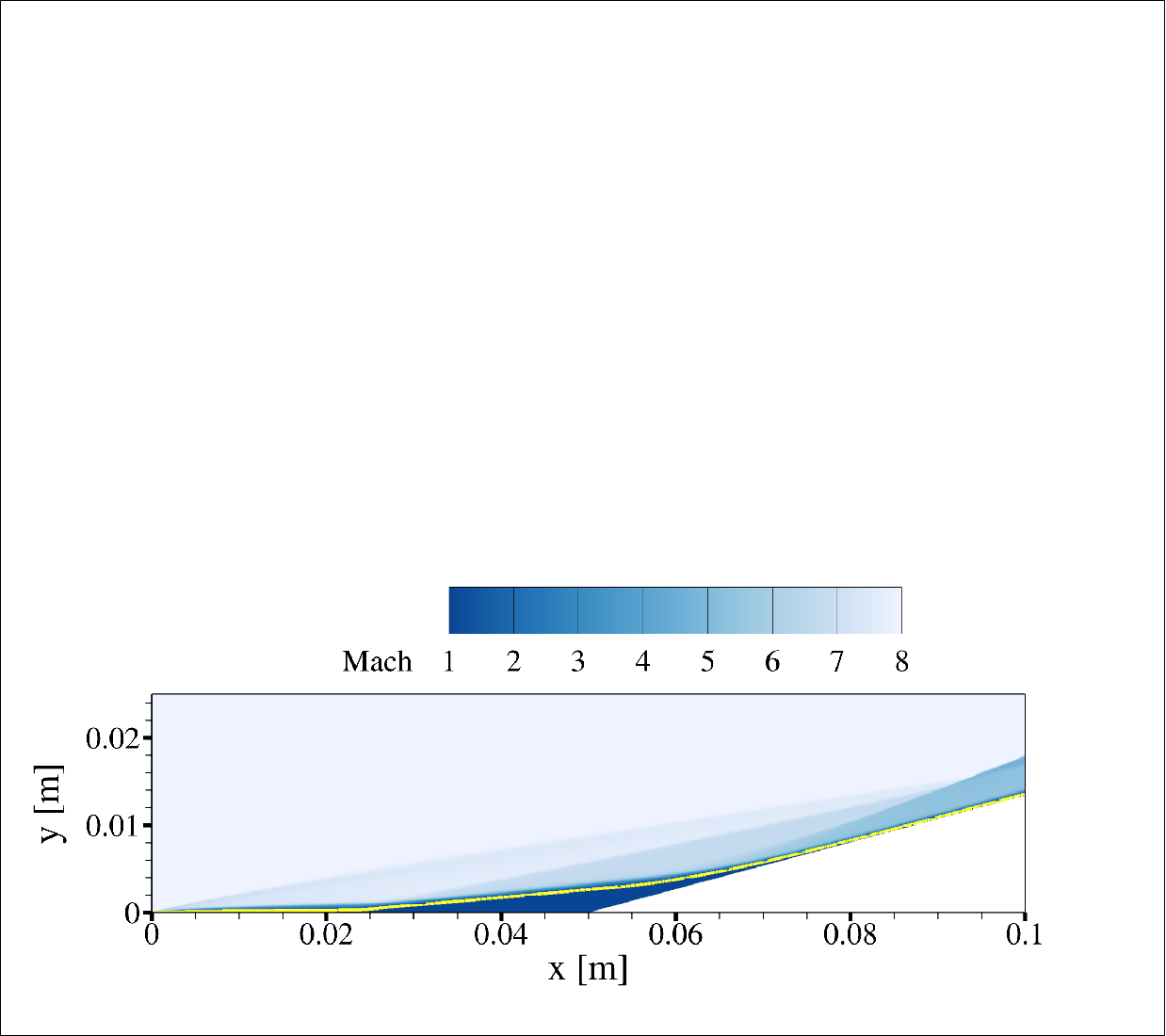}
		\put(-225,70){(a)}
	\end{subfigure}
	\begin{subfigure}[t]{.49\textwidth}\centering
		\includegraphics[clip,trim = 0.1cm 0.1cm 2cm 12.2cm,width=0.999\textwidth]{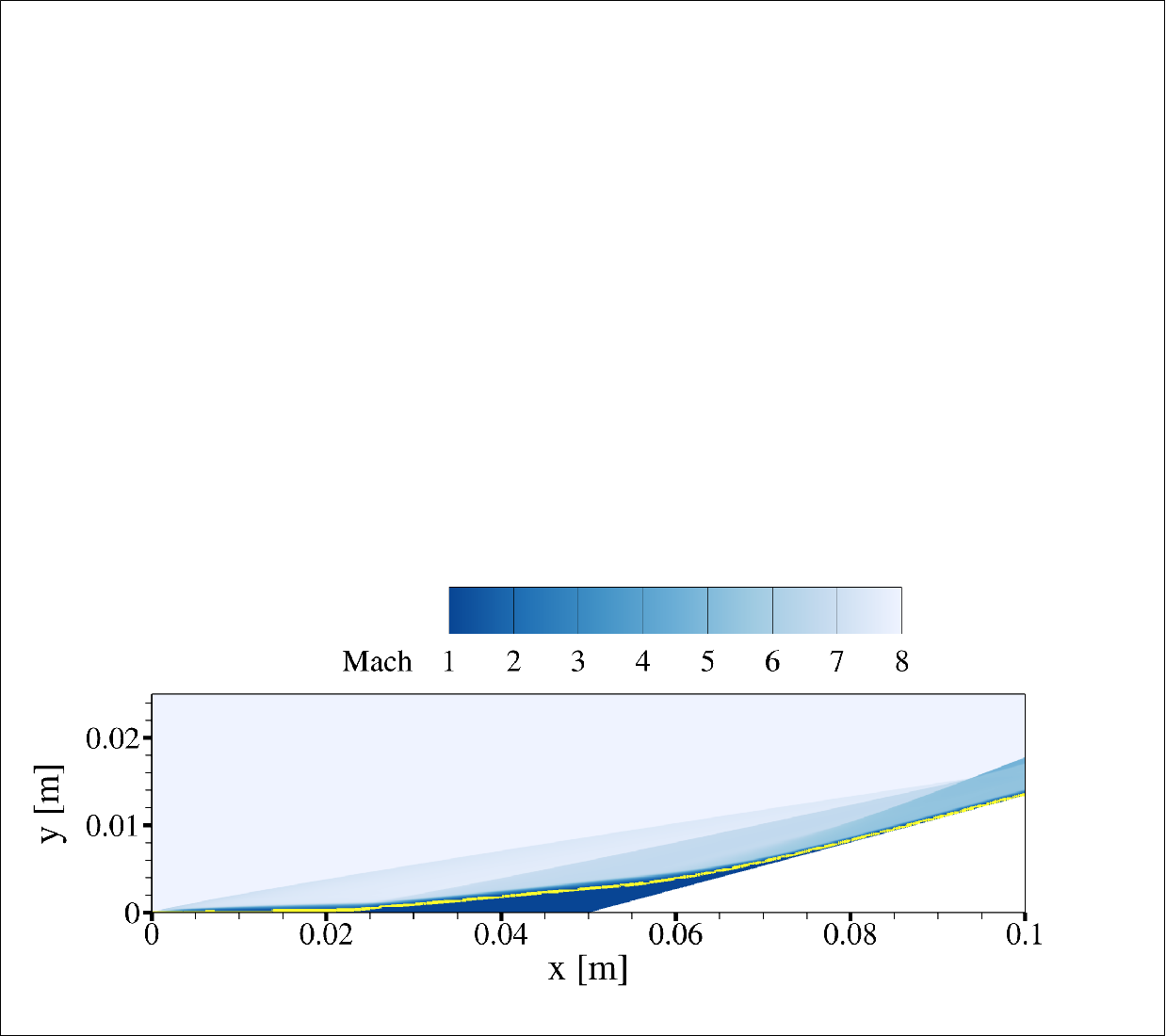}
	\end{subfigure}
\includegraphics[clip,trim = 1.5cm 6.2cm 1.5cm 10cm,clip,width=0.5\textwidth]{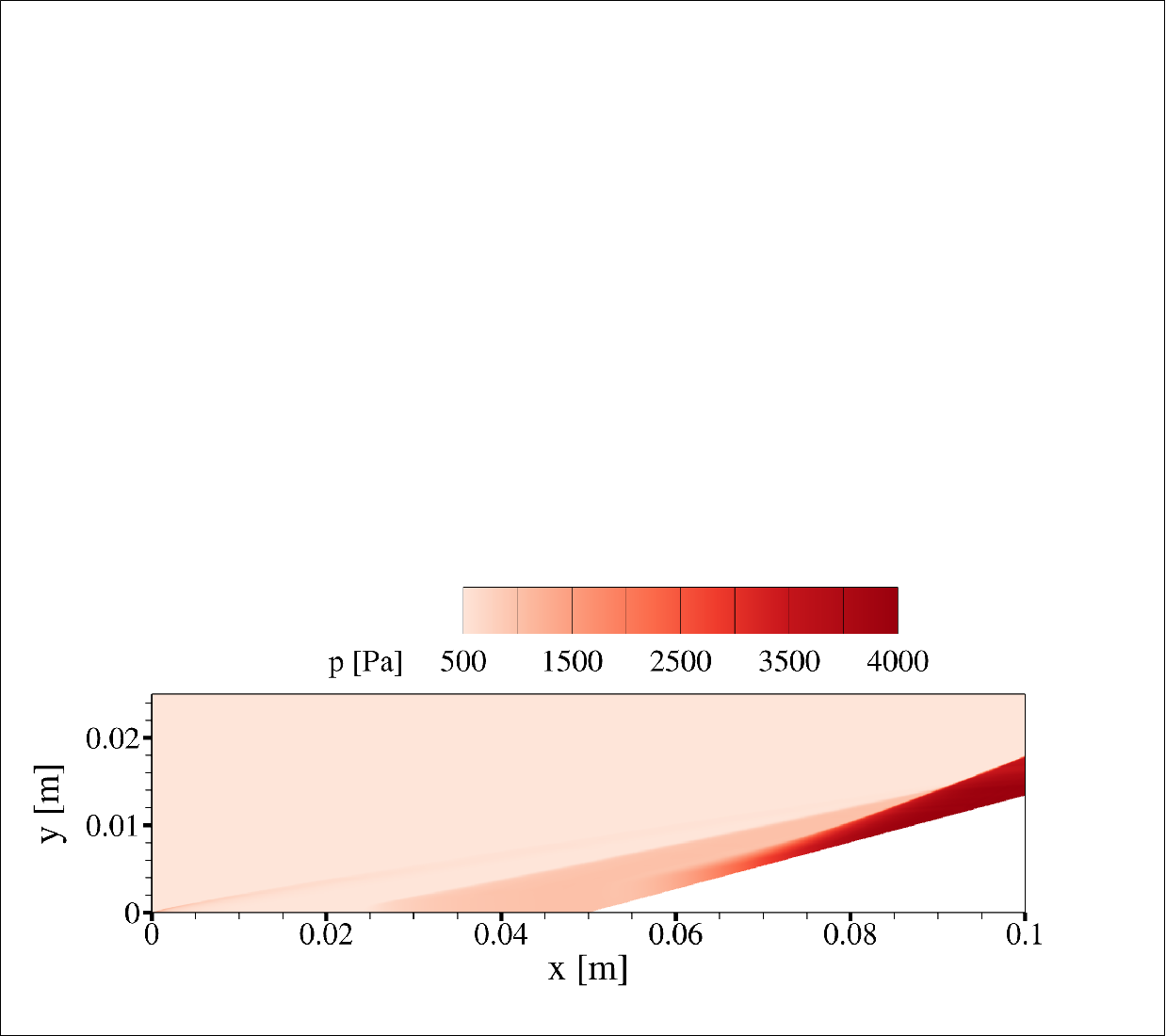} \\
	\begin{subfigure}[t]{.49\textwidth}\centering
		\includegraphics[clip,trim = 0.1cm 0.1cm 2cm 12.2cm,clip,width=0.999\textwidth]{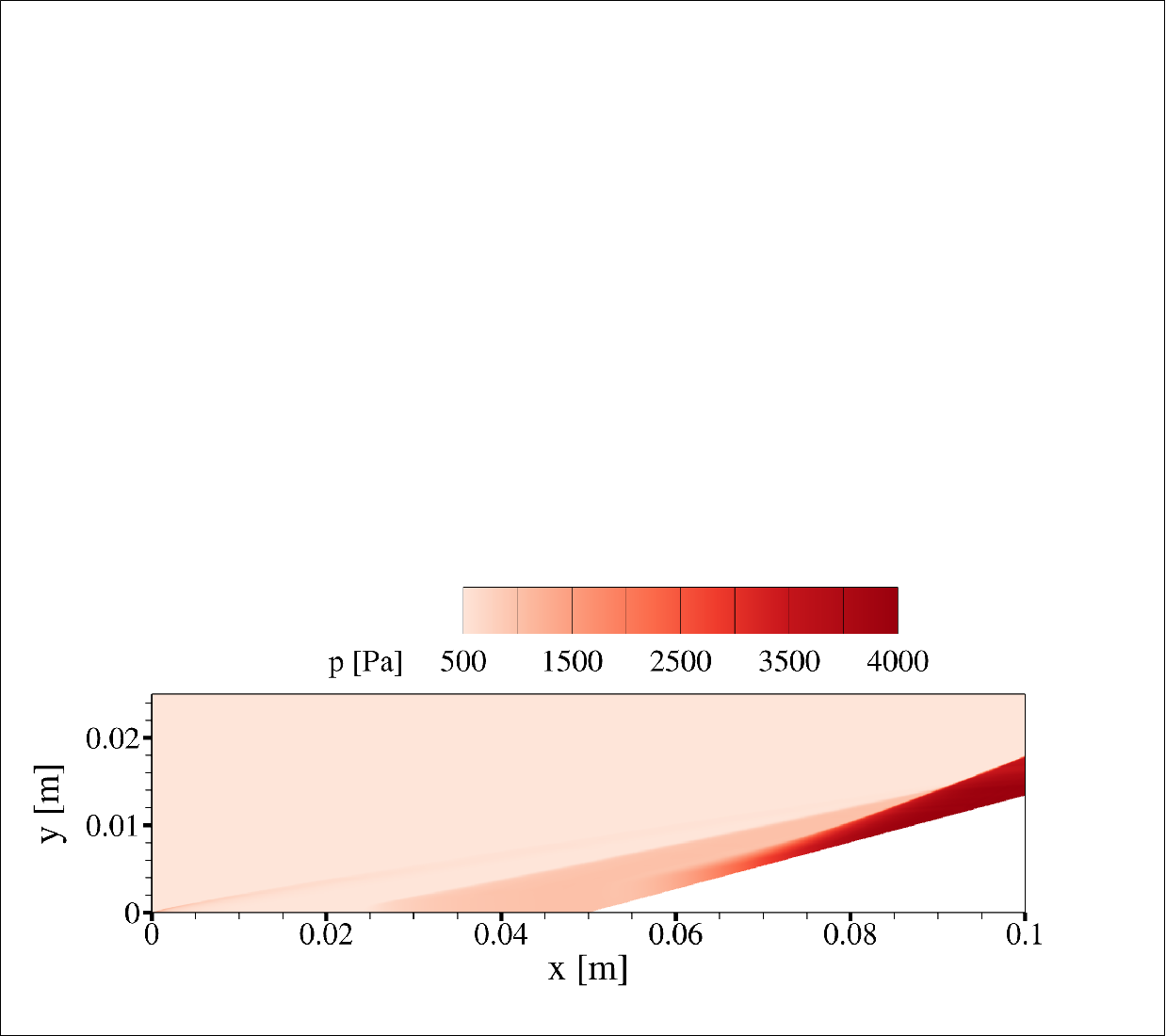}
		\put(-225,70){(b)}
	\end{subfigure}
	\begin{subfigure}[t]{.49\textwidth}\centering
		\includegraphics[clip,trim = 0.1cm 0.1cm 2cm 12.2cm,width=0.999\textwidth]{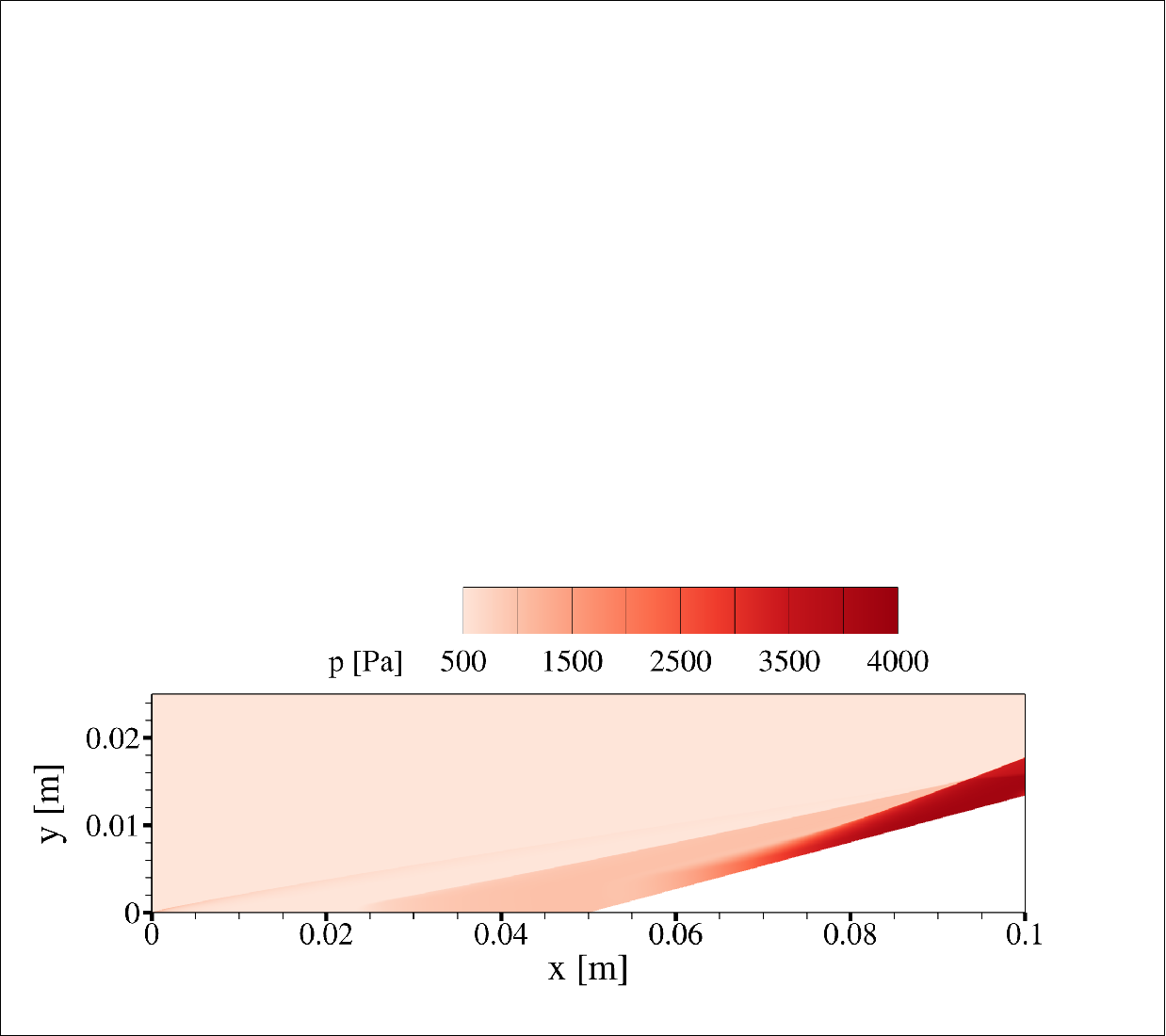}
	\end{subfigure}
\includegraphics[clip,trim = 1.25cm 6.2cm 1.75cm 10cm,clip,width=0.5\textwidth]{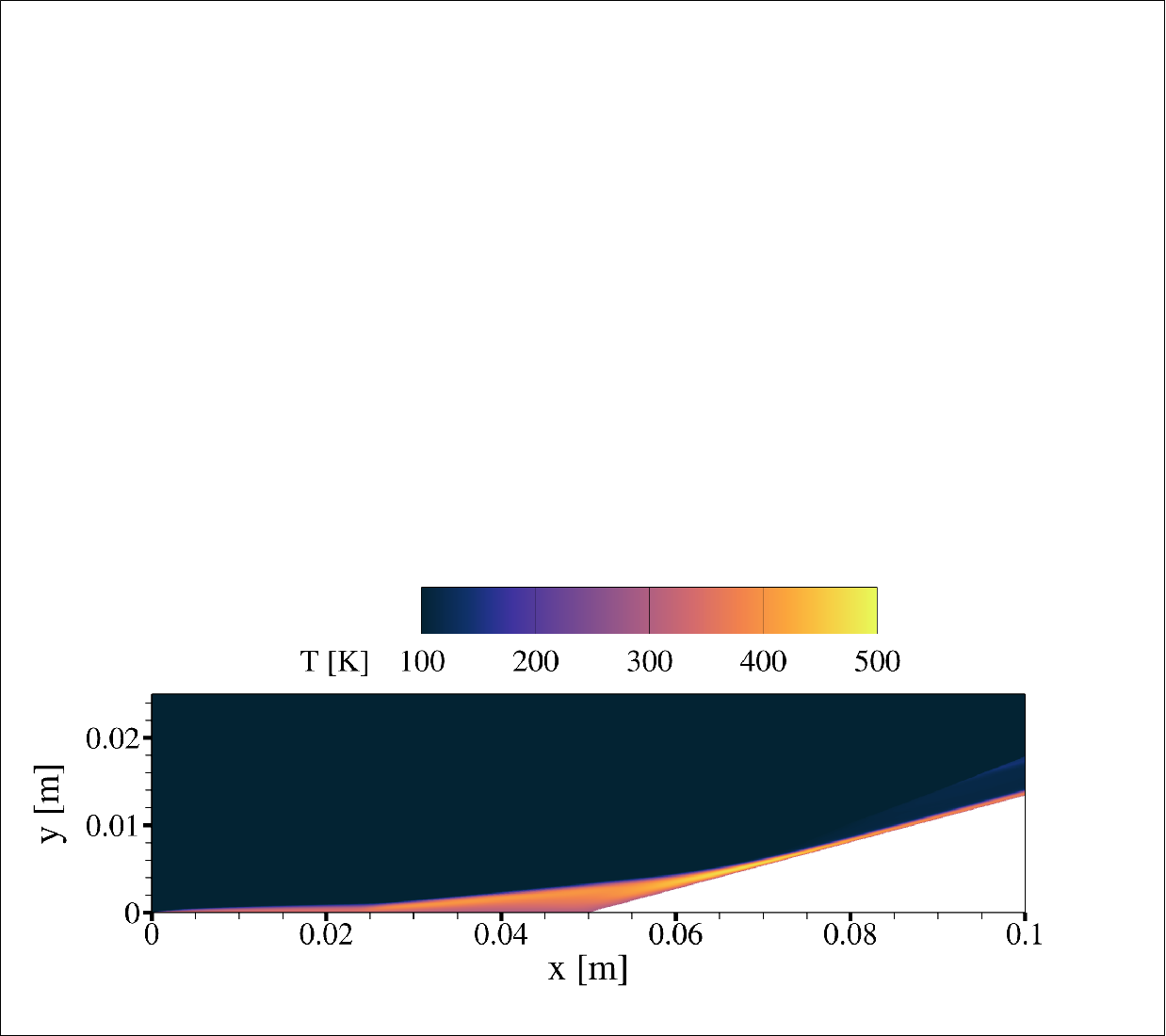} \\
	\begin{subfigure}[t]{.49\textwidth}\centering
		\includegraphics[clip,trim = 0.1cm 0.1cm 2cm 12.2cm,clip,width=0.999\textwidth]{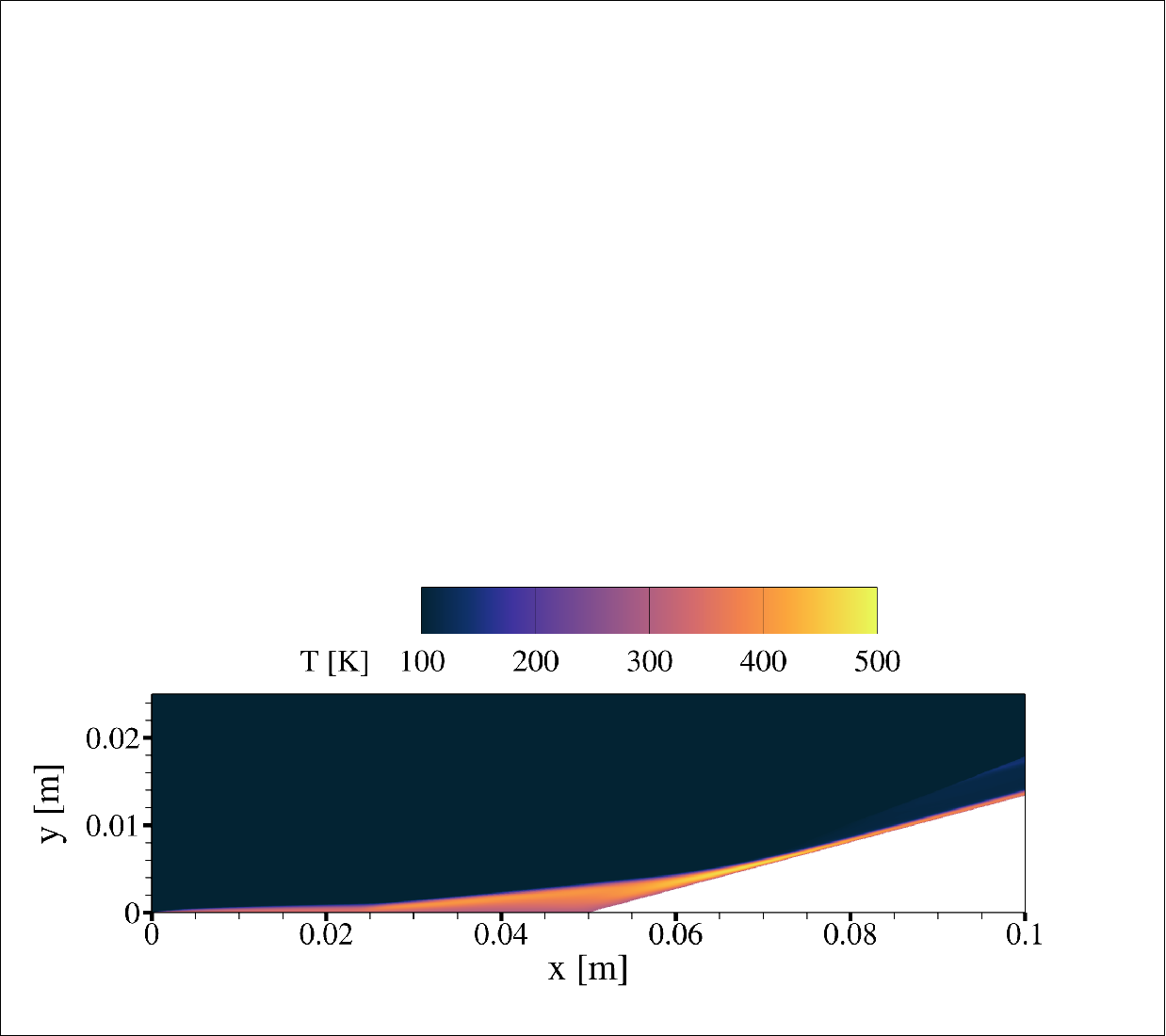}
		\put(-225,70){(c)}
	\end{subfigure}
	\begin{subfigure}[t]{.49\textwidth}\centering
		\includegraphics[clip,trim = 0.1cm 0.1cm 2cm 12.2cm,width=0.999\textwidth]{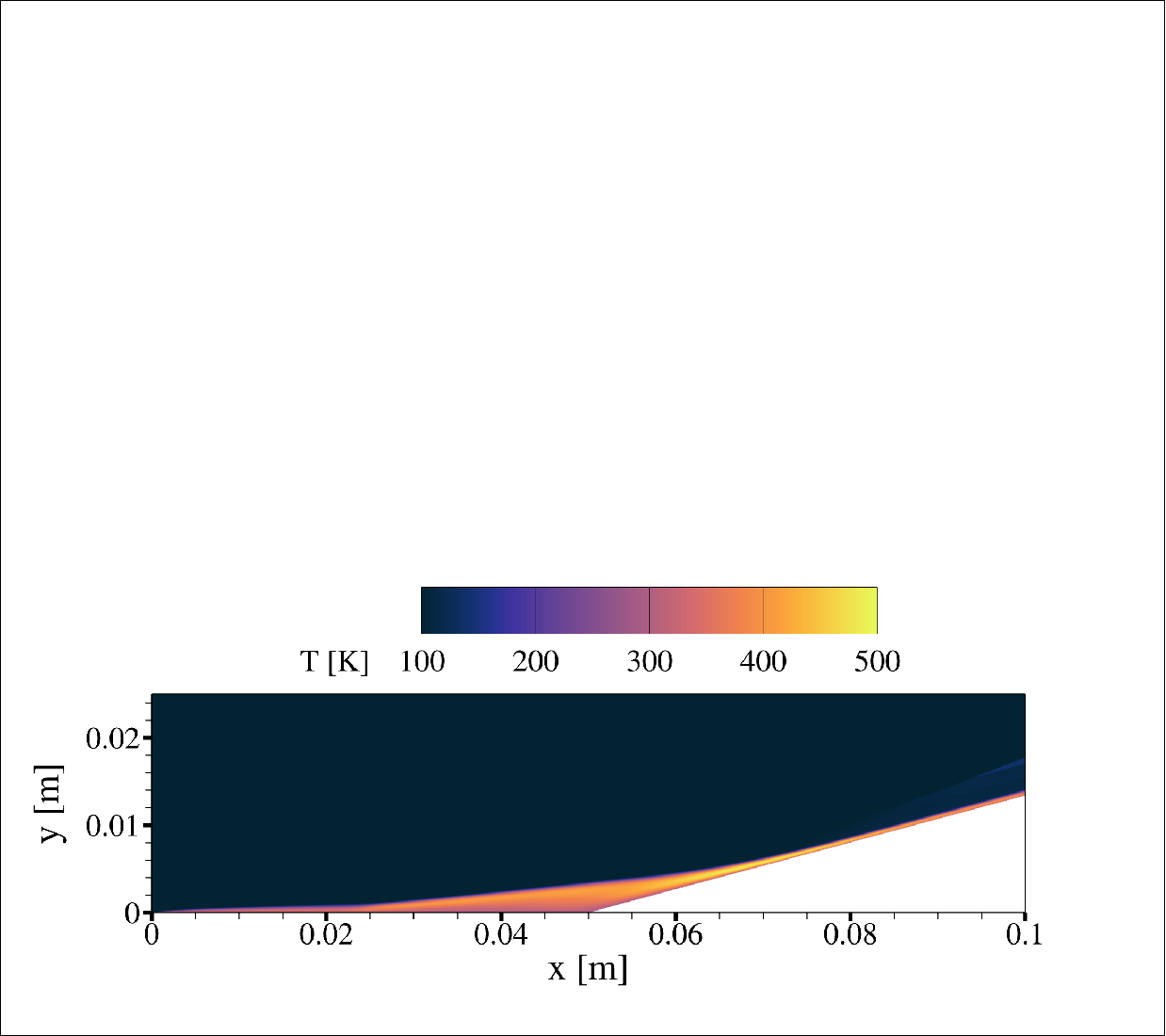}
	\end{subfigure}
	\caption{Contour plots of (a) Mach number with the sonic line, (b) pressure, and (c) temperature from INCA (left) and CHAMPS (right) for the inert compression ramp base flow.}
    \label{fig:baseflow_all}
\end{figure}

The steady, laminar solution to~\autoref{eq:navier_collapsed} for the given freestream conditions (\autoref{table:freestream}) is computed using both the INCA and the CHAMPS solvers. 
Mach number, pressure, and temperature fields are presented in~\autoref{fig:baseflow_all}. 
Good agreement between the solvers can be seen from their successful prediction of the Mach wave emanating from the leading edge, the corner separation, the pressure jump downstream of the reattachment shock, and the peak temperatures predicted beyond the reattachment shock.

\begin{figure} [tb] 
\centering
	\begin{subfigure}[t]{.49\textwidth}\centering
		\includegraphics[clip, trim=0.1cm 6cm 1.5cm 6cm, 
		width=0.999\columnwidth]{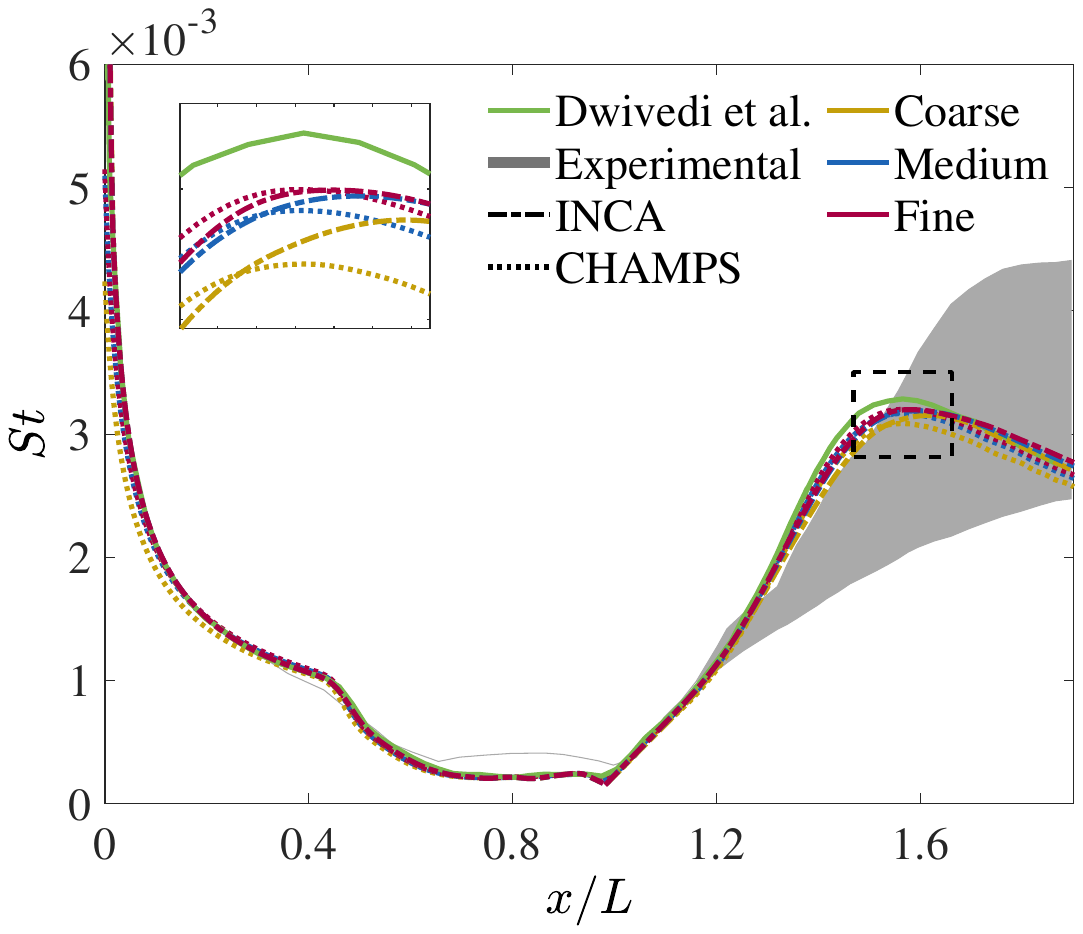}
		\put(-225,167){(a)}
	\end{subfigure}
 \hspace{0.1cm}
	\begin{subfigure}[t]{.49\textwidth}\centering
		\includegraphics[clip, trim=0.1cm 6cm 1.5cm 6cm, 
		width=0.999\columnwidth]{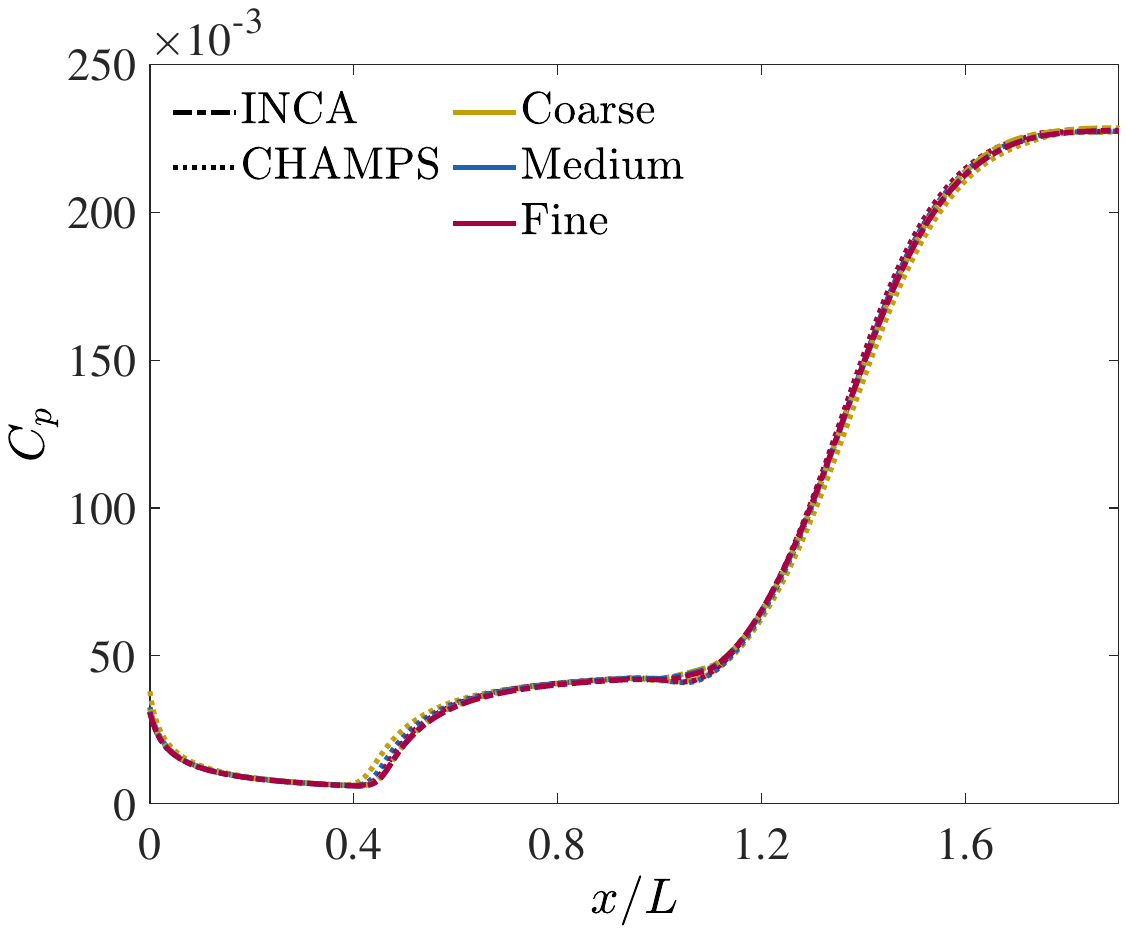}
		\put(-227,167){(b)}
	\end{subfigure}
	\caption{Comparison of (a) Stanton numbers and (b) pressure coefficients over the ramp surface for the inert base flow. \revadd{Experimental results are taken from~\citet{chuvakhov2017}.}}
	\label{fig:2d_ramp_St_Cp}
\end{figure}

For a more quantitative comparison, the Stanton number
\begin{equation}
    St = \frac{q_w}{\rho_{\infty} U_{\infty} c_p (T_{r}-T_{w})}.
\end{equation}
and the pressure coefficient 
\begin{equation}
    C_p = \frac{p_w - p_\infty}{\tfrac{1}{2}\rho_\infty U_\infty^2}
\end{equation}
are calculated along the wall. 
A recovery factor of 0.846 was used in the calculation of the recovery temperature, $T_r$, based on adiabatic wall simulations, and the specific heat at constant pressure is taken as $c_p = 1005$ J/(kg$\cdot$K) for the Stanton number.
Good agreement between the solvers for both of these quantities is presented in~\autoref{fig:2d_ramp_St_Cp} compared with the numerical results of Dwivedi et al.~\cite{Dwivedi2019} \revadd{and the experimental results of~\citet{chuvakhov2017}} for three grid resolutions.
When comparing the numerical results, minor variations arise near peak heat transfer, see zoomed insert, where the boundary layer reattaches on the ramp. 
\revadd{Although these results agree well with the upper limit of the experimental results until the peak predicted by the numerical simulations, the experimentally measured peak value is substantially higher.
These differences are attributed to the inherently three-dimensional nature of the experiments.}
Overall, all of the solvers show an excellent agreement for the inert base flow results.

\subsection{Base Flow with Ablative Blowing}
\label{sec:base_abla}

The inert isothermal ramp is replaced with an ablative material to compare the predicted mass blowing rates.
At this stage, the ramp is not yet allowed to recess and remains a rigid flat surface.
The surface material of the ramp is assumed to be made of solid camphor and only the sublimation reaction
\begin{equation}
    \text{C$_{10}$H$_{16}$O}_{(\text{solid})} \rightarrow \text{C$_{10}$H$_{16}$O}_{(\text{gas})} \:,
\end{equation}
is considered. Vapor pressure and material properties for the air-camphor mixture are taken to be the same as in the work of Rotondi et al.~\cite{rotondi2022numerical}. Surface balances are solved to obtain blowing speeds, surface densities, and wall temperatures.

\begin{figure} [tb]
\centering
\begin{subfigure}[t]{.49\textwidth}\centering
		\includegraphics[clip, trim=0.1cm 6cm 1.5cm 6cm, 
		width=0.999\textwidth]{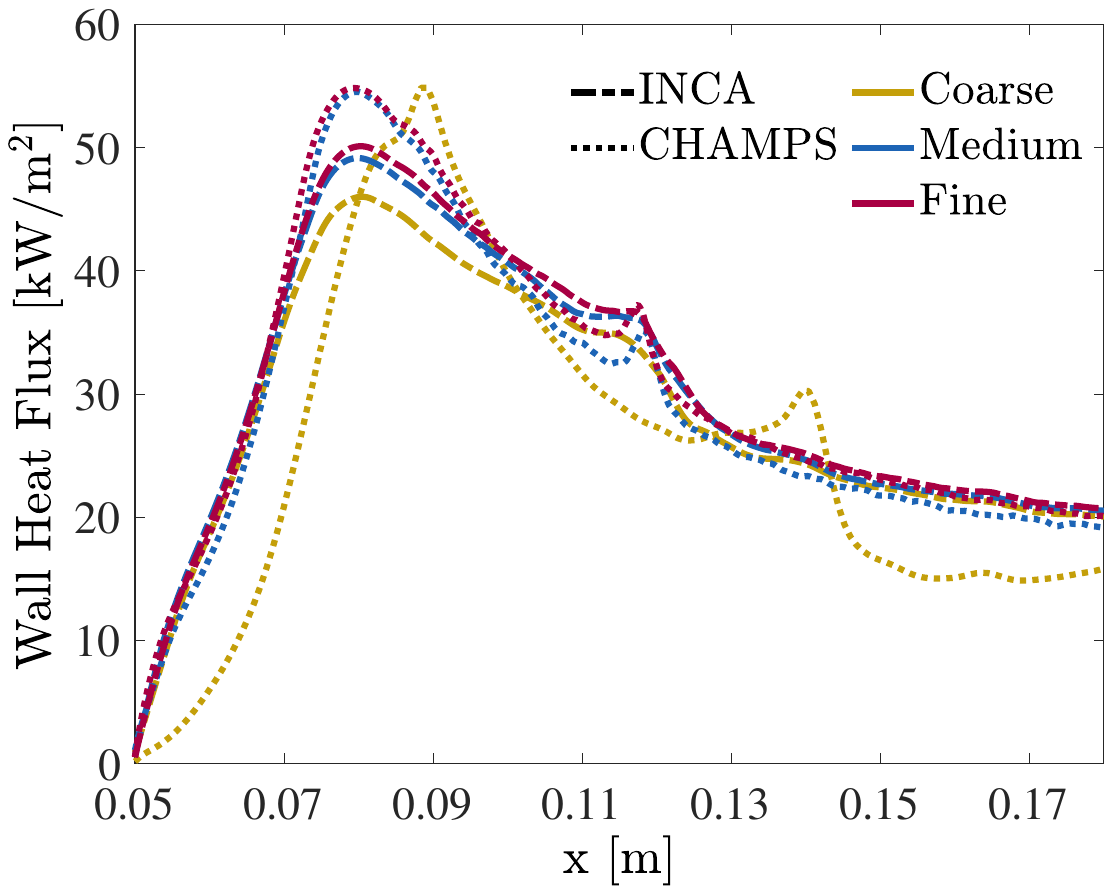}
		\put(-230,167){(a)}
	\end{subfigure}
	\begin{subfigure}[t]{.49\textwidth}\centering
		\includegraphics[clip, trim=0.1cm 6cm 1.5cm 6cm, 
		width=0.999\textwidth]{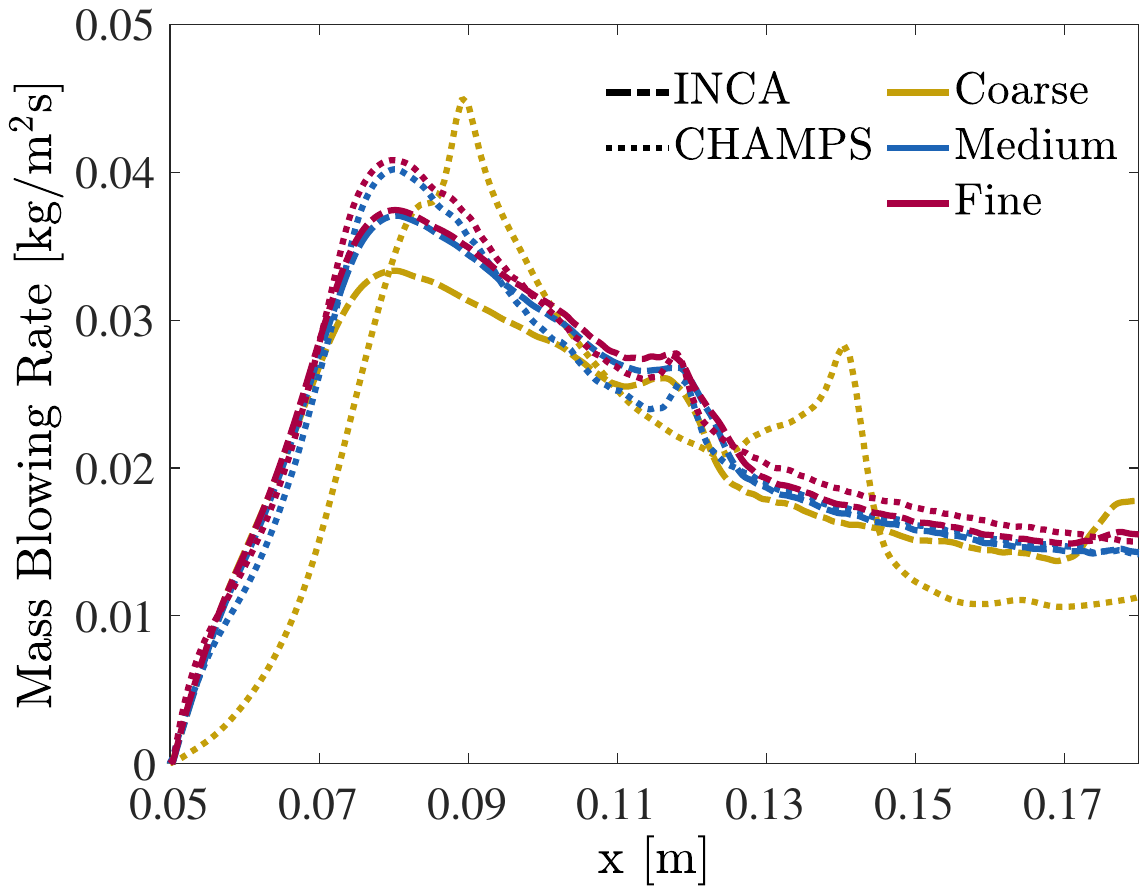}
		\put(-230,167){(b)}
	\end{subfigure}
	\begin{subfigure}[t]{.49\textwidth}\centering
		\includegraphics[clip, trim=0.1cm 6cm 1.5cm 6cm, 
		width=0.999\textwidth]{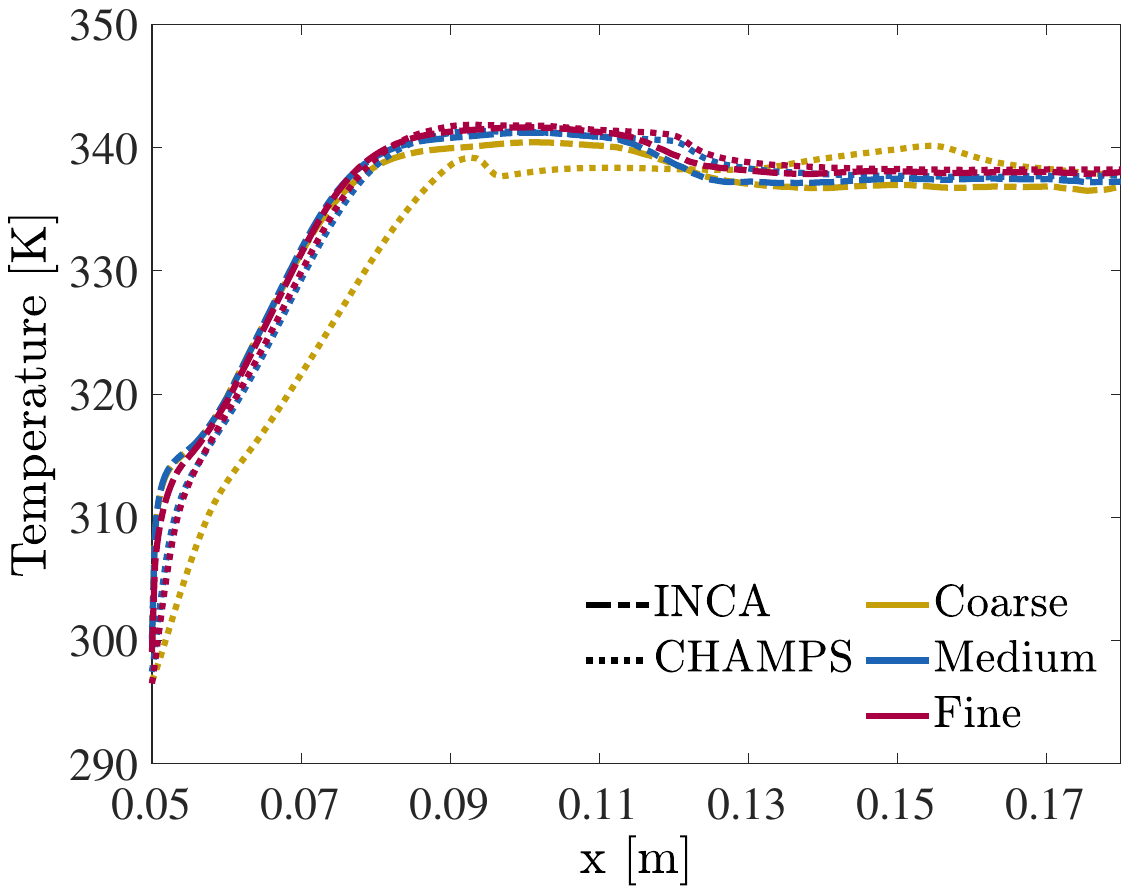}
		\put(-230,167){(c)}
	\end{subfigure}
	\begin{subfigure}[t]{.49\textwidth}\centering
		\includegraphics[clip, trim=0.1cm 6cm 1.5cm 6cm, 
		width=0.999\textwidth]{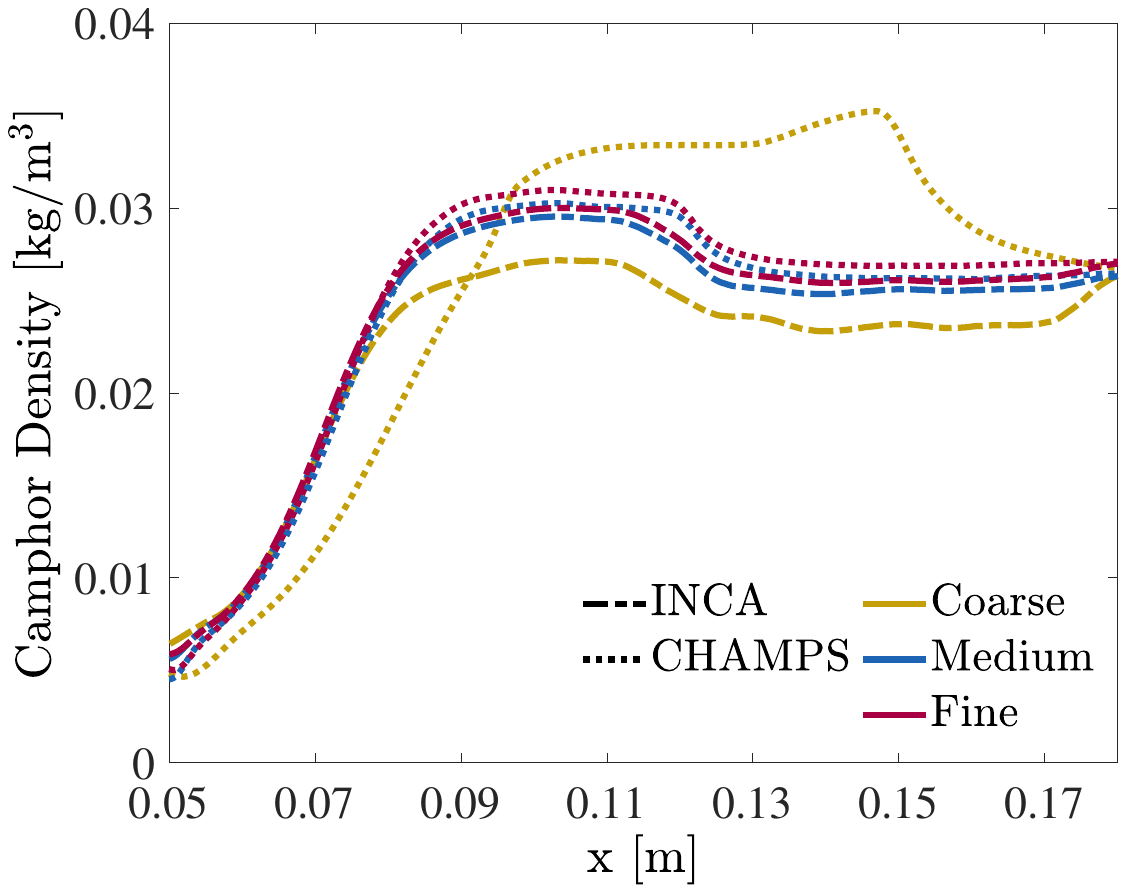}
		\put(-230,167){(d)}
	\end{subfigure}
\caption{Comparison of (a) heat fluxes, (b) mass blowing rates, (c) temperatures, and (d) camphor densities over the ramp surface for the ablative base flow.} 
	\label{fig:2d_ramp_mdot}
\end{figure}

Prediction of surface heat fluxes, mass blowing rates, wall temperatures, and camphor densities at the wall by both solvers at three successively refined grid resolutions are compared in~\autoref{fig:2d_ramp_mdot}. 
Both solvers show good grid convergence, as such, the medium resolution with $ [ \Delta h_x, \Delta h_y ] = [ 3.2 \times 10^{-5}, 1.65 \times 10^{-5} ] $ m is selected for the remainder of this study. At this grid resolution, both INCA and CHAMPS predict similar profiles for all of the quantities. Both at the corner of the ramp and downstream towards the end of the domain the agreement is excellent. Slight variations are present near reattachment and at $x=0.12$ m where the leading edge shock impinges back on the ramp. This is attributed to the differences in the numerical schemes between the solvers and the slight differences in the grid topologies leading to different resolutions in capturing the leading edge shock.

\subsection{Preliminary Speed-Up Factor Analysis}
\label{sec:pre-speed-up}

The validity for using an artificial speed-up factor has been established in the literature for simulation of problems with large separations of time scales. For example, the slow-time acceleration technique \cite{Haselbacher2010} was recently used to make the accumulation of frost under a turbulent fluid boundary layer feasible \cite{Zgheib2022}. The problem considered here of quasi-steady fluid-ablation interaction shares this property. Similar to \cite{Zgheib2022}, the actual recession applied to the boundary separating the fluid and solid camphor is artificially increased by some constant. This constant should be chosen such that the quasi-steady processes involved, namely the coupling between the slow boundary motion, wall-blowing, and outer boundary layer flow, are not disrupted so much so that the resulting flow and recession state at later times are significantly different than if a smaller constant was chosen. Given the faster turnaround time, 2-D simulations allowed preliminary investigation of the effect of the ablation speed-up factor. Values for the factor, $\mathrm{F}$, were chosen based on wall recession speed, wall temperature, and the local boundary layer height. Estimates were made for the time it would take to recess one boundary layer thickness,$\tau_{\mathrm{recc}}=\delta_{99}/v_{\mathrm{wall}}$, and the characteristic diffusion time across the boundary layer, $\tau_{\mathrm{diff}}=\delta_{99}^2/\mathcal{D}$. We found $\tau_{\mathrm{recc}}/\tau_{\mathrm{diff}}\approx 5\times 10^3$.

\begin{figure} [!tb]
\centering
    \includegraphics[clip, trim=0.0cm 0.0cm 0.0cm 0.0cm, width=0.49\textwidth]{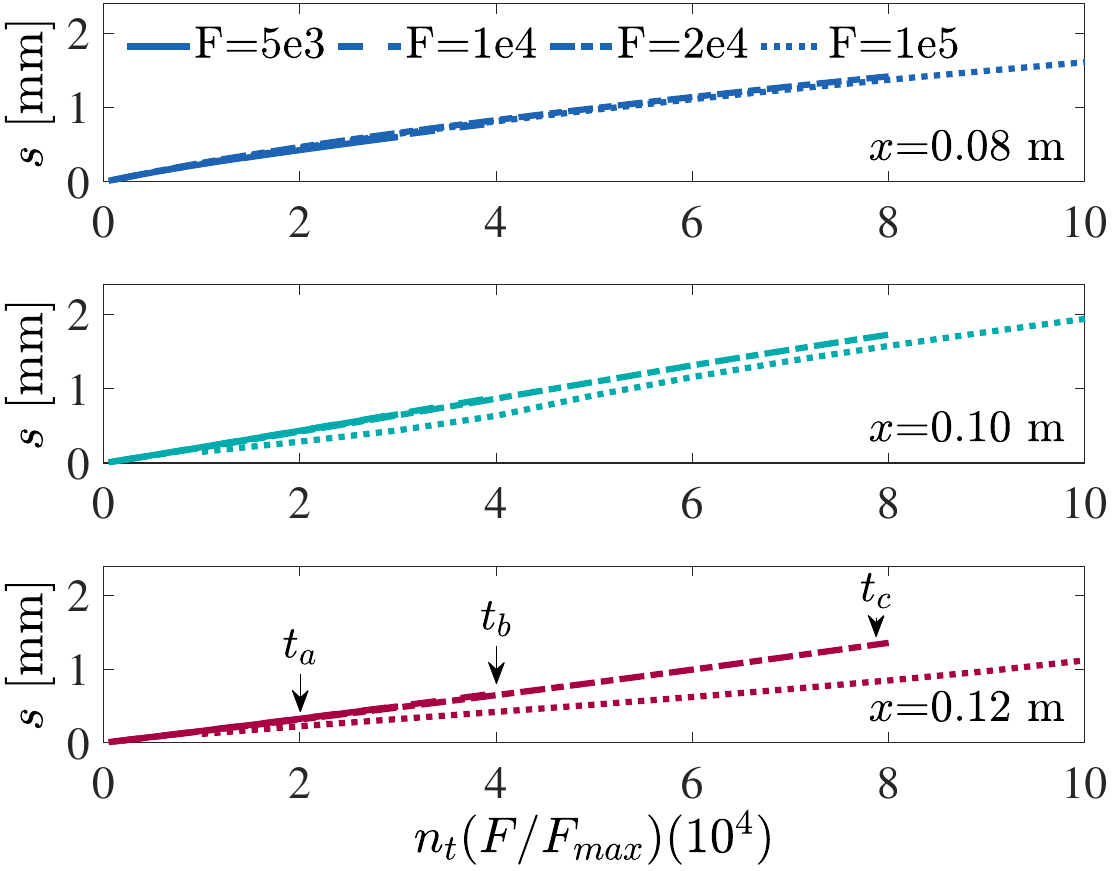}
    \includegraphics[clip, trim=0.0cm 0.0cm 0.0cm 0.0cm, width=0.499\columnwidth]{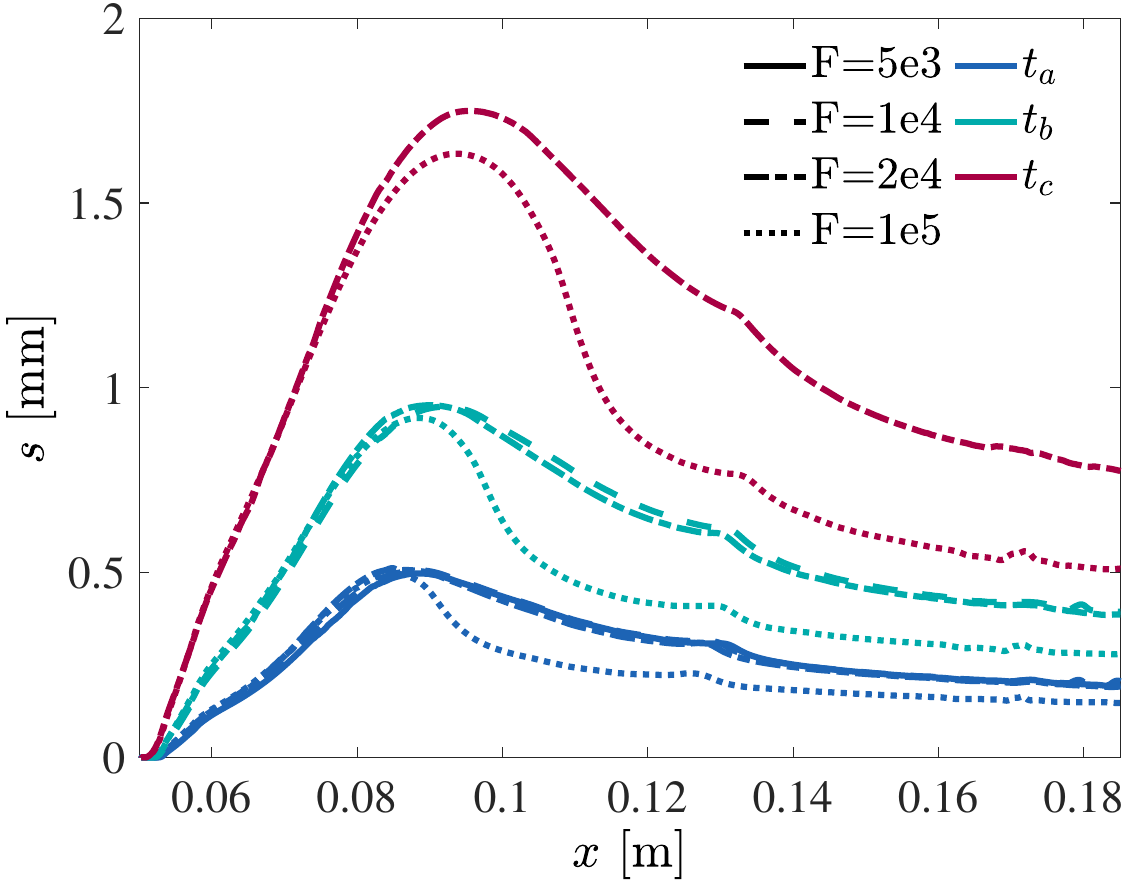}
	\caption{Comparison of recession evolution in: time for several positions along the ablating ramp surface (left), and space for several instances in time (right).}
	\label{fig:2d_recession_speedup}
\end{figure}

With this estimate in mind, four simulations were run with the speed-up factor set to $\mathrm{F}=\{5,10,20,100\}(10^3)$.
Total recession, $s$, of the ablating ramp's surface is plotted as it evolves in pseudo-time in~\autoref{fig:2d_recession_speedup} (left panel). The pseudo-time is defined with reference to the fastest speed-up factor utilized and represents the equivalent amount of simulation time from the material recession's perspective. The trajectories in pseudo-time align well for the first three speed-up factors $\mathrm{F}=\{5,10,20\}(10^3)$ over the range of overlapping pseudo-time. The $\mathrm{F}=100(10^3)$ case under predicts the recession values, indicating the threshold for maintaining correct separation of time scales has been surpassed. The spatial distributions at select instances in pseudo-time (marked with arrows in the left panel) are also shown in the right panel of~\autoref{fig:2d_recession_speedup}. The spatial distributions verify the previously mentioned agreement between the lower three speed-up factors examined at select $x$ locations holds everywhere else. The speed-up factor $\mathrm{F}=20(10^3)$ was determined to best balance accuracy and simulation turnaround time and, therefore, was utilized for further simulations. 

\subsection{Linear Stability} \label{sec:stability}

The time spectral solver in the CHAMPS framework was utilized as an inexpensive precursor calculation to study stability of the 2-D baseflow and define the 3-D setup based on these results. 
This solver, developed by Haas et al. \cite{haas2020}, solves directly for the time-spanwise periodic disturbance state for a given frequency spanwise-wavenumber pair when small perturbations are introduced into the domain. The disturbance was introduced through a volume source term limited to a small region upstream of the separation bubble.

\begin{figure} [tb]
\centering
	\begin{subfigure}[t]{.49\textwidth}\centering
		\includegraphics[clip, trim=0.0cm 0.0cm 0.0cm 0.0cm, 
		width=0.999\textwidth]{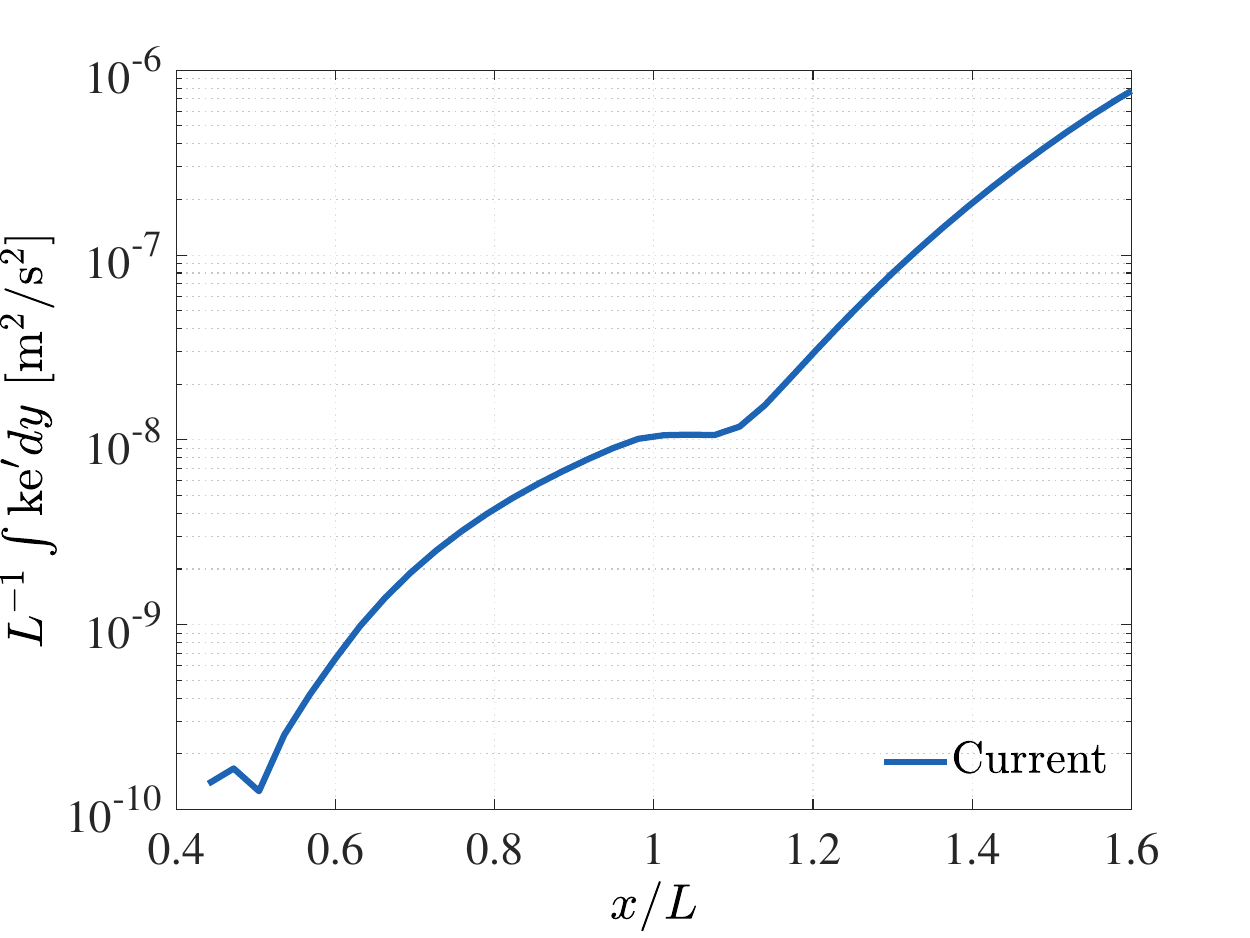}
	\end{subfigure}
    \begin{subfigure}[t]{.49\textwidth}\centering
		\includegraphics[clip, trim=0.0cm 0.0cm 0.0cm 0.0cm, 
		width=0.999\textwidth]{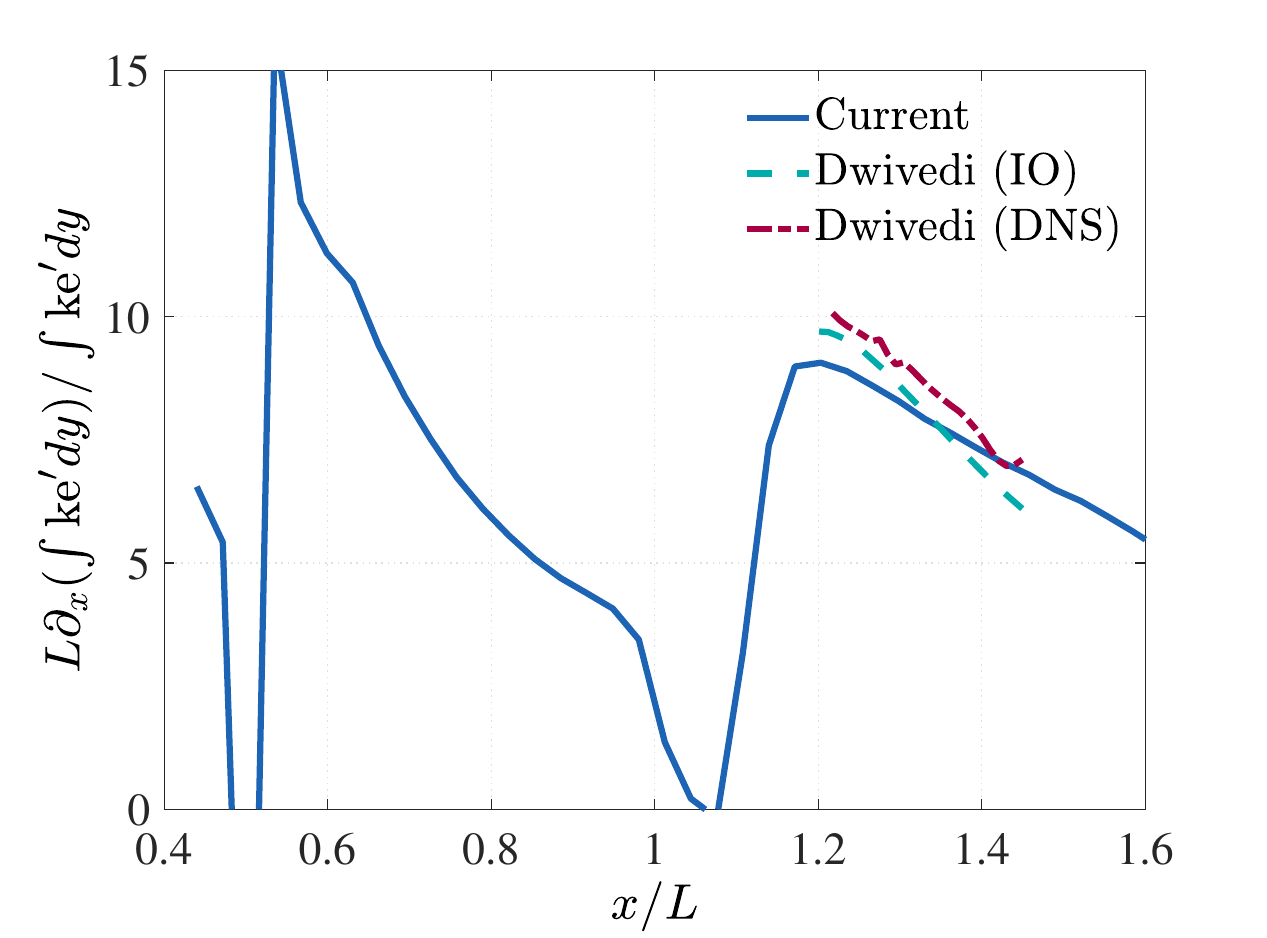}
	\end{subfigure}
\caption{\revadd{Wall-normal integrated disturbance kinetic energy (left) and streamwise growth rate of wall-normal integrated disturbance kinetic energy (right) compared with values from Dwivedi et al.~\cite{Dwivedi2019}. ``IO" and ``DNS" denote ``input/output" and ``direct numerical simulation", respectively.}}
	\label{fig:TScurves}
\end{figure}

The small-amplitude forcing \revadd{mimics the optimal vortical forcing identified by Dwivedi et al.~\cite{Dwivedi2019}.} The spatial form listed in~\cite{browne2022_candf} with near zero temporal frequency (essentially steady) generates a steady streak that undergoes very large amplification near the location where the boundary layer reattaches on the inclined ramp (i.e., reattachment). This amplification was quantified by extracting wall-normal profiles \revadd{and integrating the disturbance kinetic energy ($\mathrm{ke}'=1/2 u_{i}' u_{i}'$) for each profile}. The \revadd{streamwise} progression of this quantity is shown in the left of~\autoref{fig:TScurves}. Plotted on the log scale, one can see that \revadd{integrated} disturbance kinetic energy is amplified by roughly 4 orders of magnitude from the separation to the end of the computational domain. \revadd{The normalized growth rate of that quantity agrees well} with those of Dwivedi et al.~\cite{Dwivedi2019} \revadd{shown in the right of~\autoref{fig:TScurves}}. \revadd{The small differences are attributed to the sub-optimal forcing applied in the current case}. With these results in hand, it was clear the baseflow obtained would support the desired disturbance growth the 3-D studies sought to investigate.

\section{Fluid-Ablation Analysis on Perturbed Flow}
\label{sec:3D_ramp}

Following the agreement in the results of the 2-D base flows in~\autoref{sec:2D_ramp}, the solutions with the inert isothermal wall are extruded in the spanwise direction by \SI{0.003}{\meter} to obtain the 3-D base flows. This spanwise domain width permits one wavelength of the most amplified spanwise wavelength found by Dwivedi et al.~\cite{Dwivedi2019}. The flow is then perturbed using a steady volume force to obtain steady streaks over the ramp.
After ablation is introduced, this configuration is simulated for one flow through time without recession for the flow to adjust to the camphor blowing due to sublimation of the surface. This is performed to avoid an unphysical peak in the initial recession speeds.
Following this settling with ablative blowing, the camphor surface is allowed to decompose under ablative recession.  

This section investigates the interactions of the perturbed flow with the ablating surface.
First two headings consider the procedure to obtain the perturbed flow with ablative recession, first perturbing the steady flow in~\autoref{sec:perturbed_inert}, then introducing ablative blowing and recession in~\autoref{sec:perturbed_recession}.
Finally,~\autoref{sec:speed-up} and~\autoref{sec:influence_amplitude} examine the influence of the speed-up factor and the perturbation amplitude, respectively.

\subsection{Perturbed Flow with Steady Streaks}
\label{sec:perturbed_inert}

Following the stability analysis in~\autoref{sec:stability}, both solvers proceeded to perturb the flow by adding the forcing
\begin{equation}\label{eq:forcing}
    S(x,y,z) = f(x)\exp \left[-\left(\frac{y-y_0}{\sigma_y}\right)^2 \right]\sin\left(\frac{2\pi z}{\lambda_z}\right)
\end{equation}
to the righthand side of the vertical-momentum equation in Eq.~\ref{eq:navier_collapsed} with $y_0 = 0.5$ mm, $\sigma_y = 0.3$ mm, wavelength $\lambda_z=3.0$ mm. This spanwise wavelength was found to undergo the most amplification for this flow condition and geometry \cite{Dwivedi2019}. The forcing is applied from $x_1=0.02$ m to $x_2=0.023$ m in the streamwise direction. The streamwise shape function has vanishing derivatives at the start and end of the streamwise forcing extent and was computed as
\begin{align}
   f(x) =  \begin{cases} 
      \: 0.0 & x \leq x_1 \:\:\: \text{or} \:\:\: x \geq x_2 \:,\\
      \: \left(\Tilde{x} \frac{2r_x-\Tilde{x}}{r_x^2}\right)^3 & \Tilde{x} < r_x \:,\\
      \: 1.0 & \text{otherwise} \:,
   \end{cases} 
\end{align}
where $\Tilde{x} = \min(|x-x_1|,|x-x_2|)$ and $r_x = \frac{1}{4} (x_2-x_1)$.

\begin{figure} [!tb]
\centering
    \includegraphics[clip, trim=0.1cm 6cm 1.5cm 6cm, 
		width=0.49\textwidth]{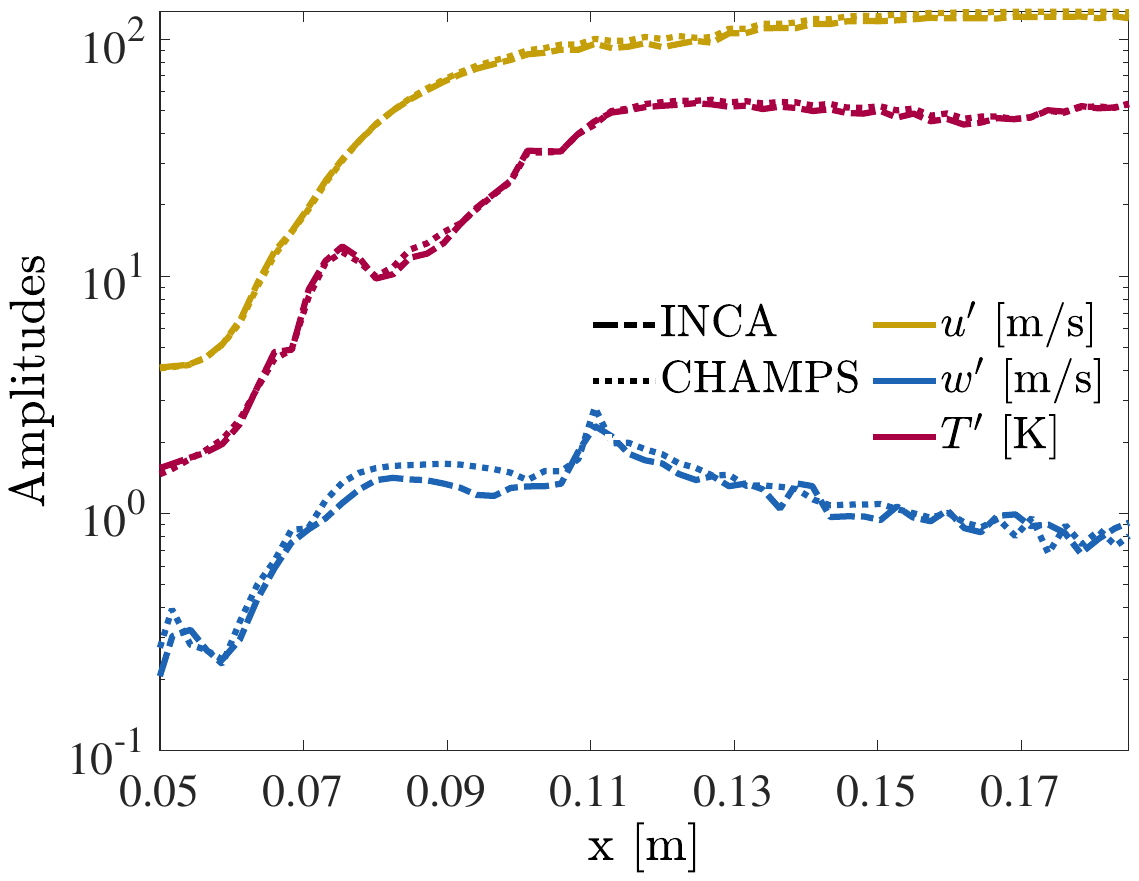}
    \includegraphics[clip, trim=0.1cm 0.1cm 0.1cm 2cm, width=0.499\columnwidth]{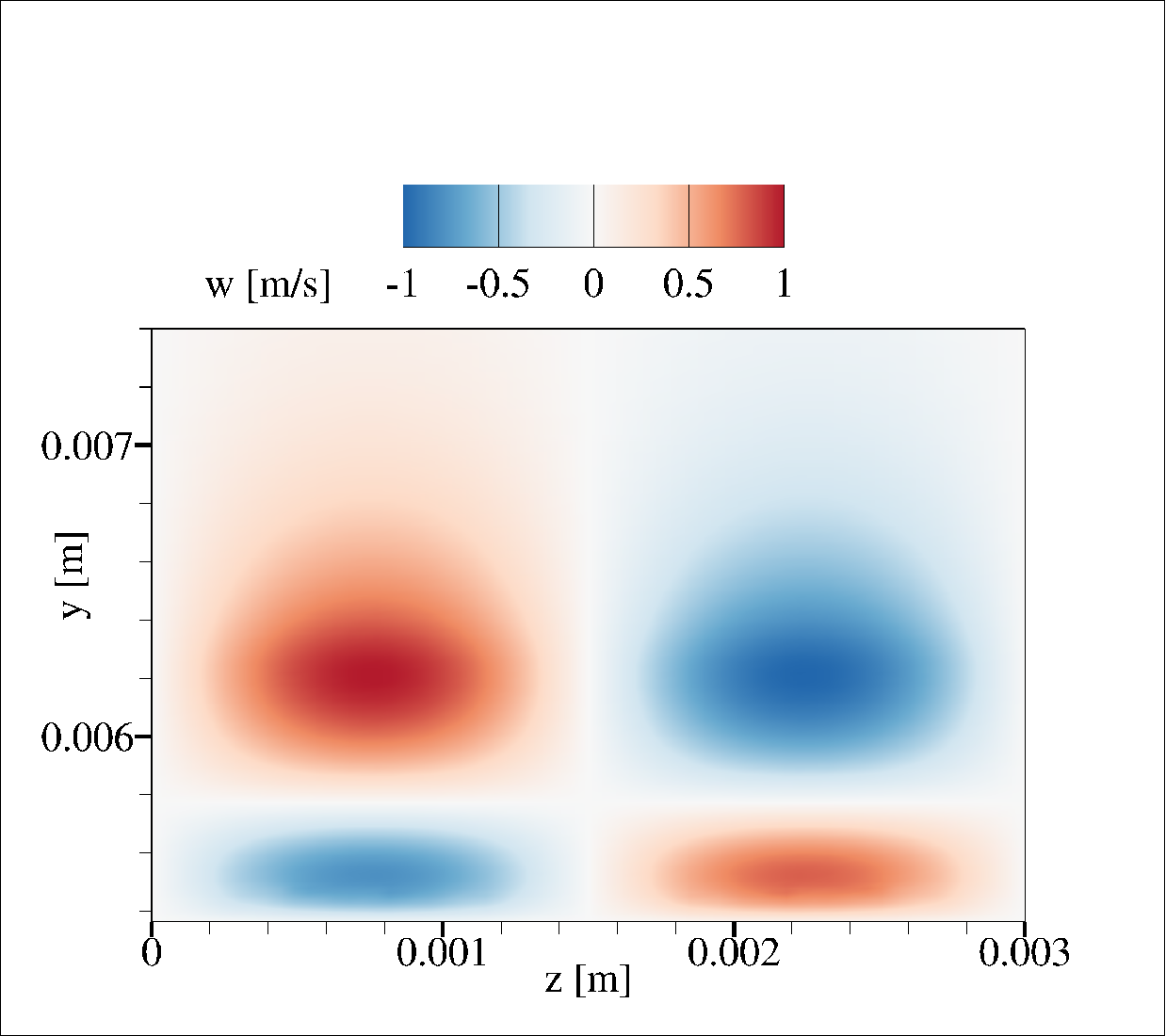}
	\caption{Comparison of streamwise velocity, spanwise velocity, and temperature perturbations (left) and the shape of the steady streak at $x=0.070$ m (right) for the perturbed flow over the inert ramp.}
	\label{fig:compare_streak_flucs}
\end{figure}

The medium resolution 2-D grid from~\autoref{sec:2D_ramp} is refined in the spanwise direction using 64 points corresponding to $ \Delta h_z = 4.7 \times 10^{-5} $ m, which is slightly coarser than the streamwise resolution.
Streamwise velocity, spanwise velocity, and temperature perturbations over the ramp are shown in~\autoref{fig:compare_streak_flucs}. The shape of the steady streak is also shown through a slice downstream of the reattachment location at $x=0.07$ m in the same figure.
The spanwise velocity contours indicate two counter-rotating vortices forming a steady streak.
Entrainment from these counter-rotating vortices leads to the formation of alternating heating and cooling streaks on the ramp surface along the spanwise direction downstream of the reattachment location.
Amplification of the spanwise velocity perturbations occurs near the reattachment point and where the leading edge shock impinges on the ramp (approximately at $x=0.11$ m).
It can be seen that the numerical predictions of both solvers are in very good agreement regarding the steady perturbations.

\subsection{Perturbed Flow with Ablative Recession}
\label{sec:perturbed_recession}

The camphor ramp is allowed to recess according to the surface reactions and the steady streaks.
For the solid density of camphor, a value of \SI{990}{\kilogram/\cubic\meter} is taken~\cite{rotondi2022numerical}, which dictates the recession speed of the solid $ u_w = \dot{m}_w \big/ \rho_s$. The recession speed acceleration factor discussed in~\autoref{sec:fai_approach} is set to \SI{2e+4}{} based on our 2-D preliminary analysis to obtain results within a reasonable number of computational hours.

\begin{figure} [tb]
\centering
	\begin{subfigure}[t]{.49\textwidth}\centering
		\includegraphics[clip, trim=0.1cm 6cm 1.5cm 6cm, 
		width=0.999\textwidth]{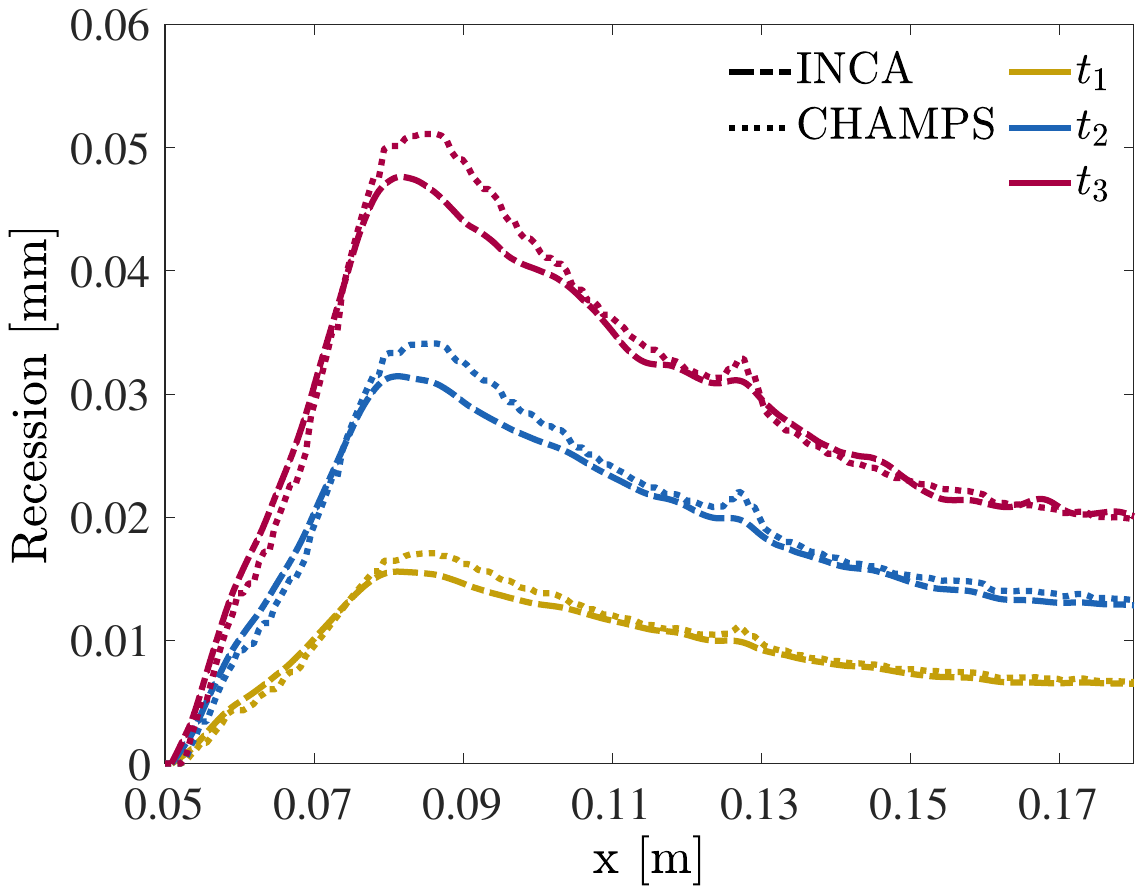}
		\put(-230,167){(a)}
	\end{subfigure}
	\begin{subfigure}[t]{.49\textwidth}\centering
		\includegraphics[clip, trim=0.1cm 6cm 1.5cm 6cm, 
		width=0.999\textwidth]{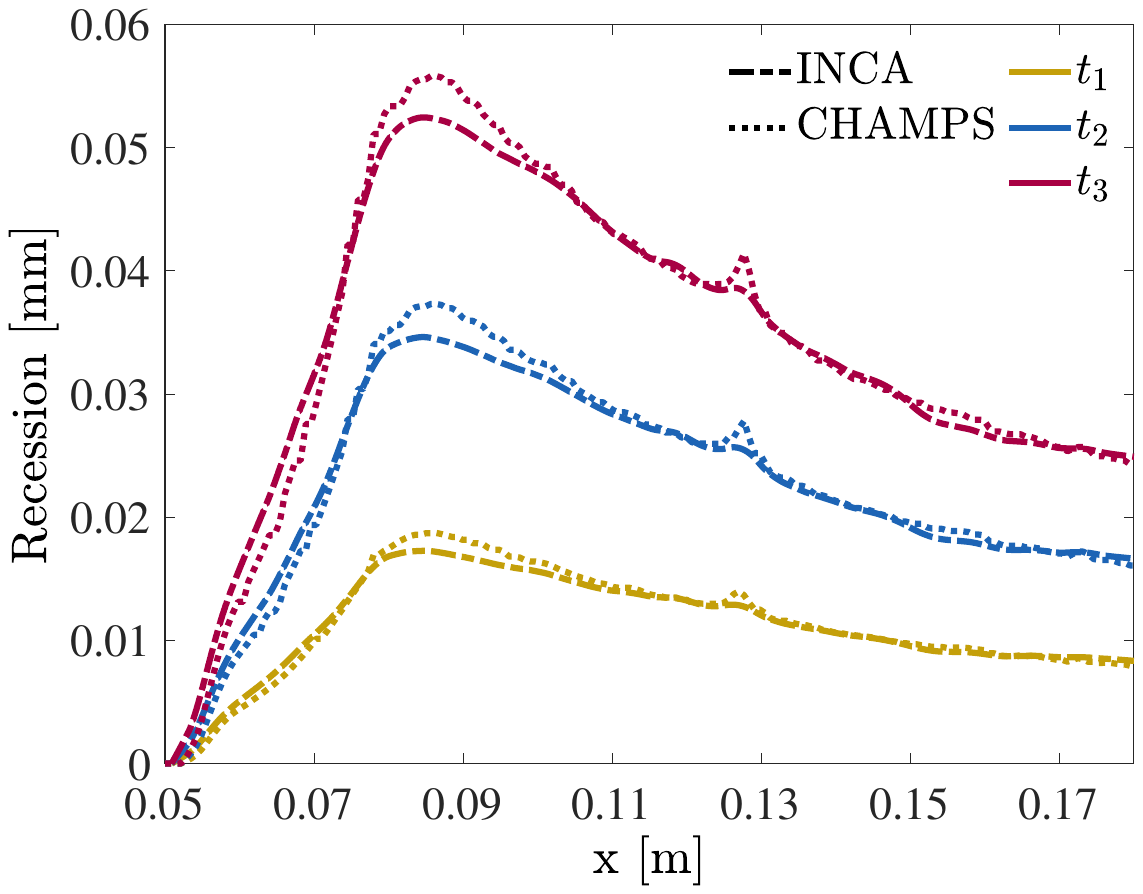}
		\put(-230,167){(b)}
	\end{subfigure}
\caption{Comparison of surface recession profiles along (a) $z=0.0$ m and (b) $z=0.0015$ m over the ablating ramp surface at three time instances.} 
	\label{fig:3D_recession}
\end{figure}

\begin{figure} [tb]
\centering
    \begin{subfigure}[t]{.49\textwidth}\centering
		\includegraphics[clip, trim=0.1cm 6cm 1.5cm 6cm, 
		width=0.999\textwidth]{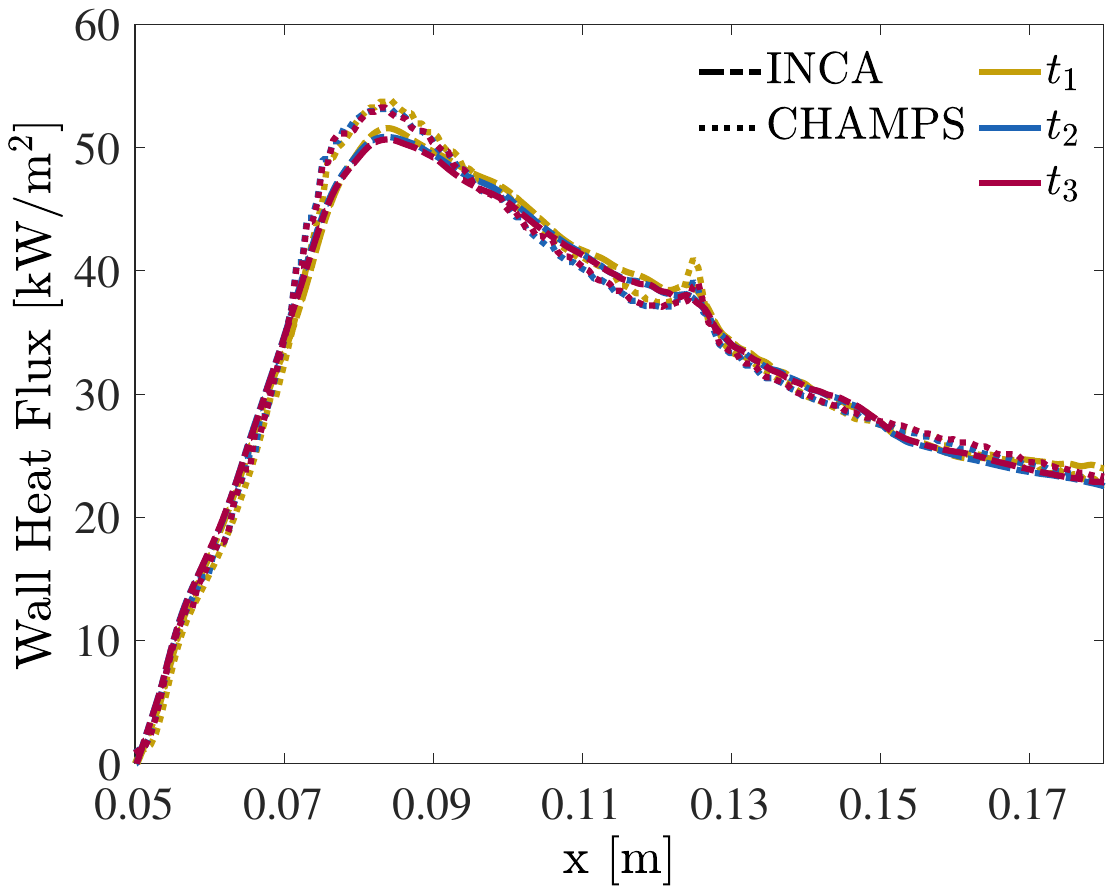}
		\put(-230,167){(a)}
	\end{subfigure}
	\begin{subfigure}[t]{.49\textwidth}\centering
		\includegraphics[clip, trim=0.1cm 6cm 1.5cm 6cm, 
		width=0.999\textwidth]{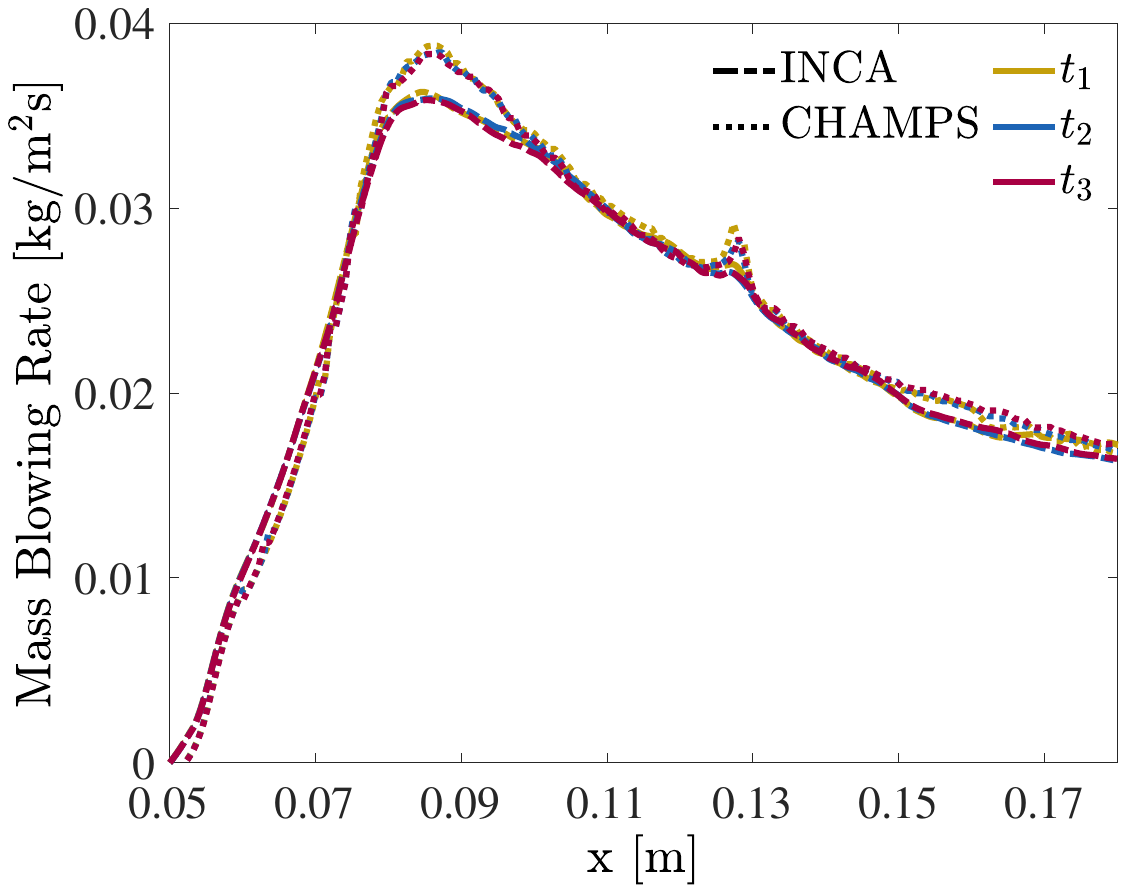}
		\put(-230,167){(b)}
	\end{subfigure}
\caption{Comparison of (a) heat fluxes and (b) mass blowing rates along the centerline ($z=0.0015$ m) over the ablating ramp surface at three time instances.} 
	\label{fig:3D_qdot_mdot}
\end{figure}

Results from three time instances are presented here: $t_1 = 0.461$ s, $t_2 = 2 t_1 = 0.922$ s, and $t_3 = 3 t_1 = 1.383$ s corresponding to successive material times of ablation. Resulting recession profiles can be seen in~\autoref{fig:3D_recession} along the edge ($z=0.0$ m) and along the centerline ($z=0.0015$ m) of the ramp. These profiles are directly linked to the distribution of surface heat fluxes and mass blowing rates plotted in~\autoref{fig:3D_qdot_mdot}, which remain mostly unchanged at these levels of recession. We would expect a constant recession speed for these surface quantities and we observe this in~\autoref{fig:3D_recession}. 
The solvers demonstrate very good agreement, consistent with prior comparisons. The minor mismatches near reattachment could lead to slight variations in recession profiles over time, but is deemed to be negligible considering the overall quality of the agreement.

\begin{figure} [tb]
\centering
    \begin{subfigure}[t]{.49\textwidth}\centering
		\includegraphics[clip, trim=0.1cm 6cm 1.5cm 6cm, 
		width=0.999\textwidth]{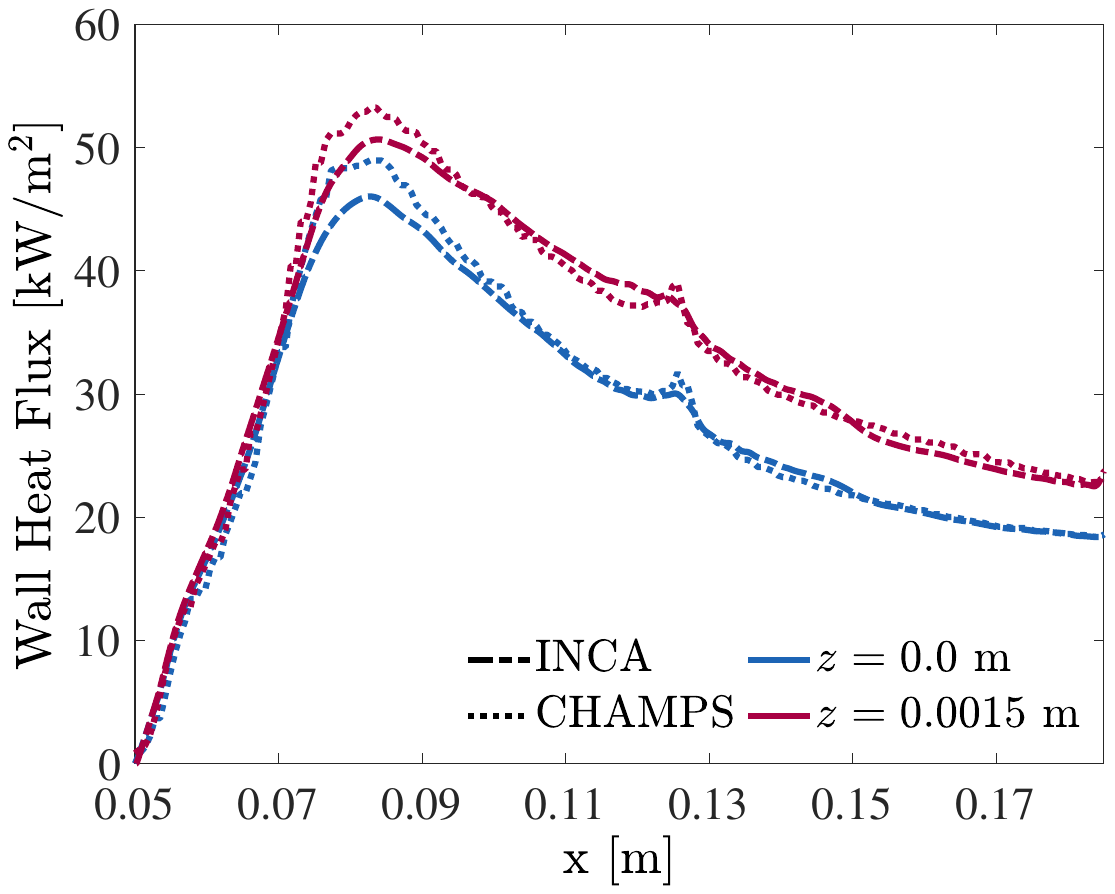}
		\put(-230,167){(a)}
	\end{subfigure}
	\begin{subfigure}[t]{.49\textwidth}\centering
		\includegraphics[clip, trim=0.1cm 6cm 1.5cm 6cm, 
		width=0.999\textwidth]{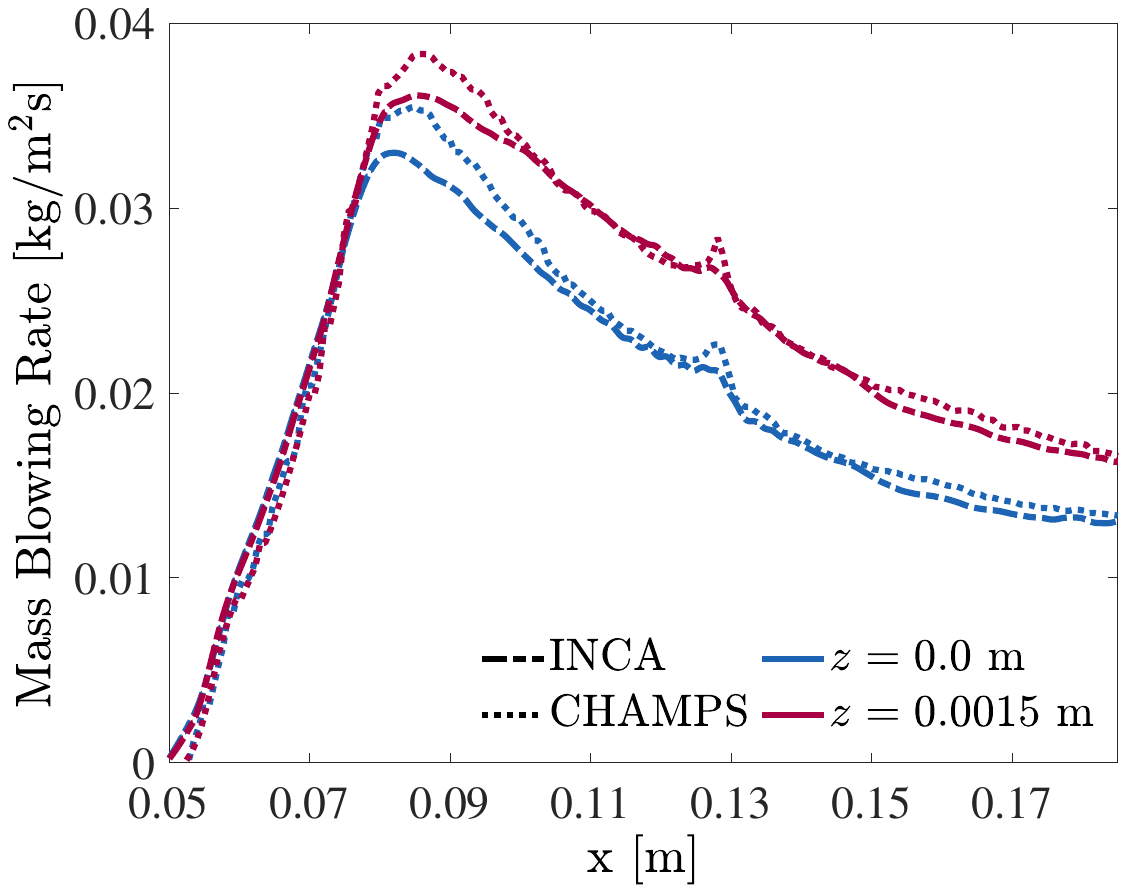}
		\put(-230,167){(b)}
	\end{subfigure}
	\begin{subfigure}[t]{.49\textwidth}\centering
		\includegraphics[clip, trim=0.1cm 6cm 1.5cm 6cm, 
		width=0.999\textwidth]{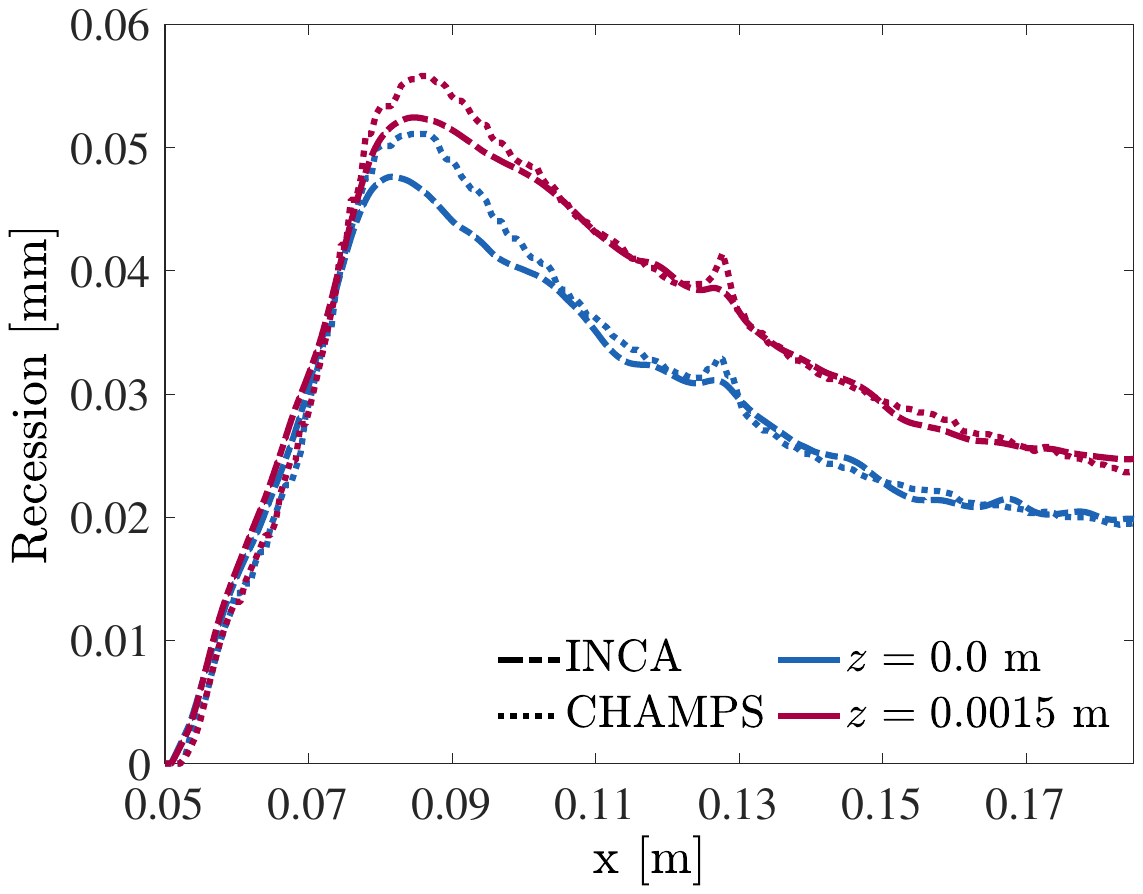}
		\put(-230,167){(c)}
	\end{subfigure}
	\begin{subfigure}[t]{.49\textwidth}\centering
		\includegraphics[clip, trim=0.1cm 6cm 1.5cm 6cm, 
		width=0.999\textwidth]{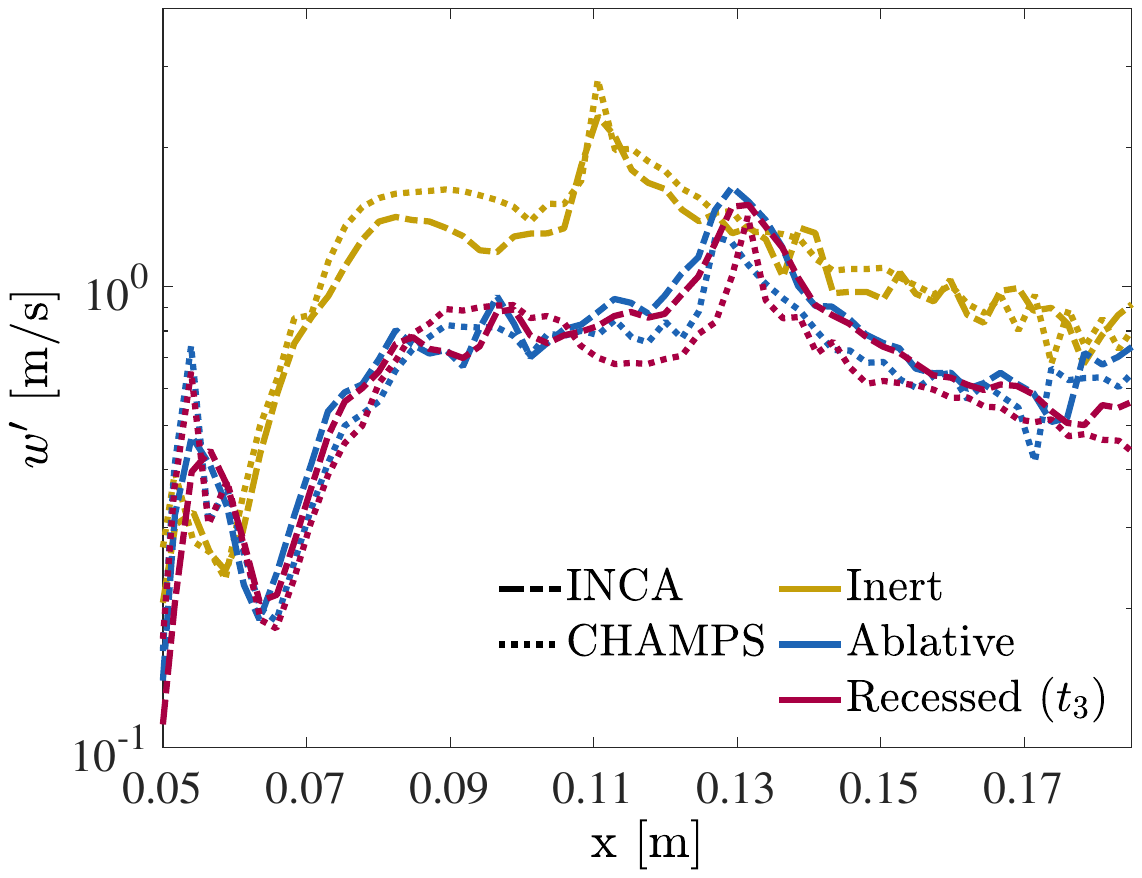}
		\put(-230,167){(d)}
	\end{subfigure}
\caption{Comparison of (a) heat fluxes, (b) mass blowing rates, (c) recession over the ablating ramp surface at time instance $t_3$, and (d) spanwise velocity perturbations for the three surface states. } 
	\label{fig:3D_rec_mid_v_side}
\end{figure}

The deeper recession along the centerline seen in~\autoref{fig:3D_recession}b can be more closely inspected in~\autoref{fig:3D_rec_mid_v_side} comparing the two streamwise distributions at time instance $t_3$. It can be seen that higher heat fluxes and mass blowing rates lead to more recession along the centerline.
\Autoref{fig:3D_rec_mid_v_side}d shows how the spanwise perturbations vary for the inert wall, the ablative rigid wall, and the wall recessing with ablation. The effect of ablation by itself lowers the amplitude of these perturbations and delays the peak value achieved. This peak is due to the impinging leading-edge shock after it intersects with the reattachment shock. Through this shock-shock interaction, the refracted leading-edge shock impacts the ramp. For the reacting wall, the reattachment shock lies closer to the ramp and therefore intersects the leading-edge shock slightly farther downstream, resulting in this shift in the peak amplitude of spanwise velocity perturbations.
The recessed surface does not affect the perturbation velocity significantly.
All of these observations are confirmed by both of the solvers.


\subsubsection{Influence of Ablative Speed-Up}
\label{sec:speed-up}

Influence of the ablative speed-up factor is further analyzed by testing faster and slower accelerations similar to the preliminary tests in~\autoref{sec:pre-speed-up}. We have performed three simulations starting from the same initial flow state and the unablated surface profile. The speed-up factor of \SI{2e+4}{} employed in earlier simulations is kept as the baseline case, while taking a faster speed-up factor of \SI{4e+4}{} and a slower one of \SI{1e+4}{}.

\begin{figure} [tb]
 	\centering
 	\includegraphics[clip, trim=0cm 0cm 0cm 0cm, 
 	width=0.49\textwidth]{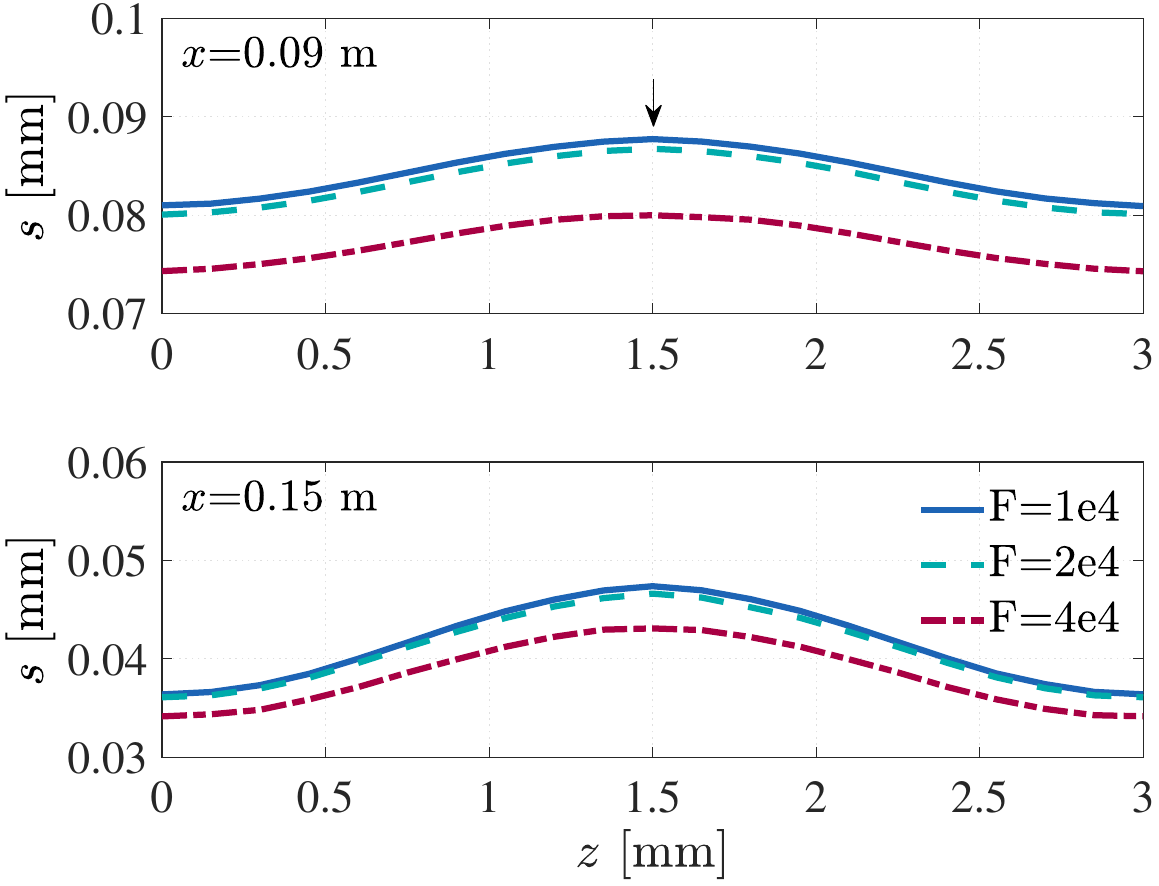}
 	\includegraphics[clip, trim=0cm 0cm 0cm 0cm, 
 	width=0.49\textwidth]{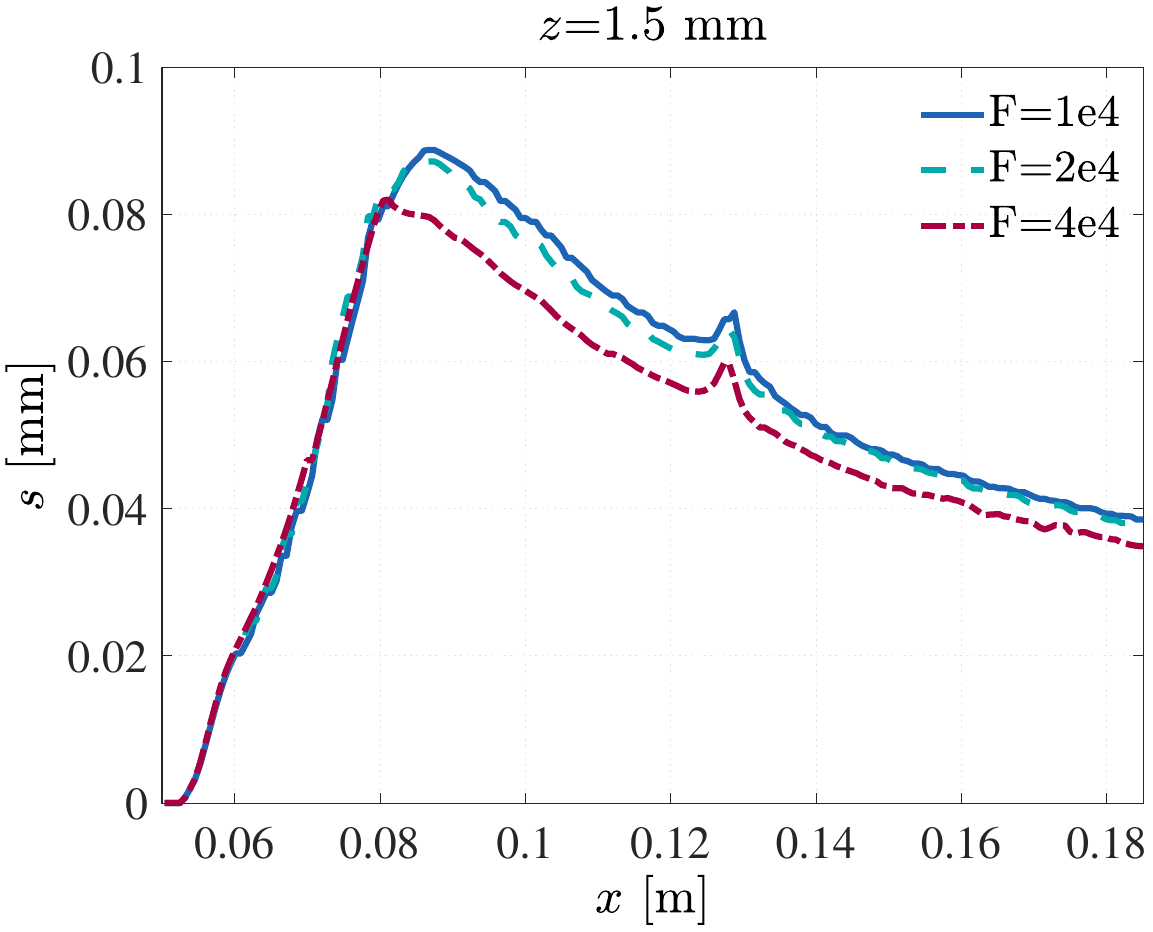}
 	\caption{Comparison at $t_{rec}=2.162$ seconds of spanwise (left) and streamwise (right) recession profiles for the low-amplitude case with an ablative recessing wall employing slow ($\mathrm{F}=\SI{1e+4}{}$), baseline ($\mathrm{F}=\SI{2e+4}{}$), and fast ($\mathrm{F}=\SI{4e+4}{}$) ablative speed-up factors.}
 	\label{fig:z_x_recession_slow_base_fast}
 \end{figure}

Similar to~\autoref{sec:pre-speed-up}, a consistent recession time $t_{rec}$, defined as the duration the fluid advances times the speed-up factor, is compared between the three cases in \autoref{fig:z_x_recession_slow_base_fast} at $t_{rec}=2.162$ seconds. The spanwise distributions of recession at select streamwise locations, namely $x=\{0.09,0.15\}$ m are shown in the left panel. The action of the steady streaks produces a spanwise modulation in the heat flux. As a result, the recession profile exhibits a maximum in the middle ($z=1.5$ mm) where peak heating is experienced. At both streamwise locations we observe good agreement of the profiles for $\mathrm{F}=\SI{1e+4}{}$ and $\mathrm{F}=\SI{2e+4}{}$. Consistent with the findings from the 2D study in~\autoref{sec:pre-speed-up}, values larger than $\mathrm{F}=\SI{2e+4}{}$, in this case $\mathrm{F}=\SI{4e+4}{}$, results in an under prediction of the final recession profile. The errors incurred by the speed-up factor manifest primarily as a consistent offset in the profiles, likely a result of the differential heating only being a small fraction of the mean heating value. This would of course change if the streak forcing amplitude were increased to yield larger amplitude heating streaks. We confirm the agreement between the $\mathrm{F}=\SI{1e+4}{}$ and $\mathrm{F}=\SI{2e+4}{}$ cases hold at all streamwise locations in the right panel of \autoref{fig:z_x_recession_slow_base_fast}. Thus, we can conclude that the baseline speed-up factor does not negatively affect the flow stability (i.e., steady streaks).

\subsubsection{Influence of Perturbation Amplitude}
\label{sec:influence_amplitude}

\begin{figure}[htb]
\centering
        \begin{subfigure}[t]{.9\textwidth}\centering
        \includegraphics[clip, trim=0.1cm 0.1cm 0.1cm 0.1cm, width=0.8\columnwidth]{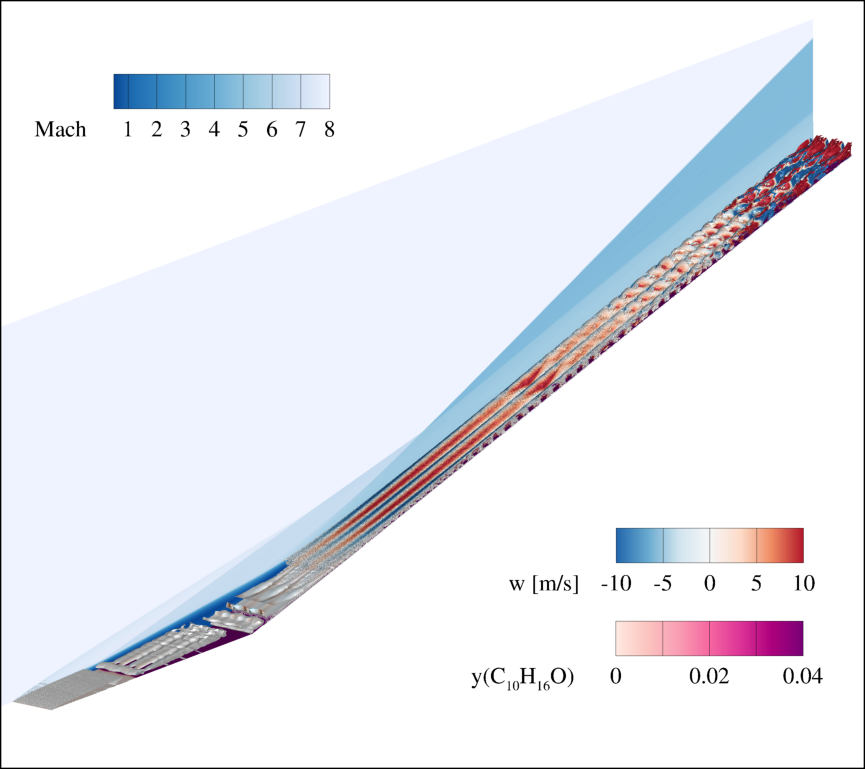}\centering
        \put(-220,50){\RN{1}}
	\put(-135,120){\RN{2}}
	\put(-50,190){\RN{3}}
        \end{subfigure} 
        \\
        \centering
        \begin{subfigure}[t]{.25\textwidth}\centering
            \includegraphics[clip, trim=7cm 4cm 3cm 6cm, width=0.99\columnwidth]{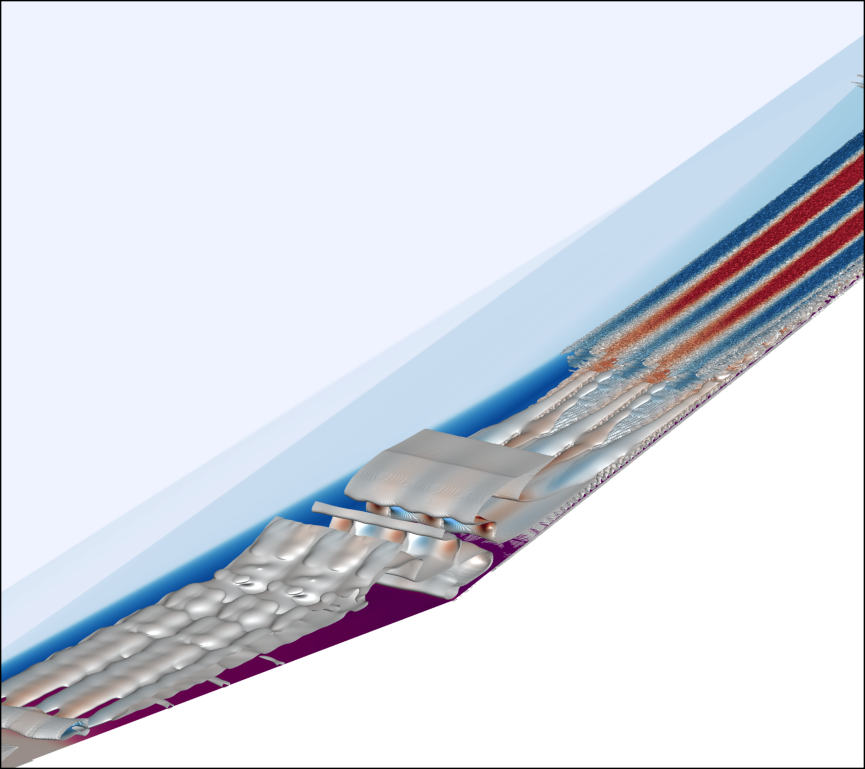}\centering
            \put(-10,5){\RN{1}}
        \end{subfigure}
        \begin{subfigure}[t]{.25\textwidth}\centering
            \includegraphics[clip, 
            trim=4cm 3cm 6cm 7cm, width=0.99\columnwidth]{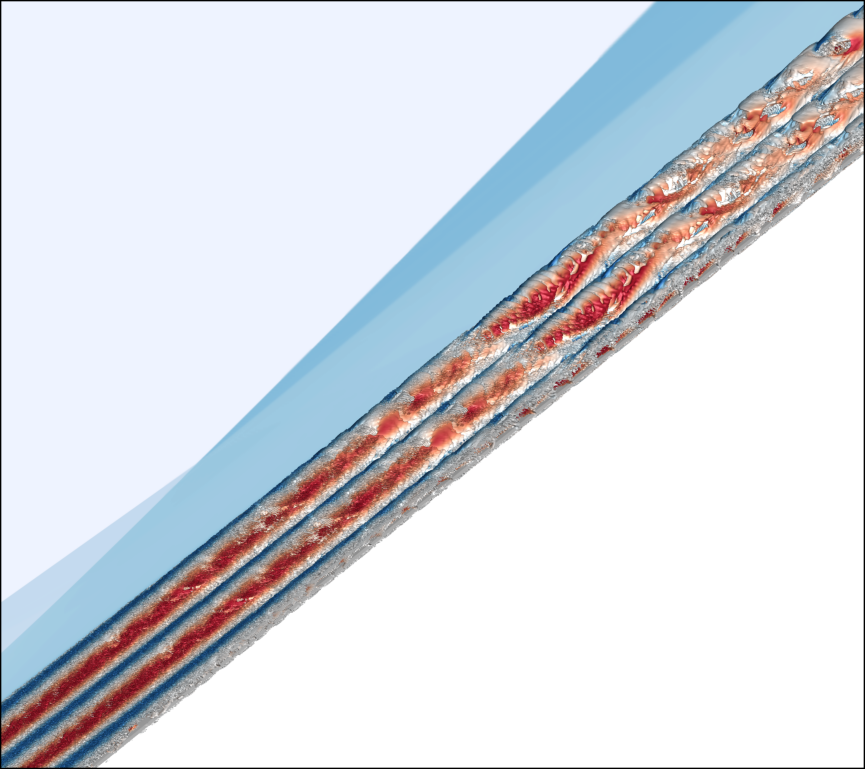}\centering
            \put(-10,5){\RN{2}}
        \end{subfigure}
        \begin{subfigure}[t]{.25\textwidth}\centering
            \includegraphics[clip, 
            trim=6cm 6cm 4cm 4cm, width=0.99\columnwidth]{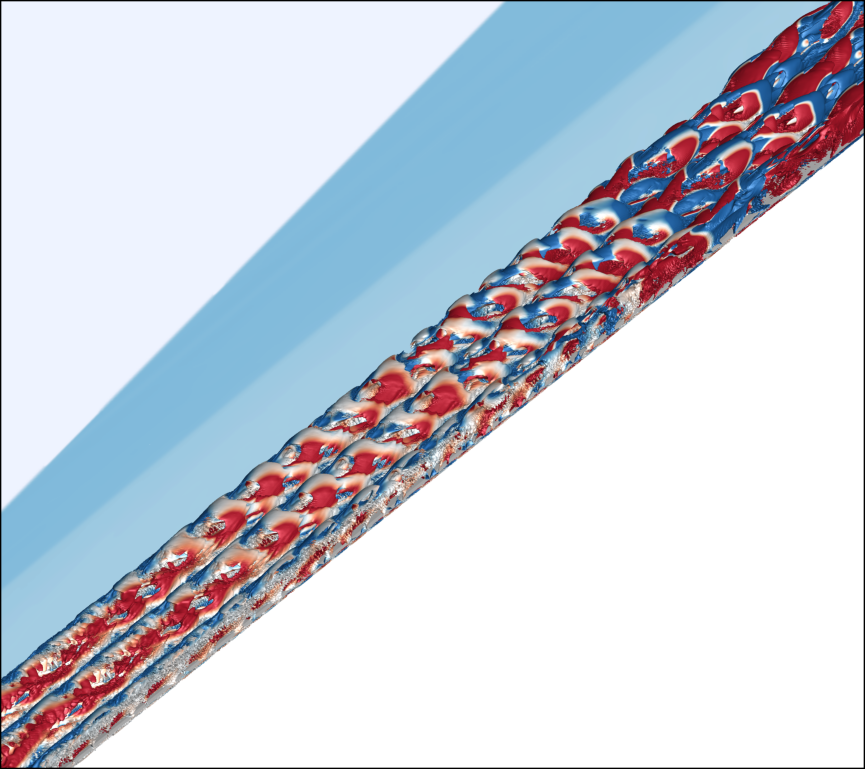}\centering
            \put(-10,5){\RN{3}}
        \end{subfigure}
        \vfill
        \vspace{2mm}
        \vfill
        \centering
        \begin{subfigure}[t]{.25\textwidth}\centering
            \includegraphics[clip, 
            trim=7cm 4cm 3cm 6cm, width=0.99\columnwidth]{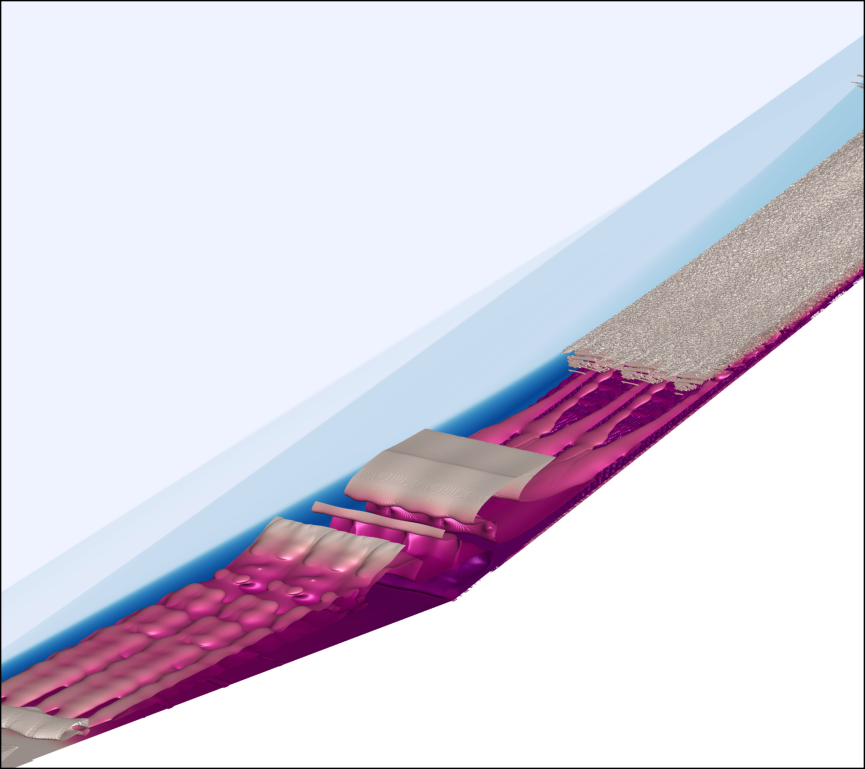}\centering
            \put(-10,5){\RN{1}}
        \end{subfigure}
        \begin{subfigure}[t]{.25\textwidth}\centering
            \includegraphics[clip, 
            trim=4cm 3cm 6cm 7cm, width=0.99\columnwidth]{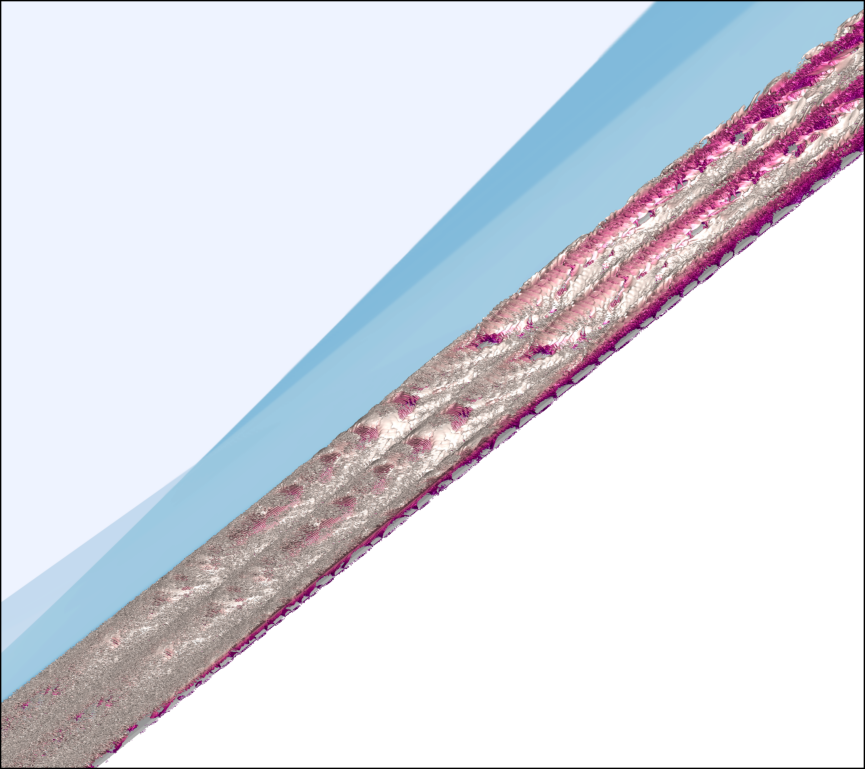}\centering
            \put(-10,5){\RN{2}}
        \end{subfigure}
        \begin{subfigure}[t]{.25\textwidth}\centering  
            \includegraphics[clip, trim=6cm 6cm 4cm 4cm, width=0.99\columnwidth]{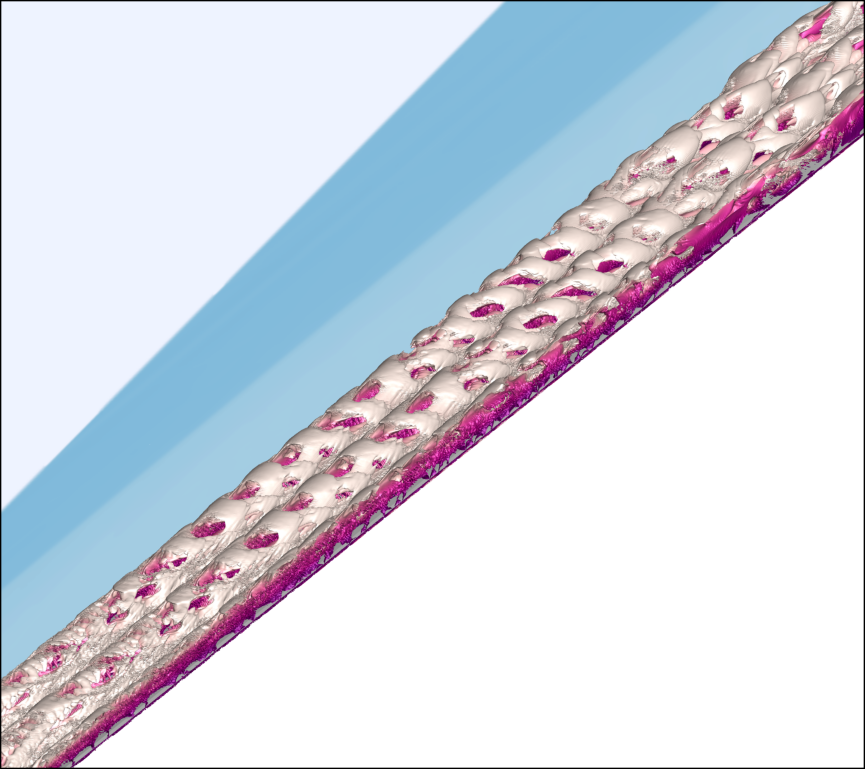}\centering
            \put(-10,5){\RN{3}}
        \end{subfigure}\\
    \caption{Flow visualization of the high-amplitude ablative-wall solution duplicated in the spanwise direction to improve clarity. Isometric view of the domain with Q-criterion iso-surfaces is contoured with the spanwise velocity. Background slice in the spanwise direction shows Mach number contours. Three close-up views focus on (\RN{1}) the steady streaks reattaching on the ramp surface, (\RN{2}) signs of transition, and (\RN{3}) the onset of turbulent breakdown. Camphor mass fractions are also shown in the second set of views.}
    \label{fig:cool_wfine_1plus3}
\end{figure}

In order to investigate the sensitivity to the forcing amplitude, we have performed simulations where the forcing amplitude is increased tenfold to generate a more significant differential between the heating-cooling streaks and to incentivize transition to turbulence within the computational domain.
Higher perturbation amplitudes are excluded from the scope of the current study due to more complex physical interactions in the boundary layer identified by Başkaya and Hickel~\cite{baskaya2025baroclinic}.
The flow field is visualized in~\autoref{fig:cool_wfine_1plus3}. 
The locations of interest, namely the reattachment of the streaks, onset of transition, and the glimpses of turbulent breakdown are identified in the close-up figures.
Camphor mass fraction contours are also provided to visualize the transport of the ablation products in the boundary layer.

\begin{figure} [tb]
\centering
    \begin{subfigure}[t]{.49\textwidth}\centering
		\includegraphics[clip, trim=0.1cm 6cm 1.5cm 6cm, 
		width=0.999\textwidth]{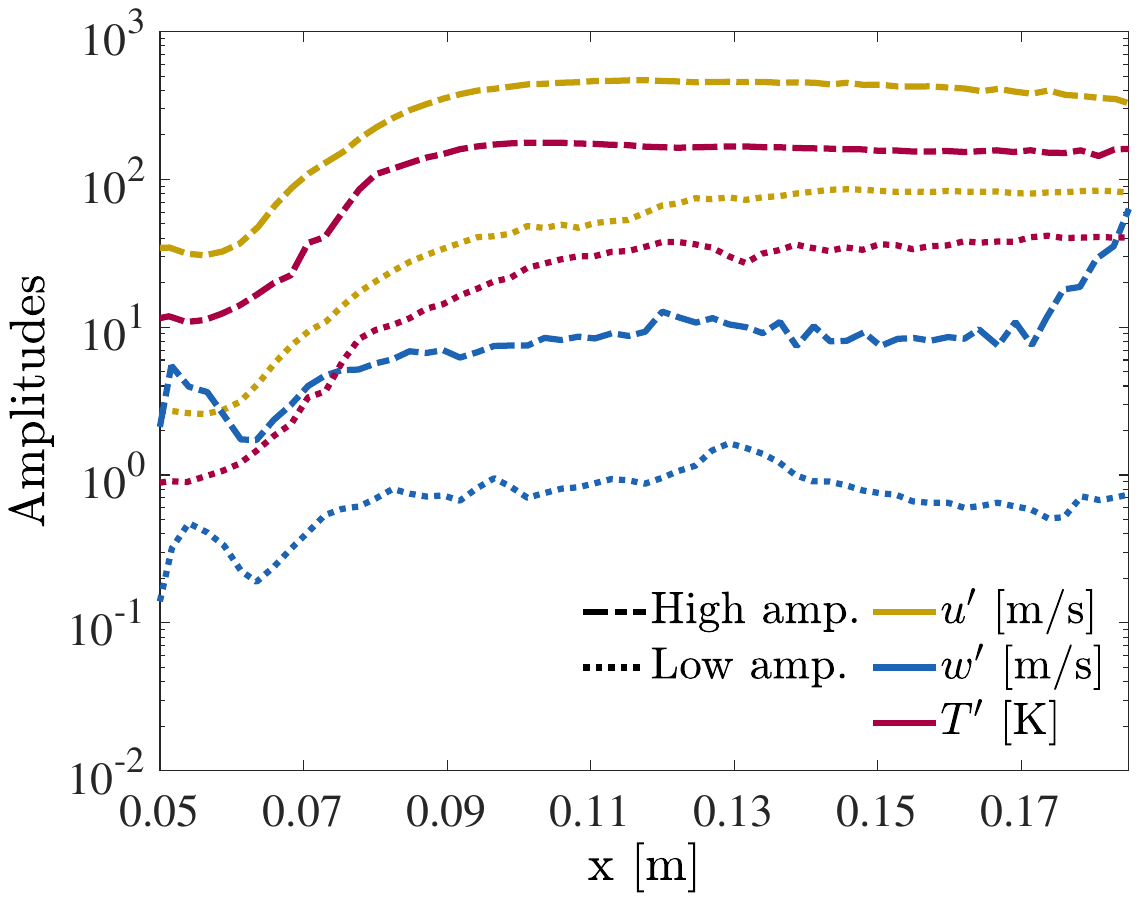}
		\put(-230,167){(a)}
	\end{subfigure}
	\begin{subfigure}[t]{.49\textwidth}\centering
		\includegraphics[clip, trim=0.1cm 6cm 1.5cm 6cm, 
		width=0.999\textwidth]{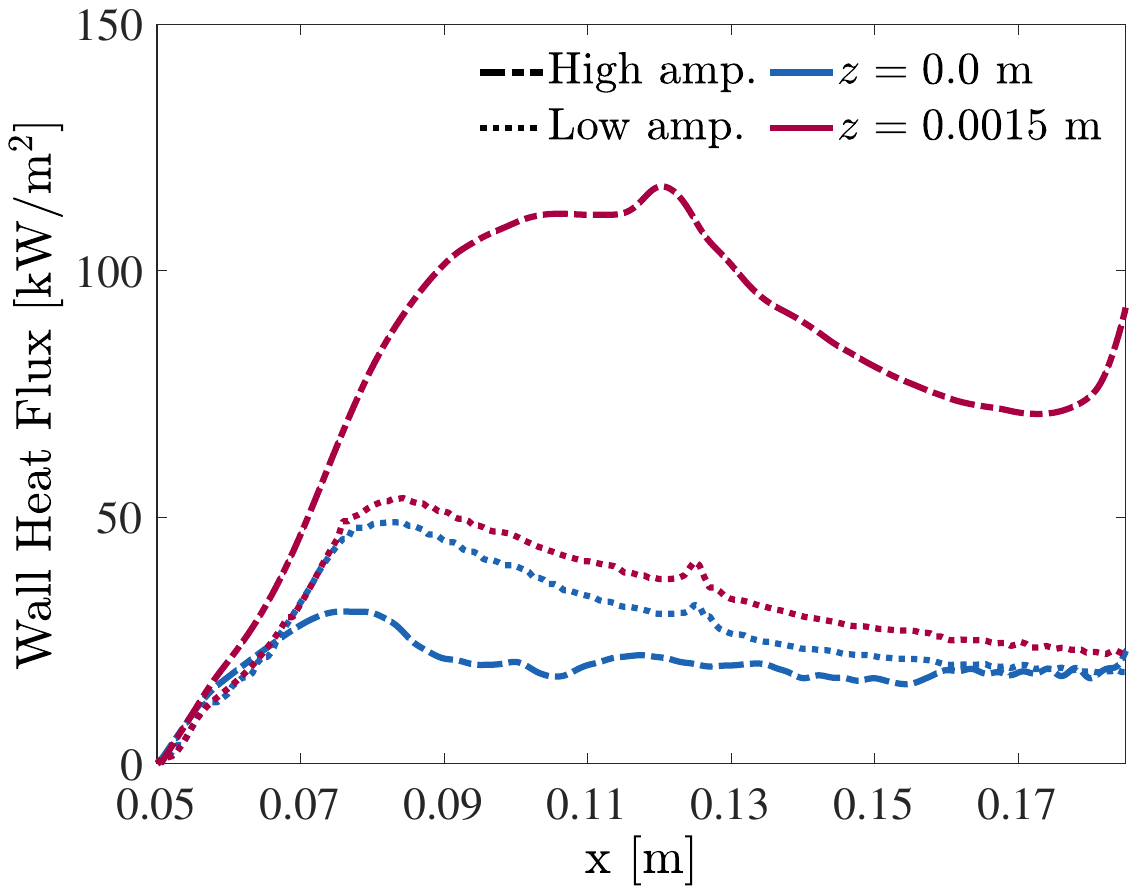}
		\put(-230,167){(b)}
	\end{subfigure}
	\caption{Comparison of (a) streamwise velocity, spanwise velocity, and temperature perturbations and (b) surface heat fluxes for the perturbed ablative compression ramp for the high and low amplitude cases.}
	\label{fig:inert-low_vs_high-amp-amps}
\end{figure}

The introduction of high-amplitude forcing generates stronger streaks. The lift-up effect of the counter-rotating vortices enhances transport, and consequently a more drastic variation in the heating-cooling streaks is obtained. This can be clearly observed in~\autoref{fig:inert-low_vs_high-amp-amps}, where perturbation data and wall heat fluxes are presented for the higher amplitude ablative results. 
In the linear stability regime, a tenfold increase in the imposed forcing amplitude directly leads to the same factor of change for the temperature perturbations.
Towards the downstream end of the domain, stronger perturbation amplification is observed for the spanwise velocity perturbations where the flow begins to destabilize.
Whereas the low-amplitude configuration produced approximately a 10\% difference in surface quantities between the centerline and the ramp edge, the elevated forcing amplitude results in a factor of five in surface heat flux difference.
The location where the leading-edge shock impinges is more apparent along the centerline with a peak in heat fluxes occurring slightly more upstream than the low-amplitude case. Higher heat fluxes are also observed towards the end of the domain, once again suggesting increased nonlinearity.

\begin{figure} [tb]
\centering
    \begin{subfigure}[t]{.49\textwidth}\centering
		\includegraphics[clip, trim=0.1cm 6cm 1.5cm 6cm, 
		width=0.999\textwidth]{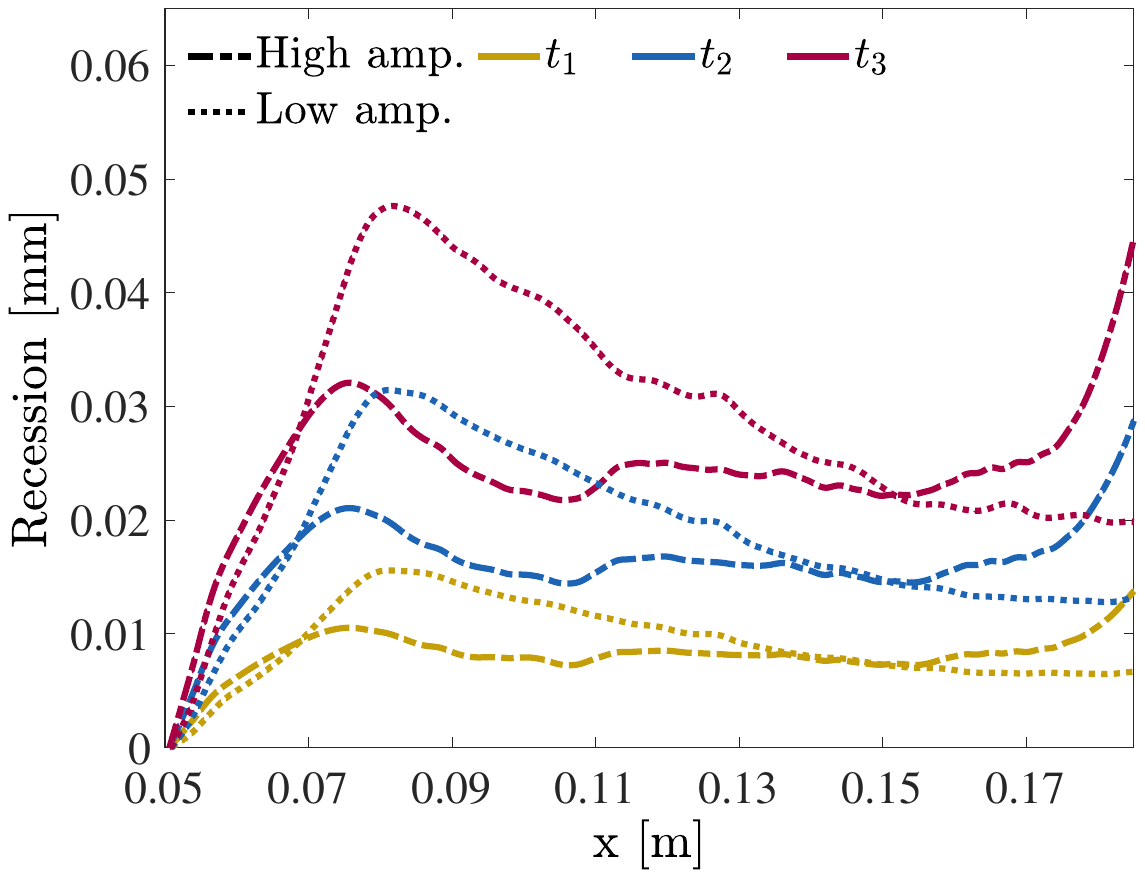}
		\put(-230,167){(a)}
	\end{subfigure}
	\begin{subfigure}[t]{.49\textwidth}\centering
		\includegraphics[clip, trim=0.1cm 6cm 1.5cm 6cm, 
		width=0.999\textwidth]{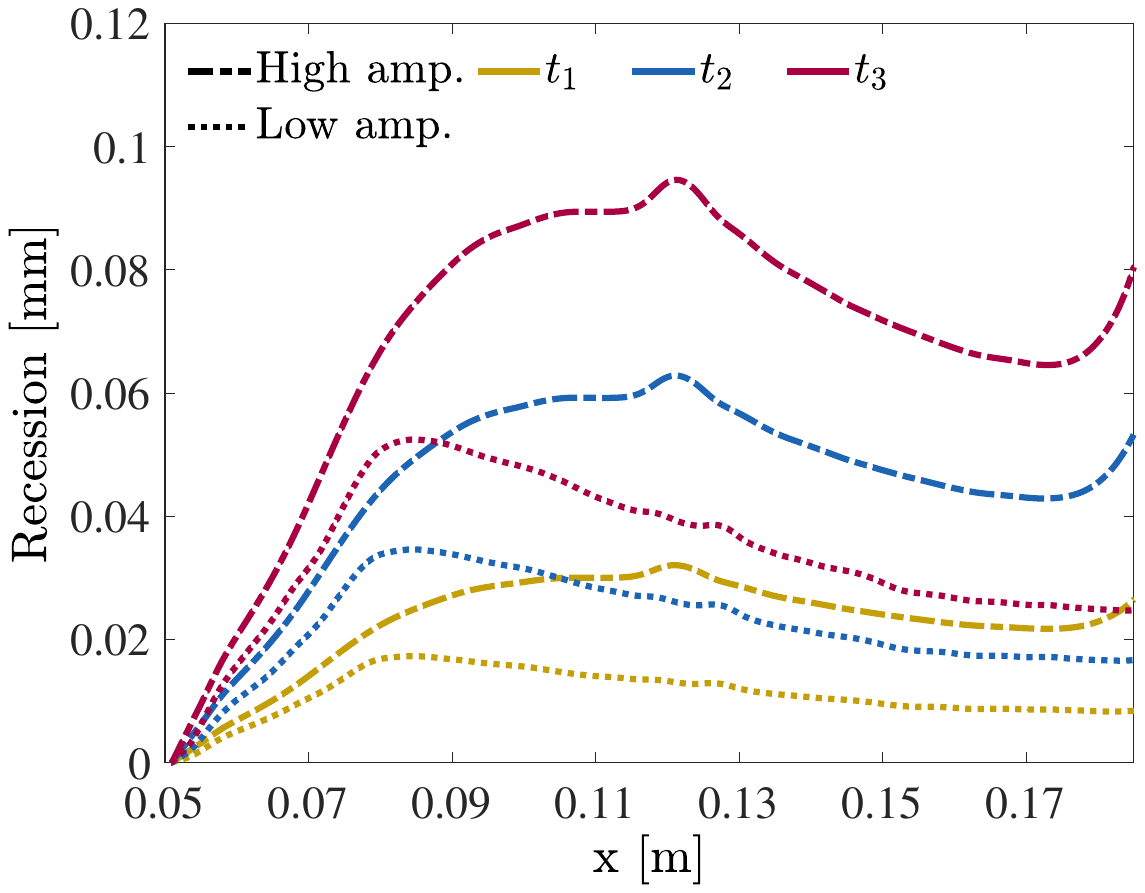}
		\put(-230,167){(b)}
	\end{subfigure}
	\caption{Comparison of surface recession profiles along (a) $z=0.0$ m and (b) $z=0.0015$ m over the ablating ramp surface at three time instances for the high and low amplitude forcings.}
	\label{fig:3D_recession_high_amp}
\end{figure}

The surface recession is shown in~\autoref{fig:3D_recession_high_amp} for the same time durations described in~\autoref{sec:perturbed_recession}.
Along the cooling streak ($z=0.0$ m), due to the lower heat fluxes and consequently lower mass blowing rates, the amount of recession is less than the low-amplitude forcing results. Towards the end of the domain, however, there is formation of deeper grooves corresponding to where the perturbations grow. 
In contrast, along the heating streak ($z=0.0015$ m) there is substantially more recession compared to the lower amplitude forcing configuration. Near reattachment and the downstream end of the domain longer exposure leads to deeper recession.
Recession speeds remain mostly constant in time as before.

\begin{figure} [tb]
\centering
    \begin{subfigure}[t]{.49\textwidth}\centering
		\includegraphics[clip, trim=0.1cm 6cm 1.5cm 6cm, 
		width=0.999\textwidth]{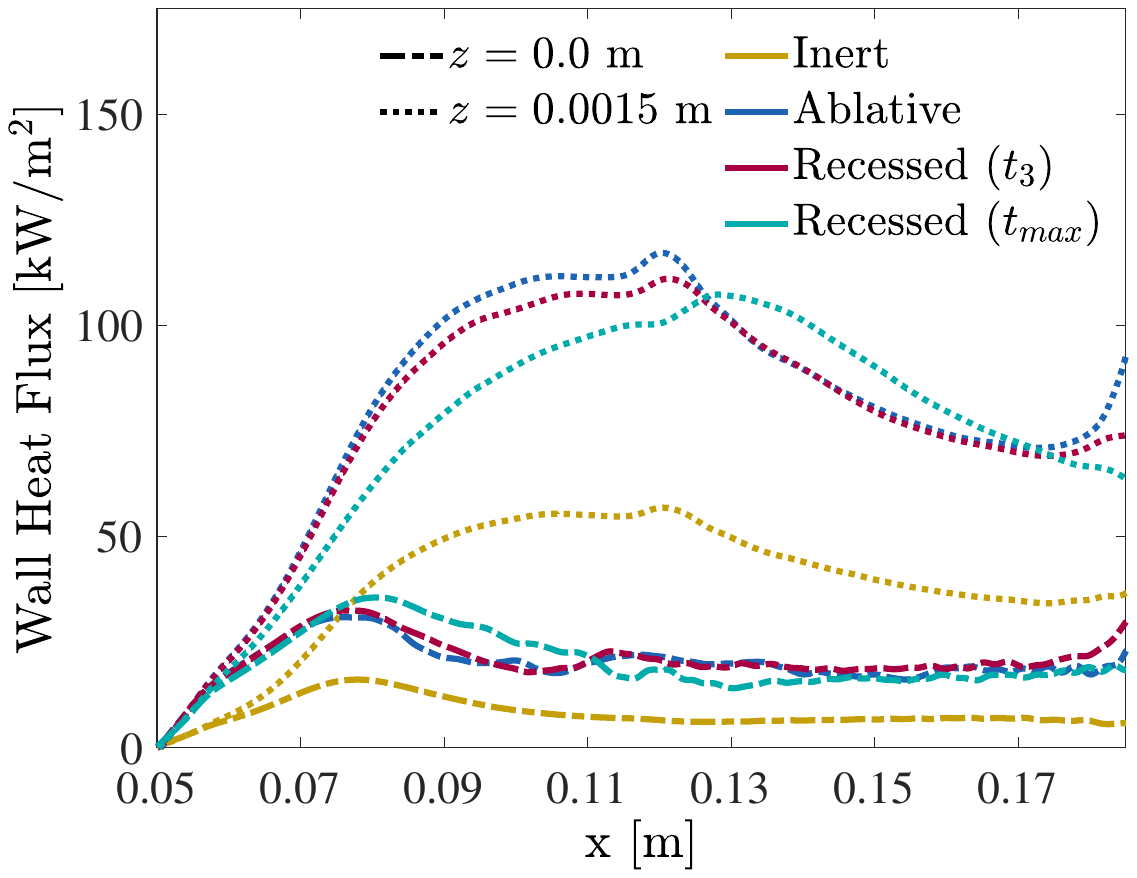}
		\put(-230,167){(a)}
	\end{subfigure}
 \hspace{0.1cm}
	\begin{subfigure}[t]{.49\textwidth}\centering
		\includegraphics[clip, trim=0.1cm 6cm 1.5cm 6cm, 
		width=0.999\textwidth]{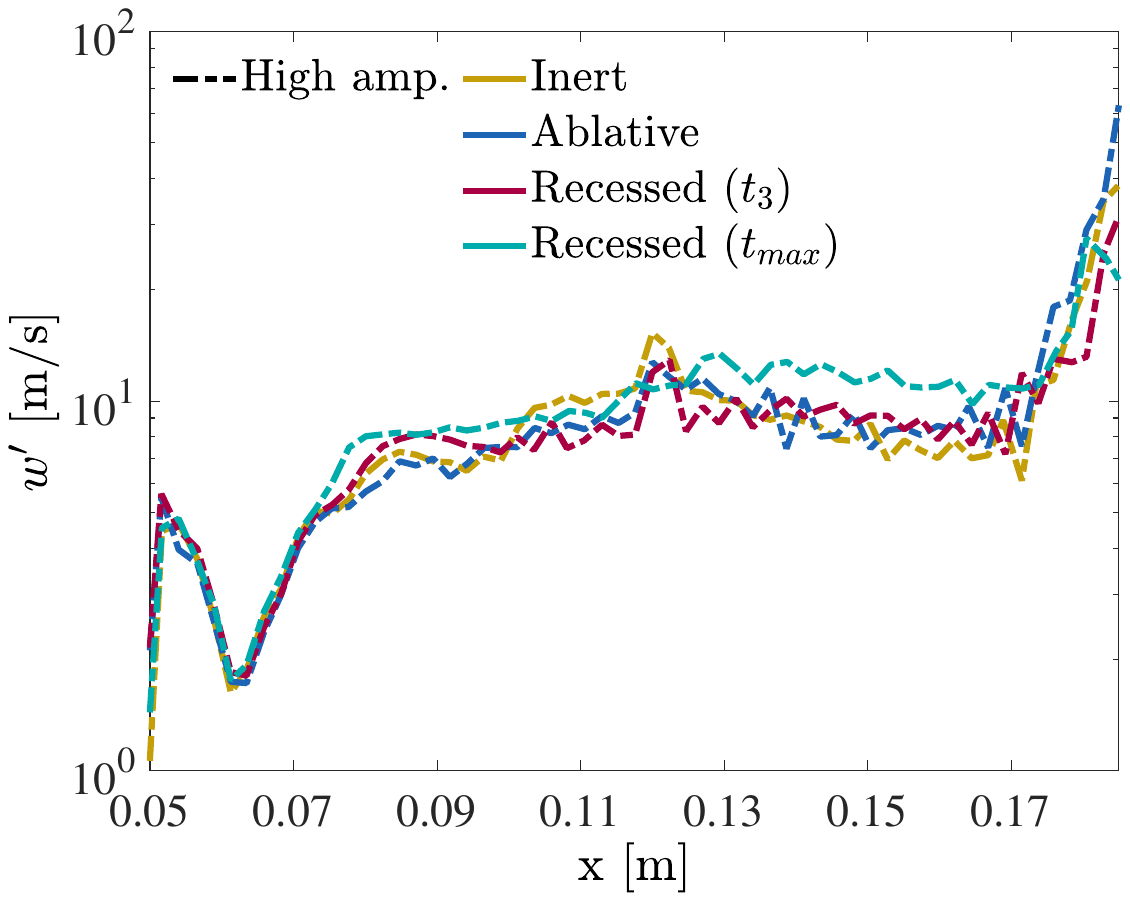}
		\put(-230,167){(b)}
	\end{subfigure}
\hspace{0.1cm}
	\begin{subfigure}[t]{.49\textwidth}\centering
		\includegraphics[clip, trim=0.1cm 6cm 1.5cm 6cm, 
		width=0.999\textwidth]{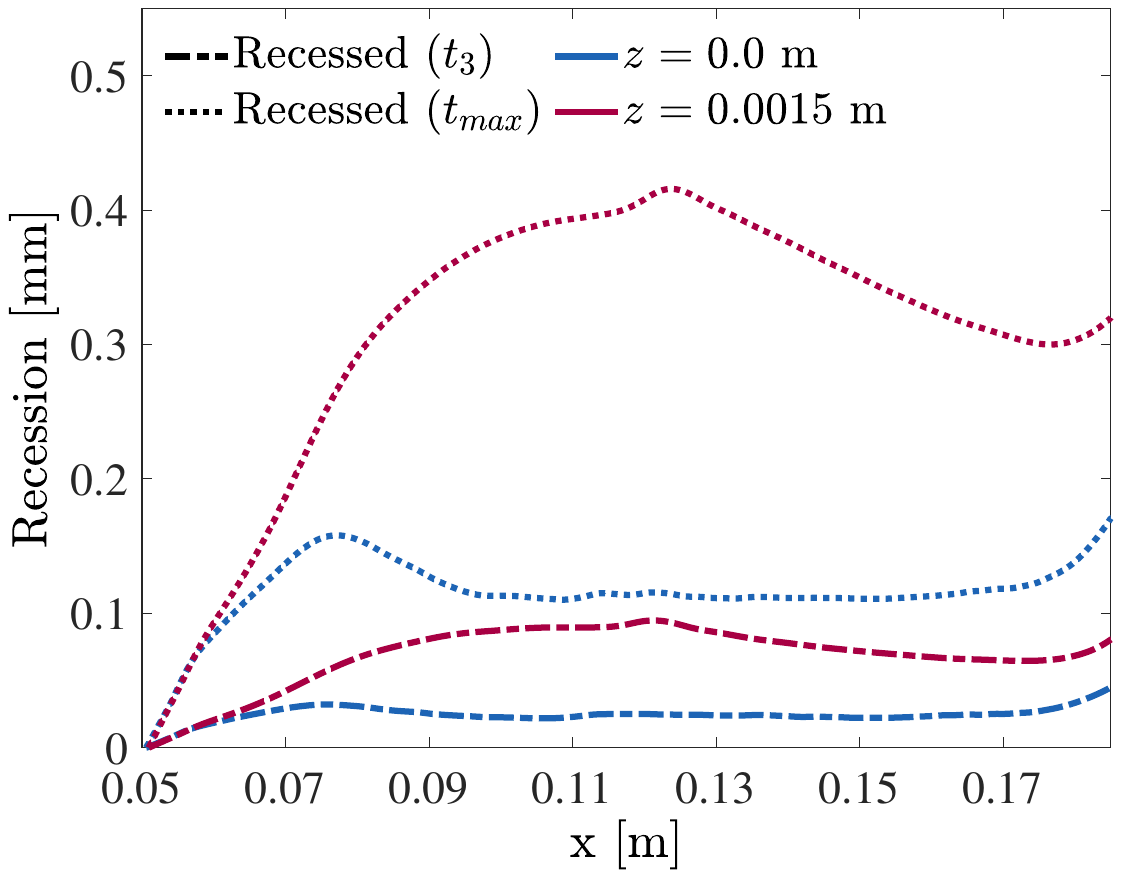}
		\put(-230,167){(c)}
	\end{subfigure}
\hspace{0.1cm}
	\begin{subfigure}[t]{.49\textwidth}\centering
		\includegraphics[clip, trim=0.1cm 6cm 1.5cm 6cm, 
		width=0.999\textwidth]{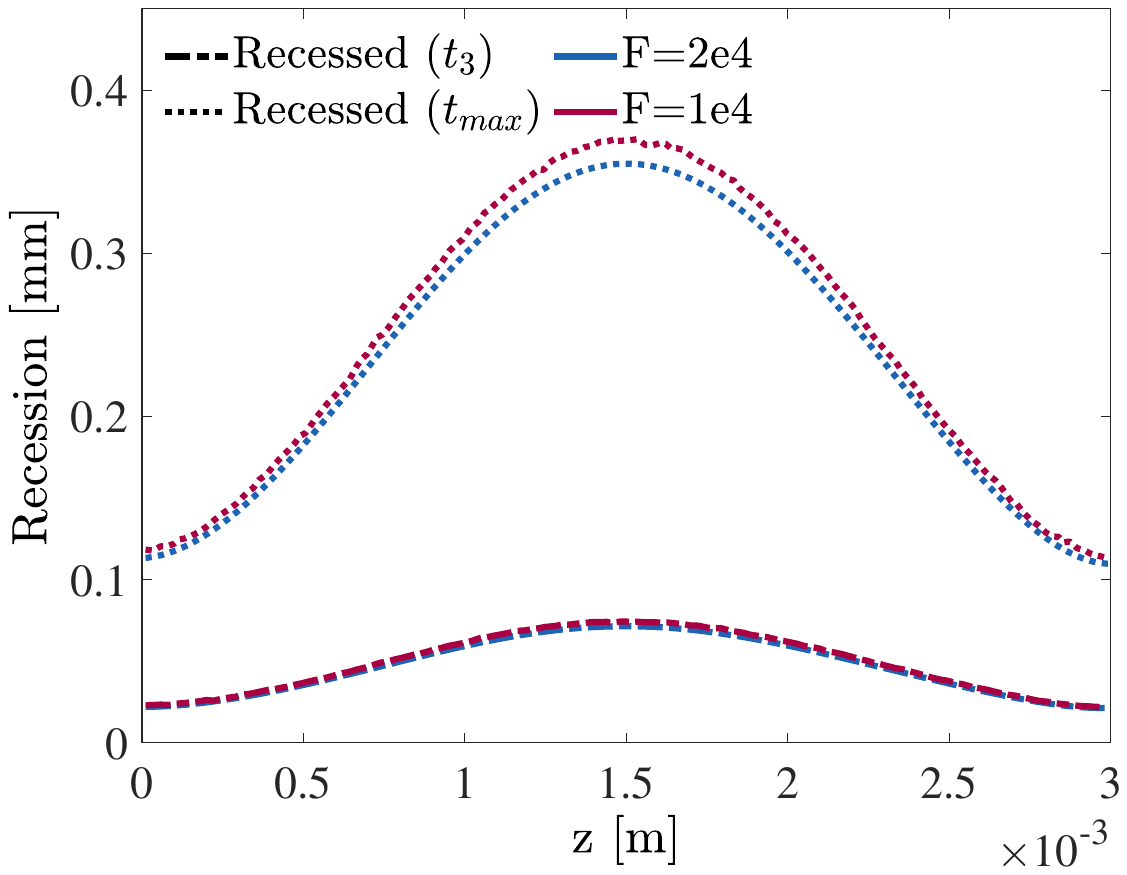}
		\put(-230,167){(d)}
	\end{subfigure}
	\caption{Comparison of (a) surface heat fluxes, (b) spanwise velocity perturbations, (c) streamwise recession profiles, and (d) spanwise recession profiles ($x = 0.15$) for the ablative compression ramp with high-amplitude forcing.}
	\label{fig:high-amp-inert_blow_rect}
\end{figure}

The aforementioned trends can also be studied in light of~\autoref{fig:high-amp-inert_blow_rect}, where surface heat fluxes, spanwise velocity perturbations, and recession profiles are presented for the flow over the ablative ramp at $t_3$ and at $t_{max}=5t_3$ of recession time with the latter included to better analyze the impact of more notable recession.
Naturally, the ablative surface conditions lead to higher total heat fluxes with the addition of the diffusive heat flux.
An interesting observation here is that the heat fluxes are slightly lower along the centerline after recession, indicating that the recession speed is beginning to slow down. 
The recession profiles confirm this as the recession at $t_{max}$ is approximately 10\% lower than if the recession rate had remained constant.
As the surface keeps recessing away, the dent created by the leading edge shock impingement shifts the location of the centerline peak heating farther downstream.

\Autoref{fig:high-amp-inert_blow_rect}b presents the variation of the spanwise velocity perturbations for the three surface states. 
Prior to flow reattachment, $x\approx0.07$, the spatial growth rate of the perturbations remains largely independent of the surface state. This behavior indicates that the disturbances still behave as small perturbations and the amplification is predominantly governed by the shear layer and the extent of the separation bubble until the nonlinear interactions begin.
The previously observed stabilization due to surface reactions is present but much less apparent than it was in the low-amplitude forcing conditions (see~\autoref{fig:3D_rec_mid_v_side}). We believe that under these high-amplitude forcing conditions, the impact of ablative blowing is relatively weak compared to the baseline perturbations and the momentum transfer of the primary streaks. 
Similar to the low-amplitude analysis, moderate recession introduces minor destabilization. Further recession at $t_{max}$ leads to a higher plateau beyond reattachment. 
We hypothesize that the trough formed by recession (visualized for clarity with a spanwise slice at $x = 0.15$ in~\autoref{fig:high-amp-inert_blow_rect}d) laterally constrains and effectively traps the vortex streak. By suppressing this spanwise spread, the trench formed by the recessed ramp surface allows the streaks to sustain an elevated saturation level.
This is followed by a sharp amplification driven by the exponential growth of these traveling secondary instabilities just before breaking down beyond the streamwise extent of the domain. The start of this growth remains mostly unaffected by the preceding plateau in the perturbations.

\revadd{\Autoref{fig:high-amp-inert_blow_rect}d additionally includes results obtained with a speed-up factor equal to half the selected baseline value to evaluate how well our earlier assessment in~\autoref{sec:speed-up} holds up at higher amplitude perturbations. The results are reported for the same $t_{rec}$. At $t_3$, the profiles are almost indistinguishable, with the maximum difference reaching approximately 4\% only at $t_{max}$. Although this value is moderately larger than those for the low-amplitude cases, no substantial anomalies attributable to the choice of the speed-up factor are observed at the recession levels considered in this study.}

\clearpage

\section{Conclusions} \label{sec:conclusion}

We have studied fluid-ablation interactions for Mach 8 transitional flow over a $15\degree$ compression ramp. The test case was first simulated in 2-D for establishing the laminar base flow. The setup was then extended to 3-D to superimpose steady perturbations comprised of streamwise counter-rotating vortices, which further downstream lead to streaks in the velocity and temperature fields.
The rigid ramp was then replaced by solid camphor, which ablates under these flow conditions and interacts with the perturbed flow through heat transfer, mass blowing, and surface recession. 

We have presented results from two independently developed solvers, CHAMPS and INCA, using the same fluid-ablation modeling interface, but employing different numerical methods.
Following a meticulous verification and validation campaign, both solvers provided results which qualitatively and quantitatively demonstrated excellent agreement for all of the varying surface states and crucial quantities such as heat fluxes, mass blowing rates, and recession profiles.
In both solvers, the recession speed was accelerated to alleviate the disparity in time scales between fluid flow and material response. The sensitivity of results to the acceleration factor was quantified, and it was confirmed that the selected speed-up factor of $2 \times 10^4$ did not introduce any unphysical anomalies to the flow field.

We have discussed the effects of fluid-ablation interactions on this streak-vortex system. Small perturbations superimposed on the baseflow generate steady streaks that imprint a spanwise varying heating-cooling pattern on the surface of a compression ramp. The ablative wall reacts to these variations in the fluid, resulting in enhanced ablation rates along heating streaks and suppressed ones along cooling streaks. Mass blowing rates and heat fluxes follow similar patterns. An increase in the initial perturbation amplitude led to a stronger lift-up effect and a greater spanwise variation. The overall impact of ablation without recession was found to be stabilizing due to the injection of camphor gas into the boundary layer. The blowing velocity itself did not cause enough of a destabilization as one might anticipate. At a higher amplitude initial perturbation, the effect was less significant, as the primary perturbation was much more dominant than the influence of ablative blowing. 

Once the surface was allowed to recess, streamwise grooves in the shape of troughs began to form as physically dictated by the surface mass blowing rates and heat fluxes. Several recession durations were considered. The heat fluxes, mass blowing rates, and consequently the recession speeds mostly remained constant over these durations.
At higher amplitudes, the differential between the heating and cooling streaks became much more significant. Longer recession durations were considered, which showed a decrease in heat fluxes over time.
These grooves constrained the spanwise growth of the streaks to allow sustaining higher levels of instabilities. This was realized as recession having a mostly destabilizing effect, promoting perturbation growth.

With these findings, this work constitutes a forward step towards predictive high-fidelity fluid-ablation interaction simulations for transitional hypersonic flows over surfaces undergoing ablative blowing and recession.
The presented results validate the two solvers for low-temperature ablators in cold hypersonic flows.
The general model framework is readily applicable also to high-enthalpy conditions for atmospheric entry applications.

\appendix
\section{Transport Coefficients} \label{sec:app_trans_coefs}

As mentioned in~\autoref{sec:numerics}, NASA polynomials are used to compute the specific heat at constant pressure and enthalpy of the gaseous species. For example, the dimensionless enthalpy is computed as
\begin{equation}
    \frac{H}{RT} = -a_1 T^{-2} + a_2 \frac{\ln(T)}{T} + a_3 + \frac{a_4}{2} T + \frac{a_5}{3} T^2 + \frac{a_6}{4} T^3 + \frac{a_7}{5} T^4 + \frac{b_1}{T}
\end{equation}
where $b_1$ is the integration constant. Specific heat at constant pressure, $c_p$, is obtained by differentiation with respect to $T$. The coefficients for air are taken from the NASA Glenn database \cite{mcbride2002nasa}. Note, below 200 K the air polynomial is reduced to $H=1002.3875 T -303277.617$. The coefficients for camphor were obtained by curve fitting a quantum chemical (QC) simulation~\cite{zibitsker2024_camphorIJHMT}. All coefficients are provided in \autoref{tab:nasa_enthalpy_coeffs} below.
\begin{table}[htpb]
    \centering
    \caption{NASA polynomial coefficients for air and camphor.}
    \label{tab:nasa_enthalpy_coeffs}
    \begin{tabular}{lcc}
        \toprule
        \textbf{Coefficient} & \textbf{Air} & \textbf{Camphor\textsuperscript{*}} \\
        \midrule
        $a_1$ & $10099.5016$ & $0$ \\
        $a_2$ & $-196.827561$ & $0$ \\
        $a_3$ & $5.00915511$ & $8.3259841$ \\
        $a_4$ & $-5.76101373 \times 10^{-3}$ & $-1.1380846 \times 10^{-2}$ \\
        $a_5$ & $1.06685993 \times 10^{-5}$ & $3.43046471 \times 10^{-4}$ \\
        $a_6$ & $-7.94029797 \times 10^{-9}$ & $-5.5344526 \times 10^{-7}$ \\
        $a_7$ & $2.18523191 \times 10^{-12}$ & $2.87556562 \times 10^{-10}$ \\
        $b_1$ & $-176.796731$ & $-36206.0071$ \\
        \bottomrule
        \multicolumn{3}{l}{\small \textsuperscript{*} Curve fit based on QC simulation.}
    \end{tabular}
\end{table}

Viscosity and thermal conductivity for camphor were computed precisely as in Ref.~\cite{zibitsker2024_camphorIJHMT} with the coefficients for the third order polynomial given in~\autoref{tab:camphor_visc_cond}. 
\begin{table}[htpb]
\centering
\caption{Camphor viscosity and thermal conductivity coefficients~\cite{zibitsker2024_camphorIJHMT}.}
\label{tab:camphor_visc_cond}
\begin{tabular}{l c c c c}
\toprule
Coefficient & $\mu$ & $k$ \\
\midrule
$c_0$ & $1.5210 \times 10^{-6}$ & $-8.4400 \times 10^{-3}$ \\
$c_1$ & $3.0470 \times 10^{-8}$ & $6.6000 \times 10^{-5}$ \\
$c_2$ & $0.0000$ & $6.8800 \times 10^{-8}$ \\
$c_3$ & $0.0000$ & $2.6600 \times 10^{-11}$ \\
\bottomrule
\end{tabular}
\end{table}

A Blottner curve fit was used to compute the viscosity of air as
\begin{equation}
    \mu(T) = 0.1 \exp \left[ \left(A \ln{T} + B \right) \ln{T} + C \right]
\end{equation}
where coefficients are listed in~\autoref{tab:blottner_coeffs}.
\begin{table}[htpb]
    \centering
    \caption{Blottner viscosity curve fit coefficients.}
    \label{tab:blottner_coeffs}
    \begin{tabular}{lc}
        \toprule
        \textbf{Coefficient} & \textbf{Value} \\
        \midrule
        $A$ & $-0.0344624$ \\
        $B$ & $1.08800$ \\
        $C$ & $-13.6715$ \\
        \bottomrule
    \end{tabular}
\end{table}
The thermal conductivity for air was computed by the following cubic polynomial in temperature based on a Blottner curve fit with coefficients given in~\autoref{tab:air_cond_coeffs}.
\begin{equation}
    k(T) = c_1 + c_2 T + c_3 T^2 + c_4 T^3.
\end{equation}
\begin{table}[htpb]
    \centering
    \caption{Polynomial coefficients for Air thermal conductivity.}
    \label{tab:air_cond_coeffs}
    \begin{tabular}{lc}
        \toprule
        \textbf{Coefficient} & \textbf{Value} \\
        \midrule
        $c_1$ & $0.00271$ \\
        $c_2$ & $9.263 \times 10^{-5}$ \\
        $c_3$ & $-6.422 \times 10^{-8}$ \\
        $c_4$ & $2.5 \times 10^{-11}$ \\
        \bottomrule
    \end{tabular}
\end{table}

The diffusivity for the air-camphor binary mixture is computed from
\begin{equation}
\mathcal{D}(T, p) = (B_1 T^2 + B_2 T + B_3)(B_4 p^{B_5}) \:,
\end{equation}
and the coefficients are given in~\autoref{tab:haskins_diff_coeffs}.
\begin{table}[htpb]
\centering
\caption{Coefficients for calculating the diffusion coefficient for air-camphor mixture valid for 200 K $\leq$ T $\leq$ 800 K and 1000 Pa $\leq$ p $\leq$ 70,000 Pa from Haskins~\cite{zibitsker2024_camphorIJHMT}.}
\label{tab:haskins_diff_coeffs}
\begin{tabular}{l c}
\toprule
Coefficient & Value \\
\midrule
$B_1$ & $0.000018912588145$ \\
$B_2$ & $0.003222763711371$ \\
$B_3$ & $-0.429233442816354$ \\
$B_4$ & $0.408695307746624$ \\
$B_5$ & $-1.000951241396257$ \\
\bottomrule
\end{tabular}
\end{table}

For calculating the sublimation source term in~\autoref{eq:KL}, the vaporization coefficient $\alpha$ is taken as $0.18$ as it was similarly used by~\citet{zibitsker2024_camphorIJHMT} and~\citet{bianchi2021numerical}.
The equilibrium vapor pressure of the ablation product is obtained from an Antoine-type correlation~\citep{zibitsker2024_camphorIJHMT} expressed as
\begin{equation}
    p_v(T) = A \exp \left[ \frac{B}{T} + C \right] \:,
\end{equation}
where the coefficients are given in~\autoref{tab:camphor_vapour}.
\revadd{Surface emissivity $\varepsilon$ in~\autoref{eq:energy_final} is taken as 0.84 for camphor.}
\begin{table}[htpb]
\centering
\caption{Camphor vapour pressure coefficients in exponential form~\cite{zibitsker2024_camphorIJHMT}.}
\label{tab:camphor_vapour}
\begin{tabular}{l c}
\toprule
Coefficients & Value \\
\midrule
$A$ & 120170 \\
$B$ & -6117.1 \\
$C$ & 12.654 \\
\bottomrule
\end{tabular}
\end{table}

\section*{Acknowledgements}
Funding provided by NSF under CBET-2146100 with Dr. R. Joslin as program manager and the ACCESS program award no. 80NSSC21K1117 with Dr. Eric Stern as program manager are gratefully acknowledged. In addition, the majority of the simulations shown here from the University of Maryland were computed on facilities provided by the University of Maryland's High-Performance Computing Cluster (Zaratan).
For simulations shown here from TU Delft, we thank the Delft High Performance Computing Centre (DHPC) for providing access to DelftBlue and the SURF Cooperative (www.surf.nl) for providing access to the National Supercomputer Snellius.

\bibliography{bib}

\end{document}